\documentclass[12pt]{article}
\usepackage[T1]{fontenc} % Use 8-bit encoding that has 256 glyphs
\usepackage{lmodern} \usepackage{microtype} % Slightly tweak font spacing for aesthetics \usepackage{soul} \usepackage{bm}
\usepackage[left=1.7cm,right=1.7cm,top=2cm,bottom=2.5 cm,columnsep=20pt]{geometry} % Document margins \addtolength{\skip\footins}{1pc plus 2pt} \usepackage{fancyhdr} % Headers and footers \pagestyle{fancy} % All pages have headers and footers \fancyhead{} % Blank out the default header \fancyfoot{} % Blank out the default footer \fancyhead[C]{} % Custom header text \renewcommand{\headrulewidth}{0pt} \fancyfoot[C]{\thepage} % Custom footer text

\usepackage{titlesec} % Allows customization of titles \titleformat{\chapter}[block]{\fontsize{30}{33.75}\bfseries\filright}{\chaptertitlename\ \thechapter\\ \vspace{0.55 cm}\noindent}{0em}{} % Change the look of the section titles \titleformat{\section}[block]{\LARGE\bfseries}{\thesection}{1em}{} % Change the look of the section titles \titleformat{\subsection}[block]{\Large\bfseries}{\thesubsection}{1em}{} % Change the look of the section titles \titleformat{\paragraph}[block]{\normalfont\bfseries}{}{1em}{}

\usepackage[dvipsnames]{xcolor} \definecolor{dark-gray}{gray}{0.20} \definecolor{gray}{gray}{0.30} \definecolor{light-gray}{gray}{0.80} \definecolor{dark-red}{rgb}{0.7,0,0} \definecolor{dark-green}{rgb}{0.1,0.4,0} \definecolor{dark-blue}{rgb}{0.3,0.3,0.7} \definecolor{light-blue}{rgb}{0.8,0.8,1}

\usepackage{cite} % [nosort] to keep the chronological order
\usepackage[pdfencoding=auto, pagebackref, psdextra]{hyperref} \hypersetup{ unicode=true, colorlinks=true, linkcolor=dark-blue, citecolor=dark-red, urlcolor=dark-green, linktoc=page
}
\usepackage{enumerate} \usepackage{graphicx} \usepackage[percent]{overpic} \usepackage{booktabs} \usepackage{bbm}
\usepackage{tikz} \usetikzlibrary{positioning, arrows.meta}
\usepackage{amsmath,amssymb,slashed,amsthm}    \numberwithin{equation}{section} \usepackage{slashed}
\usepackage{mathrsfs} \usepackage[normalem]{ulem} \usepackage{dsfont} %\usepackage{bbold}

\usepackage{xcolor} \newcommand{\blue}[1]{\textcolor{blue}{#1}} \usepackage{subcaption} \graphicspath{ {./Figures/} }

\newcommand{\cmt}[1]{}
\newcommand{\q}[1]{``#1''}
\title{}

\begin{document}

\begin{center}
{\LARGE\textbf{A class of half-BPS boundary conditions for \\ \vspace{3mm} $A_{K-1}$ circular quivers}}
%{\Large \textbf{Half-BPS Boundary Conditions and Dualities \\ \vspace{6mm} in $A_{K-1}$ circular quivers}}
\end{center}

\vspace{1 cm}

\begin{center}
Davide Bason$^a$ and Roberto Valandro$^{b,c}$

\vspace{1 cm}
{

${}^a\!\!$ {\em Yau Mathematical Sciences Center, Tsinghua University, \\ Jingzhai, Haidian District, Beijing, 100084, China}

\vspace{.3cm}

${}^b\!\!$ {\em Dipartimento di Fisica, Universit\`a di Trieste, \\ Strada Costiera 11, I-34151 Trieste, Italy}

\vspace{.3cm}

${}^c\!\!$ {\em INFN, Sezione di Trieste, Via Valerio 2, I-34127 Trieste, Italy}
}

\vskip .7cm
{\scriptsize \tt davidebason at mail dot tsinghua dot edu dot cn \\ roberto dot valandro at ts dot infn dot it
    }

\vskip 2cm
%      	{\bf Abstract }
% \vskip .1in
 \end{center}

\begin{abstract}
We study a string-motivated class of $\tfrac12$-BPS boundary conditions for 4d
$\mathcal N=2$ $A_{K-1}$ circular quiver gauge theories with gauge group
$SU(N)^K$, engineered by D4-branes suspended between NS5-branes on a circle. 
For D4-branes ending on a boundary
D6-brane sent to infinity, a single-pole ansatz reduces the BPS equations to a
constrained algebraic problem. We characterize the structure of its solutions, which
exhibit a winding phenomenon with no analogue for linear quivers, and solve
several cases explicitly in closed form. A brane-duality argument suggests
looking for the S-dual of the pure Neumann boundary condition within this
class. We show that the residual gauge symmetry can be discrete only under a
condition fixed by tracelessness, in which case it is the centre
$\mathbb Z_N$, and that solutions with this property exist for every $N$ and
$K$.
\end{abstract}

\newpage

\tableofcontents

%\newpage

\section{Introduction}

For four-dimensional gauge theories with exactly marginal couplings, bulk dualities
induce nontrivial transformations on \(\tfrac12\)-BPS boundary conditions. Making
this action explicit, and in particular identifying the S-dual of an elementary
boundary condition, is a well-posed and largely open question. The canonical example is
four-dimensional \(\mathcal N=4\) super Yang--Mills theory, where the action of
S-duality on \(\tfrac12\)-BPS boundary conditions was analyzed systematically by
Gaiotto and Witten~\cite{Gaiotto:2008sa,Gaiotto:2008ak}. One lesson of
that analysis is that duality images of elementary boundary conditions need not remain
elementary: they can involve singular boundary data, as in the Nahm-pole boundary
conditions arising when D3-branes end on D5-branes
\cite{Diaconescu:1996rk,Gaiotto:2008sa,Gaiotto:2008ak}.

In this paper we study an analogous problem for four-dimensional
\(\mathcal N=2\) \(A_{K-1}\) circular quiver gauge theories. These theories are
engineered by D4-branes suspended between NS5-branes on a
circle~\cite{Witten:1997sc,Kapustin:1998xn} and have \(K\) exactly marginal
gauge couplings. In the lift to M-theory the system is described by a single
M5-brane wrapping a punctured torus, whose modular transformations act on the
couplings~\cite{Witten:1997sc,Halmagyi:2004ju}, and in the T-dual Type~IIB
frame part of the strong--weak duality group becomes geometric.
We analyze a class of singular \(\tfrac12\)-BPS boundary conditions suggested by
the brane construction, characterize the structure of its solutions, and
determine which of them break the gauge symmetry down to a discrete subgroup.

The brane picture gives the guiding principle for the class we study. In the
Type~IIA setup the D4 segments can terminate on boundary branes, and we
consider the two simplest pure configurations, in which they end on a single
boundary NS5-brane or on a single boundary D6-brane. The former realizes a
Neumann-type boundary condition for the gauge field. For the latter, the boundary condition for the bifundamental hypermultiplets
is expected to depend on the position of the D6-brane, and we focus on the
limit in which it is sent to infinity. In this limit D4\('\)-branes are created
by the Hanany--Witten effect~\cite{Hanany:1996ie}, and the colour branes can
reach the boundary D6-brane only by recombining across the bulk NS5-branes.
This motivates singular, pole-like boundary data, which we refer to as the
D4\('\)-type boundary condition. In the T-dual Type~IIB frame, the configuration
with a single D4\('\)-brane is locally a \(\mathbb Z_K\) orbifold of D3-branes
ending on a D5-brane, the system that gives the Nahm pole in \(\mathcal N=4\)
super Yang--Mills theory.

The strong--weak duality of the circular quiver, also called S-duality in
field theory, is realized in the brane picture by a chain of string dualities.
Under this chain the boundary NS5-brane is mapped to a D4-brane that differs
from a D4\('\)-brane by a rotation acting on the Coulomb-branch scalar as the
\(U(1)_r\) symmetry. This is the brane-theoretic reason why
the D4\('\)-type single-pole class is a natural place to look for the S-dual
of pure Neumann.

The brane argument does not, however, single out a boundary condition. 
A sharper, field-theoretic requirement comes
from the \(\mathcal N=4\) case~\cite{Gaiotto:2008sa,Gaiotto:2008ak}, where the
S-dual of pure Neumann is the regular Nahm pole: this boundary condition breaks
the gauge group completely, up to its centre, which acts trivially on all
fields. The analogous requirement here is that the D4\('\)-type boundary data
have a discrete stabilizer. This is a necessary condition suggested by the
analogy, not a characterization of the dual, and, as we will see, several of
the solutions we describe satisfy it.

To make this picture concrete, we write the \(\tfrac12\)-BPS equations of the
circular quiver as a generalization of Nahm's equations and analyze them in a
single-pole sector, where the boundary condition is labelled by the residues
of the pole. The resulting algebraic system organizes the residues into
chains of joint eigenspaces, which we call \(\hat\phi\)-families. These
families can wind around the quiver, a phenomenon absent for linear quivers.
We show that a discrete stabilizer requires all family ranks to be one and
the total length of the families to saturate a bound fixed by tracelessness.
For such data the stabilizer is always the diagonal centre \(\mathbb Z_N\),
and the spectra of the Hermitian residues \(\phi_a\) are independent of the
couplings. We construct explicitly a non-winding solution, a maximal-winding
one and a maximally symmetric one, and for small \(N\) and \(K\) we enumerate
the data with discrete stabilizer and summarize the results, finding in
general several of them. For \(K\ge3\), among the data examined, only the
cyclic orbit of the maximal-winding solution has existence chambers whose
union covers the entire positive-coupling domain. Which of these solutions,
if any, is the S-dual of pure Neumann is beyond the scope of this paper:
settling the question would require the action of the duality on the fields
and couplings, as well as direct tests using protected observables.

Boundary conditions in supersymmetric gauge theories have also been studied 
directly in field theory, in particular through
hemisphere localization, gluing constructions, and exact computations of
protected BCFT data
\cite{Hama:2012bg,Dedushenko:2018aox,Dedushenko:2018tgx,Bason:2023bin,Gava:2016oep}.
The brane approach to \(\tfrac12\)-BPS boundary conditions goes back to the
D3--D5--NS5 systems~\cite{Hanany:1996ie,Gaiotto:2008sa,Gaiotto:2008ak}, and
Nahm-pole boundary conditions are the standard example of singular boundary
data governed by BPS equations
\cite{Nahm:1977tg,Diaconescu:1996rk,Gaiotto:2008sd,Mazzeo:2013zga}.
Generalized Nahm poles in affine quiver gauge theories specify the punctures
of theories of class \(\mathcal S_\Gamma\)~\cite{Heckman:2016xdl,Hassler:2017arf},
and a geometric description of brane creation through deformations of
holomorphic curves appears in~\cite{Collinucci:2026kom}. Related
lower-supersymmetry boundary conditions and their brane realizations have been
investigated recently~\cite{Huertas:2024mvy}. Other brane constructions with
eight supercharges, including M2--M5 systems and brane webs,
provide further useful points of comparison
\cite{Basu:2004ed,Berman:2009kj,Kristjansen:2021abc,Kristjansen:2024zvl,Aharony:1997bh,Hanany:1997tb}.

The paper is organized as follows. In Section~\ref{sec4dBC} we review the
construction of 4d \(\mathcal N=2\) gauge theories with \(\tfrac12\)-BPS
boundary conditions, based on the decomposition of the bulk multiplets into
3d \(\mathcal N=2\) multiplets, and specialize it to the circular quiver. In
Section~\ref{sbulk} we recall the brane engineering of the circular quiver in
Type~IIA and M-theory and in Type~IIB, together with the realization of its
strong--weak duality through the \(T_4ST_4\) chain. In Section~\ref{sboundary}
we add boundary NS5- and D6-branes, relate
them to field-theory boundary conditions, and follow them to the Type~IIB frame
and through the duality chain.
Section~\ref{sboundaryconditions} contains the field-theory analysis. After a
brief discussion of D4-branes ending on a single NS5-brane, we formulate the
D4\('\)-type boundary conditions, derive the structure of the single-pole
solutions, discuss their brane interpretation, and determine which of them
have a discrete stabilizer. We conclude in
Section~\ref{sconclusions}. Appendices~\ref{atensorconv}
and~\ref{asusyandred} collect our conventions and the supersymmetry
transformations, and Appendix~\ref{app:openstrings} reviews the open-string
analysis of branes ending on branes. Appendix~\ref{ad6bc} contains the
derivation of the structure of the single-pole solutions and the explicit
non-winding, maximal-winding and maximally symmetric solutions, together with
their stabilizers, and Appendix~\ref{achambers} determines their existence
chambers on the conformal manifold.

% 59 pages, 8 figures. v2: corrected and extended the analysis of solutions and stabilizers; revised the brane interpretation. The identification of the S-dual of pure Neumann boundary conditions remains open.

\section{4d $\mathcal{N}=2$ and $\tfrac12$-BPS boundary conditions}\label{sec4dBC}

In this section we review the construction of Lagrangian 4d $\mathcal{N}=2$ gauge theories with $\tfrac12$-BPS boundary conditions, following~\cite{Bason:2023bin} and adapting the discussion to flat space and Minkowski signature. In particular, we use the decomposition of 4d $\mathcal N=2$ vector and hypermultiplets into 3d $\mathcal N=2$ multiplets, together with the supersymmetric boundary terms and elementary Neumann/Dirichlet boundary conditions derived in~\cite{Bason:2023bin}.

Our goal is to formulate boundary conditions that preserve 3d $\mathcal{N}=2$ supersymmetry. To do so, one decomposes the bulk 4d $\mathcal{N}=2$ multiplets, restricted to the boundary, into 3d $\mathcal{N}=2$ multiplets, and then imposes boundary conditions by fixing a suitable linear combination of them to a supersymmetric background. Boundary terms must also be added so that, with the chosen boundary conditions, the action is stationary on-shell and the preserved supersymmetry is maintained.

We will only state the boundary conditions and the formulas needed below,
referring the reader to~\cite{Bason:2023bin} for further details. Since 3d
\(\mathcal N=2\) supersymmetry is kept manifest, it is convenient to use an
off-shell formalism. The corresponding on-shell description is obtained by
integrating out the auxiliary fields.

A summary of the conventions adopted throughout the paper is collected in Appendix~\ref{atensorconv}.

\subsection{Super Yang--Mills theory}

The 4d $\mathcal{N}=2$ SYM action can be written as
\begin{align}\label{eactiononflat}
S&=\frac{\text{Im}(\tau)}{4\pi}\int\,\mathcal L_{\text{YM}} + \frac{\text{Re}(\tau)}{8\pi}  \int\,\mathcal L_{\theta}~,
\end{align}
where
\begin{align}\label{eYMCSflat}
\begin{split}
&\mathcal L_{\text{YM}}=\text{Tr}\bigg(-\frac{1}{2}F^{\mu\nu}F_{\mu\nu}+4D_\mu\bar{\phi}D^\mu\phi-i\bar{\lambda}^A\bar{\slashed{D}}\lambda_A+i\lambda^A\slashed{D}\bar{\lambda}_A\\
&\ \ \ \ \ \ \ \ \ \ \ \ \ \ \ \ +2\lambda^A[\bar{\phi},\lambda_A]-2\bar{\lambda}^A[\phi,\bar{\lambda}_A]+\frac{1}{2}D^{AB}D_{AB} -4[\phi,\bar{\phi}]^2\bigg)\,,\\
&\mathcal L_{\theta}= \text{Tr}\bigg(\frac12 \varepsilon^{\mu\nu\rho\sigma} F_{\mu\nu} F_{\rho\sigma} \bigg)~,
\end{split}
\end{align}
and
\begin{align}\label{eourtau}
    \tau=\frac{4\pi}{g^2}i+\frac{\theta}{2\pi}\,.
\end{align}

The 4d $\mathcal{N}=2$ vector multiplet can be reorganized, at the boundary, into two 3d $\mathcal{N}=2$ linear multiplets. Before doing so, let us spell out the conjugation properties of the bulk fields
\begin{align}\label{econj}
\phi^*=-\bar{\phi}\,,\ A_\mu^*=A_\mu\,,\ (D_{AB})^*=-D^{AB}\,.
\end{align}

On the boundary the vector multiplet can be reorganized into two 3d $\mathcal{N}=2$ linear multiplets (see Appendix~\ref{asusyandred}), as follows
\begin{equation}\label{ecurrendDN}
\begin{aligned}
\mathcal J_{\text{D}} =&\, \Big\{2\phi_1,-i \lambda_2^+,i\lambda_1^-,-\frac{1}{2}\varepsilon_{ijk}F^{jk},-2D_\perp\phi_2-i D_{12} \Big\}\,,\\
\mathcal J_{\text{N}} =&\, \Big\{2\phi_2,\lambda_2^-,-\lambda_1^+,F_{\perp i},2D_\perp\phi_1\Big\}\,.
\end{aligned}
\end{equation}
Here, in order to be consistent with~\cite{Hama:2012bg}, we have defined the following quantities
\begin{equation}\label{ephi1phi2lamdapm}
	\phi_1 = \frac{\phi+\bar\phi}{2i}\:,\quad \phi_2 = \frac{\phi-\bar\phi}{2}\,,\quad \lambda_A^{\pm} = \frac{1}{\sqrt{2}} \left( \lambda_A \pm i  \sigma^\perp \bar\lambda_A \right)\,.
\end{equation}
Notice that, as a consequence of \eqref{econj}, the combinations $\phi_1$ and $\phi_2$ are real.
%\footnote{To determine the variations of $\mathcal{J}_{\text{N}}$, we used the fact that the equations of motion for $D_{11}$ and $D_{22}$, which relate these fields to bilinears of the hypermultiplets, vanish at the boundary as a consequence of the hypermultiplet boundary conditions. This remains true in general, even in the presence of additional boundary interactions, since the boundary condition preserves the Cartan subgroup of the $SU(2)_{\mathcal{R}}$ symmetry under which these components are charged; consequently, they cannot appear in the linear multiplet.}

The 3d $\mathcal{N}=2$ covariant boundary conditions are formulated by fixing a linear combination of these multiplets to a supersymmetric background 3d $\mathcal{N}=2$ linear multiplet. 
In the present setting, the most general such background takes the form
\begin{align}\label{eJbg}
\mathcal{J}_{\text{bg}}=\{a,0,0,0,0\}\,,
\end{align}
where $a$ is a Hermitian matrix that fixes the boundary value of the bottom component of the fixed linear multiplet. Since this field is the boundary restriction of a bulk scalar, turning on $a$ amounts to imposing a constant asymptotic value that extends into the bulk solution.\footnote{As discussed in~\cite{Dedushenko:2018aox, Dedushenko:2018tgx}, the auxiliary fields should be understood as taking their on-shell values, determined by the bulk equations of motion, when restricted to the boundary.}

The simplest supersymmetric boundary conditions are the Dirichlet (D) and Neumann (N) ones,
\begin{equation}\label{ebcvectors}
	\text{D}\,: \quad \mathcal J_{\text{D}}\big|_\partial = \mathcal{J}_{\text{bg}}\,,\qquad
	\text{and}
	\qquad
	\text{N}\,: \quad  \mathcal J_{\text{N}} + \gamma\mathcal{J}_{\text{D}}\big|_\partial = \mathcal{J}_{\text{bg}}\,,
\end{equation}
where
\begin{equation}\label{Eq: definition of gamma}
	\gamma = \frac{\text{Re}(\tau)}{\text{Im}(\tau)}\,.
\end{equation}
The non-periodic dependence on the $\theta$ angle is justified by the fact that, in the presence of a boundary, the theory is no longer periodic under $\theta\to\theta+2\pi$.

These boundary conditions are the supersymmetric extensions of the ordinary Dirichlet and Neumann boundary conditions for the gauge field. Indeed, they imply
\begin{equation}\label{Eq: bndry conditions for fluxes}
	\text{D}\,: \quad F_{ij} = 0
	\qquad
	\text{and}
	\qquad
	\text{N}\,: \quad F_{\perp i} - \frac{1}{2}\gamma \varepsilon_{ijk} F^{jk}=0\,.
\end{equation}

There are additional boundary terms that are required to preserve 3d $\mathcal N=2$ supersymmetry~\cite{Bason:2023bin}
\begin{equation}\label{eactiononHS4}
	S_{\partial} =  \frac{\text{Im}(\tau)}{4\pi} \int_\partial\, \mathcal{L}_{\text{YM}}^{\partial}
	+ \frac{\text{Re}(\tau)}{8\pi}  \int_\partial\,\mathcal{L}_{\theta}^{\partial}~,
\end{equation}
where
\begin{equation}\label{eYMbr}
\begin{aligned}
&\mathcal L_{\text{YM}}^\partial = \text{Tr}\Big[8 \phi_2 \left( D_\perp \phi_2 + \tfrac{i}{2} D_{12}\right) - \lambda_{1} \lambda_2 - \bar\lambda_1 \bar\lambda_2\Big]\:,\\
&\mathcal L_{\theta}^{\partial} = 2\text{Tr}\Big[ 8 \phi_1 \left(D_\perp \phi_2+\tfrac{i}{2} D_{12}\right) +i \lambda_1 \lambda_2 -i \bar\lambda_1 \bar\lambda_2 -\lambda_1 \sigma^\perp \bar\lambda_2+ \lambda_2\sigma^\perp\bar\lambda_1 \Big]\:.
\end{aligned}
\end{equation}
As shown in~\cite{Bason:2023bin}, once these boundary terms are included and the
boundary conditions make the action stationary on-shell, compatibility with the
3d \(\mathcal N=2\) multiplet structure is equivalent to the vanishing of the
supersymmetry variation of the action, that is, to the vanishing at the
boundary of the normal component of the supercurrent of the preserved
supercharges. This is the criterion used in~\cite{Gaiotto:2008sa,Gaiotto:2008ak}.

The elementary boundary conditions above admit further supersymmetric deformations. An important example is provided by Neumann boundary conditions coupled to additional boundary degrees of freedom carrying a conserved current, so that the boundary value of the bulk gauge field gauges a flavor symmetry of the boundary theory.
%Concretely, one may
% consider a canonical 3d matter action coupled to 3d gauge fields
%~\cite{Closset:2012ru}, treating the latter as the restriction of the 4d
% gauge multiplet:\footnote{The multiplet $\mathcal{J}_{\text{N}}$ can be
% seen as the dualization of this vector multiplet through the map
% \eqref{evectortolinear}.}
% \begin{align}\label{e3dmapvector}
% \begin{split}
% &A_i=A_i\,,\\
% &\sigma=-i(\phi+\bar{\phi})=2\phi_1\,,\\
% &\lambda=-\frac{1}{\sqrt{2}}(\lambda_1-i\sigma^\perp\bar{\lambda}_1)\,,\qquad
% \tilde{\lambda}=-\frac{1}{\sqrt{2}}(\lambda_2+i\sigma^\perp\bar{\lambda}_2)\,,\\
% &D=-iD_{12}-D_\perp(\phi-\bar{\phi})=-iD_{12}-2D_\perp\phi_2\,.
% \end{split}
% \end{align}
% This coupling gauges the corresponding boundary flavor symmetry, and the
% Neumann boundary condition in \eqref{ebcvectors} is accordingly shifted by
% the 3d $\mathcal{N}=2$ linear multiplet associated with the boundary
% current.

\subsection{Adding hypermultiplets}

The action for the hypermultiplets can be written in terms of
the scalars $q_{IA}$ and fermions $\psi_I$, where
$I,J,\ldots=1,\ldots,2N_F$ are indices in the fundamental
representation of $USp(2N_F)$, lowered by the symplectic form
$\Omega_{IJ}=-\Omega_{JI}$ and raised by $\Omega^{IJ}$,
with $\Omega_{IJ}\Omega^{JK}=\delta_I^K$.

In order to couple the hypermultiplets to the gauge fields, we gauge a subgroup of $\mathfrak{usp}(2N_F)$. The commutant is the new flavor symmetry group. We take the gauge-algebra matrices to be embedded in $\mathfrak{usp}(2N_F)$, denoting them, for instance, as $\phi_I^{\ J}$. As elements of $\mathfrak{usp}(2N_F)$, they satisfy
\begin{align}\label{euspn}
\phi^I_{\ J}\doteq\Omega^{IK}\Omega_{JL}\phi_K^{\ L}=\phi_J^{\ I}\,.
\end{align}
In the case where we aim to preserve only half of the bulk supercharges (see \eqref{ehyperssusyflat} and the related comments), the action for hypermultiplets can be written in an off-shell formalism as
\begin{align}\label{ehyperaction4dflat}
\begin{split}
S_{\text{hyp}}=\int &-\frac{1}{2}D_\mu q^{IA} D^\mu q_{IA}+q^{IA}\{\phi,\bar{\phi}\}_I^{\ J}q_{JA}-\frac{i}{2} q^{IA}D_{AB,I}^{\ \ \ \ \ J} q_J^B+\frac{i}{2}\bar{\psi}^I \bar{\sigma}^\mu D_\mu\psi_I \\
&+\frac{1}{2}\psi^I\phi_I^{\ J}\psi_J-\frac{1}{2}\bar{\psi}^I\bar{\phi}_I^{\ J}\bar{\psi}_J+q^{IA}\lambda_{A,I}^{\ \ \ J}\psi_J-\bar{\psi}^I\bar{\lambda}_{A,I}^{\ \ \ J} q^A_J+\frac{1}{2}F^{IA}F_{IA}~.
\end{split}
\end{align}
For later use, let us also record the conjugation properties of the hypermultiplet fields
\begin{align}\label{escalarconjugancy}
(q_{IA})^*=q^{IA}\,,\ (F_{IA})^*=-F^{IA}\,.
\end{align}
The $F_{IA}$ do not play any role in this context, but they become important in the presence of bulk--boundary interactions involving the hypermultiplets.

Let us focus on a theory with only a single 4d hypermultiplet. Its restriction to the boundary is organized into a 3d $\mathcal{N}=2$ chiral multiplet ${\Phi}_{11}$ and an anti-chiral multiplet ${\Phi}_{12}$, the conjugate of the chiral multiplet ${\Phi}_{21}$,
\begin{align}
\begin{split}\label{eq:bchyp}
{\Phi}_{11}&=\Big\{q_{11},\ -\frac{i}{2}(\psi_1+i\sigma^\perp\bar{\psi}_1),\ D_\perp q_{12}+2\phi_2 q_{12}-i F_{11}\Big\}\equiv {\Phi}_{22}^\dagger~,\\
{\Phi}_{12}&=\Big\{q_{12},\ \frac{i}{2}(\psi_1-i\sigma^\perp\bar{\psi}_1),\ -D_\perp q_{11}+2 \phi_2 q_{11}-i F_{12}\Big\}\equiv - {\Phi}_{21}^\dagger~,
\end{split}
\end{align}
where $\phi_2$ is defined in \eqref{ephi1phi2lamdapm}.

The boundary conditions require setting either ${\Phi}_{11}$ or ${\Phi}_{12}$ to the following background chiral multiplet
\begin{align}\label{ephibg}
\Phi_{1\bullet}\big|_\partial=\Phi_{\text{bg}}=\{b,0,0\}\,,
\end{align}
for $b\in\mathbb{C}$. We refer to these as boundary conditions $1$ and $2$, respectively. The generalization to \(N_F\) hypermultiplets is obtained by choosing \(N_F\) components labelled by \(\tilde I\) among the \(2N_F\) components labelled by \(I\), and by fixing, for each such component, either \[ \Phi_{\tilde I 1}=(\Omega^{\tilde I J}\Phi_{J2})^\dagger \qquad \text{or} \qquad \Phi_{\tilde I 2}=-(\Omega^{\tilde I J}\Phi_{J1})^\dagger \] to a background.

The boundary term that needs to be added is~\cite{Bason:2023bin}
\begin{align}\label{ehyperboundary}
S_{\text{hyp},\partial}=\int d^3x\sqrt{h}\bigg[-\frac{1}{8}\psi^I\psi_I+\frac{1}{8}\bar{\psi}^I\bar{\psi}_I+\frac{i}{4}\psi^I\sigma^\perp\bar{\psi}_I-2q_{A=1}^I(\phi_2)_I^{\ J}q_{J,A=2}\bigg]\,.
\end{align}

The boundary conditions of the hypermultiplets can also be deformed by interactions. These deformations can be implemented by coupling the fluctuating chiral multiplet above to boundary degrees of freedom. A key difference with respect to purely three-dimensional chirals is that, at the UV fixed point, these fields originate from four-dimensional bulk fields and therefore have dimension \(1\) rather than \(\tfrac12\). Consequently, the renormalizable deformations are only superpotential couplings, which we denote by \(W_{\mathrm{3d}}\). Such couplings shift the boundary condition in \eqref{ephibg} by the chiral multiplet associated with \(\partial W_{\mathrm{3d}}/\partial q\), where \(q\) is the bottom component of the three-dimensional chiral multiplet obtained from the bulk hypermultiplet.

\subsection{Relation to the $A_{K-1}$ circular quivers}

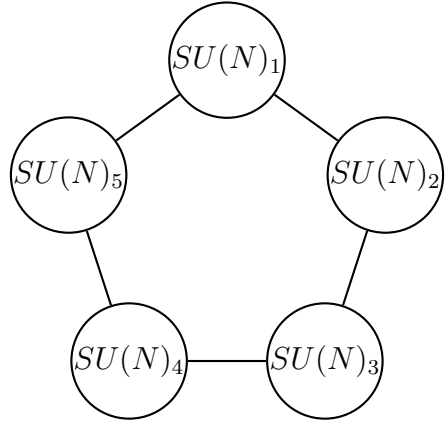
\begin{figure}[h]
\centering
\begin{tikzpicture}[
  gauge/.style={circle, draw, thick, inner sep=0pt, minimum size=1.1cm},
  bifund/.style={thick},
  every label/.style={font=\small}
]
  \def\K{5}        % number of nodes
  \def\R{2.2cm}    % ring radius

  % Place nodes on a circle
  \foreach \a in {1,...,\K}{
    \pgfmathsetmacro{\angle}{90 + 360/\K * (1 - \a)}
    \node[gauge] (N\a) at (\angle:\R)
      {$SU(N)_{\a}$};
  }

  % Draw bifundamentals between adjacent nodes (cyclically)
  \foreach \a in {1,...,\K}{
    \pgfmathtruncatemacro{\b}{mod(\a,\K)+1}
    \draw[bifund] (N\a) -- (N\b);
  }
\end{tikzpicture}
\caption{The $A_{K-1}$ circular quiver ($K=5$). Each node carries an $SU(N)$ $\mathcal{N}=2$ vector multiplet. Each link between the $a$-th and $(a+1)$-th node (with $a\sim a+K$) represents a 4d $\mathcal{N}=2$ hypermultiplet in the $(\mathbf{N}_a,\bar{\mathbf{N}}_{a+1})$ representation.}
\label{fquiverdepiction}
\end{figure}

In this paper we consider a circular quiver with $K$ nodes listed by the \textit{cyclic} index $a\in\{1,\,\hdots,\,K\}$, usually depicted by a diagram such as the one in Figure~\ref{fquiverdepiction}.

In the SYM action, there will be several terms like those in \eqref{eactiononflat} and \eqref{eactiononHS4}, each associated with its own gauge coupling $\tau_a$.

We embed the bifundamental indices of the hypermultiplets as follows. Denote by $\Psi_{i_a}$ a field in $\mathbf{N}_a$ of the $a^{\text{th}}$ quiver node, and by $\Psi^{i_a}$ a field in $\mathbf{\bar{N}}_a$, so that $(\Psi_{i_a})^*=\bar{\Psi}^{i_a}$. The indices $\tilde{I}, \tilde{J}, \dots$ are then used to collectively label all bifundamentals:
\begin{align}\label{eItildeexpansion}
\Psi_{\tilde{I}}=\begin{pmatrix}
\Psi_{i_1}^{\ i_2}\\
\vdots\\
\Psi_{i_K}^{\ i_1}
\end{pmatrix}\,.
\end{align}
By choosing
\begin{align}
\Omega_{\tilde{I}\tilde{J}}=\begin{pmatrix}
0 & -\delta_{\tilde{I}}^{\tilde{J}}\\
\delta_{\tilde{J}}^{\tilde{I}} & 0
\end{pmatrix}
\end{align}
we can write, following \eqref{escalarconjugancy},\footnote{For more general setups, such as those in~\cite{Witten:1997sc}, one may also include fundamental hypermultiplets. In that case $\Psi_{\tilde{I}}$ in \eqref{eItildeexpansion} would run also over these $\Psi_{j_a}$ fields in the fundamental.}
\begin{align}\label{ehyperbifundbase}
q_{IA}=\begin{pmatrix}
q_{\tilde{I}A}\\
\varepsilon_{AB}(q_{\tilde{I}B})^*
\end{pmatrix}\,.
\end{align}
We then fix the embedding of the generators such that the gauge-covariant derivative acts as follows:
\begin{align}
D_\mu \Psi_{i_a}^{\ i_{a+1}}=\partial_\mu\Psi_{i_a}^{\ i_{a+1}}-i (A^{[a]}_\mu)_{i_a}^{\ j_a}\Psi_{j_a}^{\ i_{a+1}}+i\Psi_{i_a}^{\ j_{a+1}} (A^{[a+1]}_\mu)_{j_{a+1}}^{\ i_{a+1}}\,,
\end{align}
where $(A^{[a]}_\mu)_{i_a}^{\ j_a}=A^{x_a} T_{x_a,i_a}^{\ \ j_a}$, with $x_a$ the adjoint index for $SU(N)_a$ and $T_{x_a,i_a}^{\ \ j_a}$ a $\mathbf{N}_a$ generator.

We thus write
\begin{align}
T_{x_a,I}^{\ \ \ \ J}=\begin{pmatrix}
T_{x_a,\tilde{I}}^{\ \ \ \ \tilde{J}} & 0\\
0 & -T_{x_a,\tilde{J}}^{\ \ \ \ \tilde{I}}
\end{pmatrix}\,,
\end{align}
which guarantees \eqref{euspn}, and
\begin{align}
T_{x_a,\tilde{I}}^{\ \ \ \ \tilde{J}}=\text{diag}\{0,\,\hdots,\,0,\,\underset{(a-1)^{\text{th}}\text{-entry}}{\underbrace{-\delta_{i_{a-1}}^{j_{a-1}}T_{x_a,j_a}^{\ \ \ i_a}}},\,\underset{a^{\text{th}}\text{-entry}}{\underbrace{T_{x_a,i_a}^{\ \ \ j_a}\delta_{j_{a+1}}^{i_{a+1}}}},\,0,\,\hdots,\,0\}\,.
\end{align}

\section{Brane engineering of $A_{K-1}$ circular quivers}\label{sbulk}

In this section we review the brane constructions of the \(A_{K-1}\) circular quiver shown in Figure~\ref{fquiverdepiction}. This theory provides a natural 4d $\mathcal N=2$ counterpart of the 4d $\mathcal N=4$ brane setups studied by Gaiotto and Witten~\cite{Gaiotto:2008sa,Gaiotto:2008ak}: it has exactly marginal gauge couplings, a nontrivial duality group, and a string-theory realization in which part of the duality action becomes geometric. We first fix the Type~IIA description of the bulk theory and then summarize its strong--weak duality in the form used later in the paper. We also briefly recall the Type~IIB orbifold frame, which provides a complementary realization of the same circular quiver.

For convenience, we denote the 10$^{\text{th}}$ direction in M-theory by \q{$\#$}; the reader is referred to Appendix~\ref{atensorconv} for the remaining conventions.

\subsection{Brane engineering in Type~IIA and M-theory}\label{stypeiiamth}

Table~\ref{tIIA} summarizes the Type~IIA construction of the theory considered in this article, depicted in Figure~\ref{ftypeIIA}.
\begin{table}[h!]
	\centering
	\begin{tabular}{c|cccccccc}
		 & $012$ & $3$ & $\mathbf{\underline{4}}$ & $5$ & $6$ & $7$ & $8$ & $9$  \\
		 \hline
		D4 & $\bullet$ & $\bullet$ & $\bullet$ & & & & & \\
		NS5 & $\bullet$ & $\bullet$ & & $\bullet$ & $\bullet$ & & &
	\end{tabular}
	\caption{The table shows the directions along which the branes in Figure~\ref{ftypeIIA} extend. The 4-direction is a circle.}
	\label{tIIA}
\end{table}
\begin{figure}[h!]
\begin{center}
\begin{tikzpicture}

  \draw[thick] (0,0) circle (2.5cm);

  % NS5-branes: 4 equally spaced (change angles for different K)
  \foreach \ang in {90, 0, 270, 180}{
    \draw[thick] (\ang:2.5cm) circle (0.28cm);
    \draw[thick] (\ang:2.5cm) +(-0.20,-0.20) -- +(0.20, 0.20);
    \draw[thick] (\ang:2.5cm) +(-0.20, 0.20) -- +(0.20,-0.20);
    \node[font=\small\itshape] at (\ang:3.1cm) {\textup{NS5}};
  }

  % D4 labels at arc midpoints
  \foreach \ang in {45, 315, 225, 135}{
    \node[font=\small\itshape] at (\ang:2.1cm) {\textup{D4}};
  }

\end{tikzpicture}
\end{center}
\caption{Stacks of $N$ D4-branes extend along 01234 and are suspended between
neighbouring NS5-branes arranged along the compact $x^4$ direction
($K=4$ in the figure). Each D4 segment supports a gauge node;
bifundamental hypermultiplets connect neighbouring segments
across the NS5-branes.}
\label{ftypeIIA}
\end{figure}
This is the standard Type~IIA realization of four-dimensional \(\mathcal N=2\) quiver gauge theories. As explained in~\cite{Witten:1997sc}, the relative \(U(1)\) factors are fixed by the finite-energy condition on the NS5-brane fluctuations, while the diagonal \(U(1)\) decouples. Therefore, the theory effectively consists of $SU(N)$ gauge nodes. Strings on the D4 segments describe the gauge degrees of freedom, while strings stretching across the NS5-branes correspond to bifundamental hypermultiplets.

Denoting by $\tau_i$ the complexified gauge coupling at each node, $\text{Im}(\tau_i)$ is proportional, in the Type~IIA limit, to the distance between two adjacent NS5-branes along the 4-direction. This relation follows from the dimensional reduction of the 5d SYM theory living on the suspended D4-branes. The corresponding real parts are most naturally understood after lifting to M-theory, to which we now turn.

\

\begin{figure}[h!]
\begin{center}
\includegraphics[scale=1]{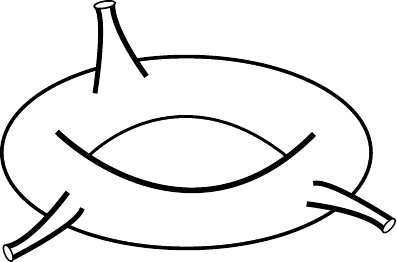}
\end{center}
\caption{The figure represents the M5-brane wrapping the \(4\#\)-torus. The spikes correspond to asymptotic regions in which the M5 leaves the torus directions. In the Type~IIA limit, these spikes become the NS5-branes, while the remaining parts project to D4 segments.}
\label{fM}
\end{figure}

Type~IIA string theory lifts to M-theory by adding an eleventh direction $x^\#$, compactified on a circle. A D4-brane is then interpreted as an M5-brane wrapped on the M-theory circle, while an NS5-brane lifts to an M5-brane localized along $x^\#$. In the present configuration the $x^4$ direction is already compact; therefore the lift naturally combines the two compact directions $x^4$ and $x^\#$ into a two-torus. The Type~IIA D4--NS5 system is consequently reinterpreted as a single M5-brane with non-trivial worldvolume, whose projection to the $(x^4,x^\#)$ torus encodes the positions of the NS5-branes and the complexified gauge couplings~\cite{Witten:1997sc}. The setup is summarized in Table~\ref{tM} and illustrated in Figure~\ref{fM}.

\begin{table}[h!]
\centering
	\begin{tabular}{c|ccccccccc}
		 & $012$ & $3$ & $\mathbf{\underline{4}}$ & $5$ & $6$ & $7$ & $8$ & $9$  & $\mathbf{\underline{\#}}$\\
		 \hline
		M5 & $\bullet$ & $\bullet$ & $\bullet$ & & & & & & $\bullet$\\
		M5 & $\bullet$ & $\bullet$ & & $\bullet$ & $\bullet$ & & & &
        \end{tabular}
	\caption{M-theory lift of the Type~IIA configuration in Table~\ref{tIIA}. From the M-theory point of view, what here appear as two families of M5s are actually a single M5, as depicted in Figure~\ref{fM}.}
	\label{tM}
\end{table}

The asymptotic positions of the M5 spikes on the complex $4\#$-torus determine the gauge couplings. Let $p_i$ denote the marked points corresponding, in the Type~IIA limit, to the NS5-branes. We normalize the torus so that
\begin{align}\label{etorusidentification}
    p_i\sim p_i+1\sim p_i+\tau\,,
\end{align}
and choose a fundamental domain together with an ordering such that
\begin{align}\label{eorderpi}
\text{Im}(p_{i+1})>\text{Im}(p_i)\,,
\end{align}
with
\begin{align}\label{eperiodicitycondp}
p_{i+K}=p_i+\tau\,.
\end{align}
The complexified couplings are then
\begin{align}\label{ecomplexcouplingquiver}
\tau_i=p_{i+1}-p_i\,,
\end{align}
and therefore
\begin{align}\label{esumoftau}
\tau=\overset{K}{\underset{i=1}{\sum}}\tau_i\,.
\end{align}
In terms of the physical circle coordinates this can be written as
\begin{align}\label{eNS5position}
p_i=\frac{x^{\#}_i}{L^{(\#)}}+\Big(\frac{i}{g_s}+\frac{\theta}{2\pi}\Big)\frac{x^{4}_i}{L^{(4)}}\,,
\end{align}
so that
\begin{align}\label{ecomplexstructure}
\tau=\frac{i}{g_s}+\frac{\theta}{2\pi}\,.
\end{align}
Thus the M-theory torus geometrizes both the individual gauge couplings $\tau_i$ and their total sum $\tau$.

 More generally, the punctured torus has a larger mapping class group, whose action may also move the marked points and therefore reshuffle the individual gauge couplings of the circular quiver. In this paper we will only use the modular transformation that exchanges the two torus cycles and acts as a strong--weak duality on the quiver couplings. We refer to it as the S-duality of the circular quiver.

Defining a generic point on the torus by
\begin{align}
z=\sigma_1+\tau\sigma_2\,,\qquad \sigma_i\sim \sigma_i +1\,,
\end{align}
the modular action is
\begin{align}\label{eSactiononpoints}
z\longrightarrow z'=\sigma_1'+\frac{a\tau+b}{c\tau+d}\sigma_2'=\frac{z}{c\tau+d}\,,
\end{align}
with
\begin{align}\label{egenericsl2z}
\begin{pmatrix}
    a & b\\
    c & d
\end{pmatrix}\in SL(2,\mathbb{Z})\,.
\end{align}
Specializing to the $S$ generator, we take $a=d=0$ and $b=-c=-1$.\footnote{\label{fdifferentS}Choosing instead $a=d=0,\ b=-c=1$ would lead to $p_i'=-\tfrac{p_i}{\tau}$. This differs only by conventions and does not change the physical content of the transformed theory.} If $p_i\,,\ i\in\{1,\hdots,K\}$ are taken in the chosen fundamental domain, then
\begin{align}\label{eSdualonpoints}
p_i'=\frac{p_i}{\tau}\,,\qquad p^{'}_{i+K}=p_i^{'}-\frac 1\tau\,.
\end{align}
Using \eqref{ecomplexcouplingquiver}, one finds
\begin{equation}\label{eSdualfromMtheory}
\tau_i'=\frac{\tau_i}{\tau}\,,\qquad i\in\{1,\hdots, K-1\}\qquad\mbox{and}\qquad
\tau_K'=-\frac 1\tau-\overset{K-1}{\underset{i=1}{\sum}}\frac{\tau_i}{\tau}\,.
\end{equation}
The transformation \eqref{eSdualfromMtheory} should be understood up to the
natural relabeling of the punctures (equivalently, of the quiver nodes). Indeed,
after the modular transformation the naive differences $p'_{i+1}-p'_i$ taken
with the original ordering need not all lie in the standard chamber with
$\text{Im}(\tau_i')>0$. This is not an inconsistency: one should simply re-order
the transformed marked points, or equivalently compose with a permutation in the
extended duality group~\cite{Halmagyi:2004ju}, before reading off the physical
couplings. This re-ordering is not an additional choice: it is fixed by the
imaginary parts of the transformed marked points. In particular, the last
coupling plays a distinguished role because of the periodicity condition.
When two transformed marked points have the same imaginary part, the
corresponding coupling lies at the boundary $\text{Im}(\tau_i')=0$ of the space
of couplings, whatever the ordering. This is what happens when all the
$\theta$-angles vanish: the $\tau_i$ are then purely imaginary, so that, up to a
translation, all the $p_i$ lie on the imaginary axis and all the $p_i'=p_i/\tau$
are real. Hence $K-1$ of the transformed couplings lie at
$\text{Im}(\tau_i')=0$, while the last one, which also receives the
contribution $-1/\tau$ in \eqref{eSdualfromMtheory}, remains finite.

\subsection{Brane engineering in Type~IIB}\label{stypeIIBgeneral}

It is useful to briefly recall the dual Type~IIB description, since in that frame the same transformation is simply the ordinary $SL(2,\mathbb{Z})$ duality of Type~IIB string theory.

\begin{figure}[h]
\begin{center}
\includegraphics[width=80mm]{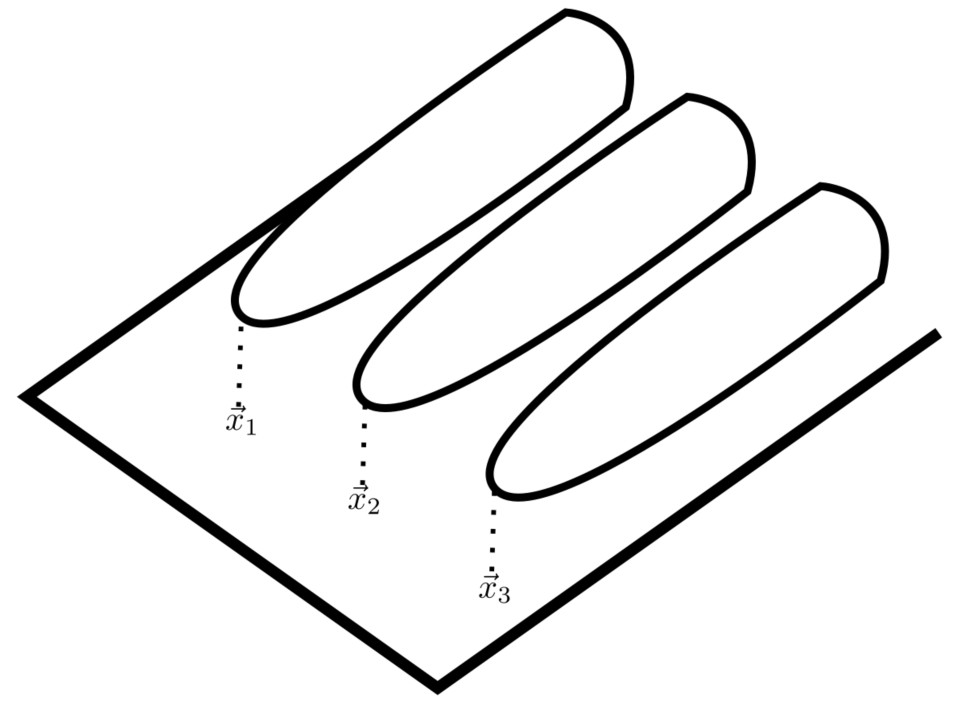}
\caption{The Taub--NUT space as an $S^1$ fibration over a plane, with several points where the fiber shrinks to zero size.}
\label{fbfluxesimage}
\end{center}
\end{figure}

T-dualizing the Type~IIA circle $x^4$ maps the NS5-branes to a Taub--NUT geometry~\cite{Gregory:1997te,Witten:2009xu},
\begin{align}
ds^2 = V(\vec{x})\, d\vec{x}^{\,2} + \frac{1}{V(\vec{x})}\left(R\, d\theta + \vec{\omega}\cdot d\vec{x}\right)^2\,,
\end{align}
with
\begin{equation}\label{eTaubNUT}
V=1+\sum_{i=1}^{K}\frac{R}{2|\vec{x}-\vec{x}_i|}\,,\qquad
\vec{\nabla}\times \vec{\omega} = \vec{\nabla} V\,.
\end{equation}
At each point $\vec{x}=\vec{x}_i$, the $S^1$ fiber shrinks to zero size. The geometry is illustrated in Figure~\ref{fbfluxesimage}.

\begin{figure}[h]
\begin{center}
\includegraphics[width=90mm]{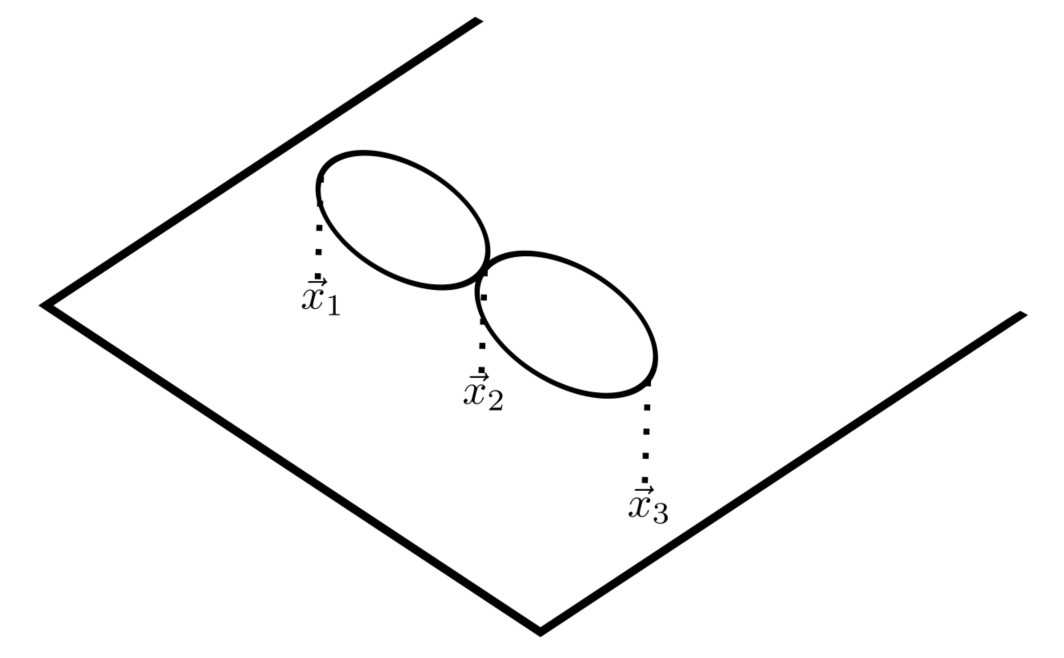}
\caption{The same Taub--NUT geometry as in Figure~\ref{fbfluxesimage}, now showing the $S^1$ fibration along a path connecting two points where the fiber collapses.}
\label{fspheres}
\end{center}
\end{figure}

The separations of the original NS5-branes along the Type~IIA circle are encoded in $B_{(2)}$ periods through suitable two-cycles. Choosing an ordering of the centers, one introduces cigars $\mathcal C_i$ ending on the degenerate fibers and identifies
\begin{align}
\mathcal{C}_{K+1} \equiv \mathcal{C}_1\,,
\end{align}
together with the two-spheres
\begin{align}\label{ebasespheres}
S^2_i = \mathcal{C}_{i+1} - \mathcal{C}_i\,, \qquad i = 1, \dots, K.
\end{align}
These are shown in Figure~\ref{fspheres}. In a suitable normalization, the $B$-field periods encode the original NS5 positions along $x^4$,
\begin{align}\label{ex4andB}
\frac{x^4_i}{L^{(4)}} = \int_{\mathcal{C}_i} B_{(2)}\,.
\end{align}
Since \(\mathcal C_{K+1}\equiv\mathcal C_1\), the two-spheres satisfy
\(\sum_{i=1}^K S^2_i=0\) in homology, and only \(K-1\) of them are
independent. Correspondingly, \eqref{ex4andB} holds modulo integers, as
periods of \(B_{(2)}\) on non-compact cycles are defined up to large gauge
transformations, consistently with \(x^4_{K+1}=x^4_1+L^{(4)}\). This is why
the last coupling in \eqref{ecouplingtypeiib} is fixed by
\(\sum_i\tau_i=\tau\) rather than by a period over \(S^2_K\).

In the ALE limit, obtained by sending the asymptotic radius of the Taub--NUT fiber to infinity, the geometry near coincident Taub--NUT centers reduces locally to the orbifold $\mathbb{C}^2/\mathbb{Z}_K$. Placing $N$ D3-branes at the singular point yields the same $A_{K-1}$ circular quiver.\footnote{The perturbative open-string description gives \(U(N)^K\)~\cite{Douglas:1996sw}. The diagonal \(U(1)\) decouples, while the relative abelian factors are broken by couplings to twisted RR fields in the Type~IIB orbifold description, equivalently by a Green--Schwarz/Stueckelberg mechanism~\cite{Morrison:1998cs}. Thus the interacting massless gauge sector is effectively \(SU(N)^K\), in agreement with the Type~IIA finite-energy argument for freezing the relative abelian factors.}

It is useful to distinguish the local data that determine the quiver theory from the global Taub--NUT asymptotics that make certain duality frames available. The local ALE limit does not change the low-energy quiver theory on D3-branes at the singular point: the gauge group and complexified couplings are still encoded in the periods of $B_{(2)}$ and $C_{(2)}$ through the vanishing two-cycles. By contrast, the Taub--NUT asymptotics are important for the global duality dictionary, since the asymptotic circle fiber is what makes it possible to relate the Type~IIB description back to the Type~IIA/M-theory frames by T-duality.

In this frame the gauge couplings are expressed in terms of the twisted periods of $B_{(2)}$ and $C_{(2)}$ through the vanishing cycles~\cite{Cachazo:2001gh},
\begin{align}\label{ecouplingtypeiib}
\begin{split}
\tau_i&=\tau\underset{S^2_i}{\int}B_{(2)}+\underset{S^2_i}{\int}C_{(2)}\,,\qquad \forall i\in\{1,\hdots,K-1\}\,,\\
\tau_K&=\tau-\overset{K-1}{\underset{i=1}{\sum}}\bigg(\tau\underset{S^2_i}{\int}B_{(2)}+\underset{S^2_i}{\int}C_{(2)}\bigg)\,,
\end{split}
\end{align}
where
\begin{align}
\tau = \frac{i}{g_s} + \frac{\theta}{2\pi}\,.
\end{align}
This matches the identification of $\tau$ already obtained in the M-theory frame.

Type~IIB string theory has its usual non-perturbative $SL(2,\mathbb Z)$ duality,
\begin{equation}\label{e:Sduality_doublet}
\tau\ \mapsto\ \frac{a\tau+b}{c\tau+d}\,,\qquad
\binom{C_{(2)}}{B_{(2)}}\ \mapsto\
\begin{pmatrix} a & b\\ c & d \end{pmatrix}
\binom{C_{(2)}}{B_{(2)}}\,,
\qquad
\begin{pmatrix} a & b\\ c & d \end{pmatrix}\in SL(2,\mathbb{Z})\,.
\end{equation}
For the $S$ generator we take $a=d=0$ and $b=-c=1$, the sign being chosen so
that \eqref{eSdualTypeIIB} reproduces \eqref{eSdualfromMtheory}. This gives
\begin{equation}\label{eSdualityTypeIIB}
\tau\ \longrightarrow\ -\frac{1}{\tau}\,,\qquad
B_{(2)}\ \longrightarrow\ -C_{(2)}\,,\qquad
C_{(2)}\ \longrightarrow\ B_{(2)}\,.
\end{equation}
Substituting into \eqref{ecouplingtypeiib} gives
\begin{align}\label{eSdualTypeIIB}
\tau_i'&=\frac{\tau_i}{\tau}\,,\qquad i=1,\dots,K-1\,,\\
\tau_K'&=-\frac 1\tau-\sum_{i=1}^{K-1}\frac{\tau_i}{\tau}\,,
\end{align}
in agreement with the M-theory result \eqref{eSdualfromMtheory}.\footnote{The re-ordering of the $p_i$ discussed below \eqref{eSdualfromMtheory} has a counterpart in the Type~IIB frame, where it is realized as a flop transition \cite{Strominger:1995cz, Greene:1995hu, Ooguri:1997ih, Uranga:1998vf, Cachazo:2001sg, Cachazo:2001gh}.}

The M-theory and Type~IIB descriptions therefore give the same action on the quiver couplings. The modular transformation of the $(x^4,x^\#)$ torus and the Type~IIB $S$-duality action on the twisted periods both send \(\tau_i \mapsto \tau_i/\tau\) for \(i=1,\ldots,K-1\), with the last coupling fixed by \(\sum_i \tau'_i=-1/\tau\).

From the Type~IIA viewpoint, the same operation is realized as the \
\begin{equation}\label{Eq:TST}
T_4 \circ S\circ  T_4
\end{equation}
chain. The first T-duality along the compact \(x^4\) direction brings the system to the Type~IIB orbifold/Taub--NUT frame, where the exchange of the two periods is implemented by the standard ten-dimensional Type~IIB \(S\)-duality. The second T-duality is needed to return to a Type~IIA brane engineering description. Equivalently, this chain is the string-theoretic way of performing the M-theory modular transformation and then reducing back to Type~IIA using the dual circle as the M-theory circle. This is the main role of the Type~IIB frame in what follows: it gives a concrete representative of the duality that can also be applied to the boundary branes.

\section{Boundary brane engineering and boundary conditions}%
\label{sboundary}

In this section we add boundary branes to the Type~IIA realization of the
\(A_{K-1}\) circular quiver and use them to identify the \(\tfrac12\)-BPS
boundary conditions studied in Section~\ref{sboundaryconditions}. Our use of
the brane construction is pragmatic: it indicates which boundary values of the
4d $\mathcal N=2$ fields are fixed and which fluctuate, and which singular
behaviour should be expected, while the boundary conditions and the BPS
equations themselves are formulated and solved in field theory.

%----------------------------------------------------------
\subsection{Boundary branes}%
\label{sboundary:setup}
%----------------------------------------------------------

\begin{figure}[h]
\begin{center}
\includegraphics[width=90mm]{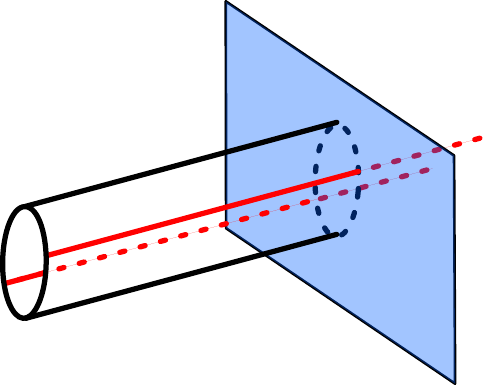}
\caption{A schematic representation of the configuration: a cylindrical stack of D4-branes intersected by two NS5-branes (in red). The D4-branes can end on boundary branes, while the NS5-branes continue straight through the intersection.}%
\label{fD4onbr}
\end{center}
\end{figure}

We start from the Type~IIA engineering of the circular quiver reviewed in
Section~\ref{sbulk}, and introduce a planar boundary at \(x^3=0\), on which the
D4 segments can terminate. This use of brane terminations as a guide to
supersymmetric boundary conditions is directly analogous to the Gaiotto--Witten
analysis of \(\tfrac12\)-BPS boundary conditions in 4d \(\mathcal N=4\) gauge
theory, based on D3-branes ending on D5- and
NS5-branes~\cite{Hanany:1996ie,Gaiotto:2008sa,Gaiotto:2008ak,Gaiotto:2008sd}.
The branes involved are listed in Table~\ref{tIIAbr}; see
Figure~\ref{fD4onbr} for a picture of the setup.

\begin{table}[h]
  \centering
  \begin{tabular}{c|cccccccc}
       & $012$ & $3$ & $\mathbf{\underline{4}}$ & $5$ & $6$
       & $7$ & $8$ & $9$ \\
  \hline
  D4   & $\bullet$ & $\bullet$ & $\bullet$ & & & & & \\
  \color{red}{NS5}  & \color{red}{$\bullet$} & \color{red}{$\bullet$} & & \color{red}{$\bullet$} & \color{red}{$\bullet$} & & & \\
  \blue{D6}  & \blue{$\bullet$} & & \blue{$\bullet$} & &
               \blue{$\bullet$} & & \blue{$\bullet$} & \blue{$\bullet$} \\
  \blue{NS5} & \blue{$\bullet$} & & \blue{$\bullet$} &
               \blue{$\bullet$} & & \blue{$\bullet$} & & \\
  \blue{D4$'$} & \blue{$\bullet$} & & & &
               \blue{$\bullet$} & \blue{$\bullet$} & & \\
  \end{tabular}
 \caption{Brane content of the boundary configuration. Bulk branes are in
 black (D4) and red (NS5), boundary branes in blue. The D4\('\)-branes are
 created when the boundary D6-brane is moved along \(x^7\) across the bulk
 NS5-branes.}%
  \label{tIIAbr}
\end{table}

Each of the boundary branes in Table~\ref{tIIAbr} preserves four supercharges
together with the bulk system, i.e.\ it realizes $\tfrac12$-BPS boundary
conditions for the 4d $\mathcal N=2$ theory on the suspended D4 segments, and
with suitable relative orientations the full set of branes in
Table~\ref{tIIAbr} preserves the same four supercharges
(Appendix~\ref{app:openstrings}). The boundary branes break the $SO(3)$
rotating $789$ to the $SO(2)$ rotating the $89$-plane, in parallel with the
breaking $SU(2)_{\mathcal R}\to U(1)_{\mathcal R}$ induced by $\tfrac12$-BPS
boundary conditions. In this paper we consider only pure configurations, in
which the colour D4-branes end either on a single boundary NS5-brane or on a
single boundary D6-brane; configurations with both kinds of boundary branes
are left for future work.

\paragraph{The boundary D6-brane and the D4\('\)-branes.}
The boundary D6-brane is transverse to \(x^3,x^5,x^7\). For the colour
D4-branes, which sit at \(x^7=x^8=x^9=0\), to end on it, it must lie at
\(x^3=0\), at the \(x^5\) position of the D4-branes, and at \(x^7=0\): it then
intersects each bulk NS5-brane along \(0126\), without ending on it. Since
\(x^7\) is the only direction transverse to both, the boundary D6-brane and a
bulk NS5-brane are a linked pair in the sense of~\cite{Hanany:1996ie}, as the
NS5- and D6-branes of~\cite{Witten:1997sc}: when the D6-brane is moved across
an NS5-brane along \(x^7\), a D4-brane extended along \(0126\) and stretched
along \(x^7\) between them is created. This is the D4\('\)-brane of
Table~\ref{tIIAbr}. It sits at the boundary in \(x^3\) and at the position of
its NS5-brane in \(x^4\), and differs from the colour D4-branes, which extend
along \(x^3\) and \(x^4\). The bulk NS5-brane, the boundary D6-brane and the
D4\('\)-brane preserve by themselves eight supercharges
(Appendix~\ref{app:openstrings}), and the creation is governed by the
conservation of the linking number~\cite{Hanany:1996ie}, as in the Type~IIA
constructions of~\cite{Elitzur:1997fh,Brandhuber:1997ta}. 
For the boundary D6-brane it reads
\begin{equation}\label{e:linkingD6}
L_{\rm D6}=\frac12\,(l-r)+\big(n_{\rm D4'}^{(r)}-n_{\rm D4'}^{(l)}\big)\,,
\end{equation}
where \(l\) and \(r\) are the numbers of bulk NS5-branes to the left and to the
right of the D6-brane along \(x^7\), and \(n^{(r,l)}_{\rm D4'}\) count the
D4\('\)-branes ending on it from the right and from the left. By the s-rule of~\cite{Hanany:1996ie}, in a
supersymmetric configuration at most one D4\('\)-brane can stretch between the
boundary D6-brane and a given bulk NS5-brane. For the D6-brane at \(x^7=0\),
\(l\) and \(r\) are not defined, and the flat configuration does not determine
\(L_{\rm D6}\). Correspondingly, the number of D4\('\)-branes left behind when
the D6-brane is sent to \(x^7\to-\infty\), \(m=L_{\rm D6}+K/2\), with
\(0\le m\le K\) by the s-rule, is a datum of the configuration.

The colour D4-branes, which stay at \(x^7=0\), cannot end on the D4\('\)-branes,
which are point-like in \(x^4\). To reach the D6-brane they must move along
\(x^7\), which a segment suspended between two NS5-branes can only do by
recombining with the adjacent segments across the NS5-branes, i.e.\ by
switching on the bifundamental hypermultiplets. When the D6-brane is sent to
\(x^7\to-\infty\), the recombined D4-branes are expected to run off along
\(x^7\), the direction of the D4\('\)-branes, near the boundary. This is the
pole-like behaviour depicted in Figure~\ref{fournahm}, and it motivates the
singular boundary conditions of Section~\ref{sendingond6}. A suitably oriented D4\('\)-brane preserves the same supercharges
as the boundary D6-brane, as checked in
Appendix~\ref{app:openstrings}.

%----------------------------------------------------------
\subsection{Relation to the boundary conditions}%
\label{srelationtostringtheory}
%----------------------------------------------------------

We now translate the brane terminations into boundary data for the 4d
$\mathcal N=2$ multiplets. The D4 segments support five-dimensional maximally
supersymmetric Yang--Mills theory, which gives the 4d $\mathcal N=2$ fields of
the circular quiver; at the boundary, their restrictions are organized into
multiplets of the preserved 3d $\mathcal N=2$ supersymmetry.

\paragraph{Vector multiplet.}
For the boundary-condition assignments below we set the gauge
theta angles to zero. We use the schematic endpoint pattern of
Appendix~\ref{app:openstrings}, equation~\eqref{endpointpattern}.
For the scalar fields, directions tangent and transverse to the
boundary brane determine the allowed and fixed displacements,
respectively. The gauge-field assignments are part of the
corresponding supersymmetric boundary conditions. Components
$A_\mu$ along the D4 worldvolume are gauge fields, while
$A_5,A_6$ are scalars describing transverse fluctuations
(see~\eqref{Eq:appC10dAM}). The entries involving the normal
gauge component $A_3$ below are understood in this schematic
multiplet description; its value alone is not an independent
gauge-invariant boundary datum.

\begin{itemize}

\item \textbf{Termination on a boundary NS5 (NS5-type).}
At the endpoint, the fixed components are
\begin{equation}\label{e:NS5type_frozen_main}
  A_3\,,\quad A_6\,,
\end{equation}
while the fluctuating boundary values include
\begin{equation}\label{e:NS5type_fluct_main}
  A_{0,1,2}\;\text{(boundary gauge field)}\,,
  \qquad A_5\;\text{(real scalar)}\,.
\end{equation}
Together with their superpartners, the fluctuating fields
organize into a 3d $\mathcal N=2$ vector multiplet. For the
bulk gauge field this corresponds to a Neumann-type boundary
condition, with its supersymmetric completion.

\item \textbf{Termination on a boundary D6 (D6-type).}
The fixed and fluctuating sets are exchanged:
\begin{equation}\label{e:D6type_frozen_main}
  A_{0,1,2}\,,\quad A_5
  \quad\text{are fixed,}
\end{equation}
while
\begin{equation}\label{e:D6type_fluct_main}
  A_3\,,\quad A_6
  \quad\text{can fluctuate.}
\end{equation}
Together with their superpartners, the latter fields organize
into a 3d $\mathcal N=2$ chiral multiplet. For the bulk gauge
field this corresponds to a Dirichlet-type boundary condition,
with its supersymmetric completion.

\item \textbf{The D6-brane at infinity (D4\('\)-type).}
We use the same vector-multiplet assignment as in the D6-type
case. Since the colour branes do not end on the D4\('\)-branes
in the naive sense, the assignment is motivated in the
T-dual frame: locally near the coincident centres, the
configuration is an orbifold of the D3--D5 termination,
with $\phi_1$ fixed and $\phi_2$ in the Nahm triple
(Section~\ref{secTNframe}).

\end{itemize}

The NS5-type condition fixes the Neumann linear multiplet
to a background value,
\begin{equation}\label{e:JNbg_brane_main}
  \mathcal{J}_{\mathrm N}\big|_{\partial}
  =\mathcal{J}_{\mathrm{bg}}\,,
\end{equation}
whose scalar component prescribes the boundary value of
$\phi_2$. The D6-type and D4\('\)-type conditions fix instead
the Dirichlet linear multiplet,
\begin{equation}\label{e:JDbg_brane_main}
  \mathcal{J}_{\mathrm D}\big|_{\partial}
  =\mathcal{J}_{\mathrm{bg}}\,,
\end{equation}
whose scalar component prescribes the boundary value of
$\phi_1$. Transformations with nontrivial boundary values
that preserve the prescribed Dirichlet data act as boundary
global symmetries.

% \begin{table}[t]
%   \centering
%   \renewcommand{\arraystretch}{1.25}
%   \begin{tabular}{c|c|c|c}
%     & \textbf{Brane termination}
%     & \textbf{Vector multiplet}
%     & \textbf{Hypermultiplets} \\
%   \hline
%   \textbf{NS5-type}
%     & D4 ends on \textcolor{blue}{NS5}
%     & $\mathcal{J}_{\mathrm N}$ fixed (Neumann-type)
%     & one chiral fixed (not selected) \\
%   \textbf{D6-type}
%     & D4 ends on \textcolor{blue}{D6} at $x^7=0$
%     & $\mathcal{J}_{\mathrm D}$ fixed (Dirichlet-type)
%     & deformation of $q_{\tilde I1}=0$ (not determined) \\
%   \textbf{D4$'$-type}
%     & \textcolor{blue}{D6} at $x^7\to-\infty$
%     & $\mathcal{J}_{\mathrm D}$ fixed (Dirichlet-type)
%     & $q_{\tilde I1}=0$, eq.~\eqref{e:D4pbc_hyper} \\
%   \end{tabular}
%   \caption{Schematic relation between boundary branes and
%   boundary conditions at vanishing gauge theta angles.
%   For the vector multiplet see also
%   \eqref{Eq: bndry conditions for fluxes}.}
%   \label{t:bc_summary_vector}
% \end{table}

\paragraph{Hypermultiplets.}
For bifundamental hypermultiplets the open-string endpoint argument does not
apply, since these fields are not perturbative open strings of a single
suspended D4 segment (see e.g.~\cite{Hanany:1996ie} and
Appendix~\ref{app:openstrings}). The brane picture is nevertheless informative
once the hypermultiplet scalars are related to the positions of the recombined
D4-branes in \(789\). Denote by \(X_a\) the boundary value of \(q_{\tilde I1}\)
on the link between the nodes \(a\) and \(a+1\), an \(N\times N\) matrix in
\((\mathbf N_a,\bar{\mathbf N}_{a+1})\), and by \(Y_a=-(q_{\tilde I2})^\dagger|_a\) the bottom component of the 3d chiral multiplet in the conjugate representation built from \(q_{\tilde I2}\),
as follows from \eqref{ehyperbifundbase} and \eqref{eq:bchyp} with
\(\varepsilon_{12}=-1\). With respect to the \(U(1)_{\mathcal R}\) preserved by
the boundary, \(X_a\) and \(Y_a\) are chiral of the same charge.

The hyperk\"ahler moment maps at the nodes,
\(\mu_{\mathbb C,a}=X_aY_a-Y_{a-1}X_{a-1}\) and
\(\mu_{\mathbb R,a}=X_aX_a^\dagger-Y_a^\dagger Y_a-X_{a-1}^\dagger X_{a-1}+Y_{a-1}Y_{a-1}^\dagger\),
are the F- and D-terms of the quiver. The
positions of the branes are read from link quantities on the regular-brane locus of the
auxiliary \(U(N)^K\) hyperk\"ahler quotient~\cite{Douglas:1996sw}. There the
colour segments recombine into \(N\) regular D4-branes, which can move along
\(x^7,x^8,x^9\), and one may take \(X_a=\mathrm{diag}(\alpha_i)\) and
\(Y_a=\mathrm{diag}(\beta_i)\) on every link, with
\begin{equation}\label{e:linkpositions}
x^8+ix^9\ \leftrightarrow\ \text{eigenvalues of }X_aY_a=\alpha_i\beta_i\,,\qquad
x^7\ \leftrightarrow\ \text{eigenvalues of }X_aX_a^\dagger-Y_a^\dagger Y_a=|\alpha_i|^2-|\beta_i|^2\,,
\end{equation}
up to normalization. The identification \eqref{e:linkpositions} holds on the regular-brane locus,
where the full \(U(N)^K\) moment maps vanish. In the \(SU(N)^K\) theory of
Section~\ref{sboundaryconditions} only their traceless parts are imposed,
since the relative \(U(1)\) factors are frozen (see
Section~\ref{stypeIIBgeneral}). Its Higgs branch is therefore larger, and
contains directions, such as baryonic ones, along which the link quantities in
\eqref{e:linkpositions} are not positions of regular branes. We use
\eqref{e:linkpositions} to interpret the boundary conditions geometrically,
not to parametrize the moduli space.

\begin{itemize}
\item \textbf{D4\('\)-type.} The D4\('\)-branes extend along \(x^7\) but not along
\(x^8,x^9\). The colour branes should therefore remain free only along \(x^7\),
and only on the side of the D4\('\)-branes. This is achieved by fixing one half
of the hypermultiplet chirals,
\begin{equation}\label{e:D4pbc_hyper}
X_a\big|_\partial=q_{\tilde I1}\big|_\partial=0\,.
\end{equation}
By \eqref{e:linkpositions}, this sets \(X_aY_a=0\), so that the recombined
branes do not move in the \(89\)-plane, and their positions along \(x^7\) are
the eigenvalues of \(-Y_a^\dagger Y_a\), which is negative semidefinite: the
branes are confined to the half-line \(x^7\le0\), along which they can run off
to the D6-brane at \(x^7\to-\infty\). As for a standard brane termination, the
directions tangent to the D4\('\)-branes remain free and the transverse ones are
fixed.

\item \textbf{D6-type.} The boundary D6-brane at \(x^7=0\) extends along
\(x^8,x^9\), so that the recombined branes can move along it in the
\(89\)-plane and \(X_aY_a\) need not vanish at the boundary. We therefore expect
the boundary condition to be a deformation of \eqref{e:D4pbc_hyper} by boundary
interactions, depending on the position of the D6-brane along \(x^7\), on its
Wilson line along \(x^4\) and on its linking number, which reduces to
\eqref{e:D4pbc_hyper} when the D6-brane is sent to infinity. We do not attempt
to determine it here.

\item \textbf{NS5-type.} The boundary NS5-brane extends along \(x^7\) but not
along \(x^8,x^9\), so it also sets \(X_aY_a=0\); unlike the D4\('\)-branes,
however, it extends along the whole \(x^7\) axis. For a single recombined
brane, with \(X_a=\alpha\) and \(Y_a=\beta\), the condition \(\alpha\beta=0\)
has two branches: \(\alpha=0\), along which \(x^7=-|\beta|^2\le0\), and
\(\beta=0\), along which \(x^7=|\alpha|^2\ge0\). Together they reproduce the
\(x^7\) axis spanned by the NS5-brane. An elementary boundary condition, which
fixes one of the two chiral multiplets, imposes only one of the two branches,
and hence captures only one half of the axis; the brane picture does not
select which one.
\end{itemize}

%----------------------------------------------------------
\subsection{The Type~IIB frame and the duality chain}
\label{secDualBoundaryConfigurations}
%----------------------------------------------------------

From the Type~IIA viewpoint, the strong--weak duality of the circular quiver is
realized by the chain \(T_4ST_4\) of \eqref{Eq:TST}. We first use the
intermediate Type~IIB frame, where the creation of D4\('\)-branes and the
D4\('\)-type configuration acquire a geometric description, analogous to the
one of~\cite{Collinucci:2026kom} for a different brane system, and
then complete the chain for the boundary NS5-brane.

%----------------------------------------------------------
\subsubsection{The boundary branes in the Taub--NUT frame}
\label{secTNframe}
%----------------------------------------------------------

Under T-duality along \(x^4\), the bulk NS5-branes become the Taub--NUT space
\(\mathrm{TN}_K\) of Section~\ref{stypeIIBgeneral}, a circle fibration with
fibre \(S^1_{\tilde 4}\) over \(\mathbb R^3_{789}\) with centres at the
positions of the NS5-branes, and the colour D4-branes become D3-branes at the
centres. The boundary D6-brane wraps \(x^4\) and becomes a D5-brane, while the
D4\('\)-branes, which are point-like in \(x^4\), become D5-branes wrapping the
fibre over a half-line in \(x^7\). The resulting configuration is summarized in
Table~\ref{tIIBbr}.

\begin{table}[h]
  \centering
  \begin{tabular}{c|cccccccc}
       & $012$ & $3$ & $\tilde 4$ & $5$ & $6$ & $7$ & $8$ & $9$ \\
  \hline
  D3   & $\bullet$ & $\bullet$ & & & & & & \\
  \color{red}{$\mathrm{TN}_K$} & & & \color{red}{fibre} & & & \color{red}{base} & \color{red}{base} & \color{red}{base} \\
  \blue{D5} {\small (from D6)} & \blue{$\bullet$} & & & & \blue{$\bullet$} & & \blue{$\bullet$} & \blue{$\bullet$} \\
  \blue{NS5} {\small (from NS5)} & \blue{$\bullet$} & & \blue{$\bullet$} & \blue{$\bullet$} & & \blue{$\bullet$} & & \\
  \blue{D5$'$} {\small (from D4$'$)} & \blue{$\bullet$} & & \blue{$\bullet$} & & \blue{$\bullet$} & \blue{$\bullet$} & & \\
  \end{tabular}
  \caption{The configuration of Table~\ref{tIIAbr} after T-duality along
  \(x^4\). The bulk NS5-branes are replaced by the Taub--NUT geometry
  \(\mathrm{TN}_K\), with fibre \(S^1_{\tilde 4}\) over \(\mathbb R^3_{789}\).
  The D5-brane obtained from the boundary D6-brane extends along \(x^8,x^9\)
  only asymptotically: it wraps a holomorphic curve in \(\mathrm{TN}_K\), see
  \eqref{e:D5curveS}.}
  \label{tIIBbr}
\end{table}

\paragraph{A geometric description of the D6 limit.}
To describe the limit, first separate the centres in the
$89$-plane, with distinct $z_a=x^8_a+ix^9_a$. In the complex
structure selected by the boundary brane conditions, the Taub--NUT
space is described by
\[
w_+w_-=\prod_{a=1}^K(z-z_a).
\]
The boundary D6-brane wraps $x^4$ and becomes a D5-brane
wrapping a holomorphic section over the $z$-plane. A useful
family of such sections is
\begin{equation}\label{e:D5curveS}
\Sigma_S:\qquad
w_+=\nu\prod_{a\notin S}(z-z_a),\qquad
w_-=\nu^{-1}\prod_{a\in S}(z-z_a),
\qquad
S\subset\{1,\ldots,K\},\quad \nu\in\mathbb C^*.
\end{equation}
Over \(z=z_a\) the fibre degenerates into two cigars, \(\{w_+=0\}\) below and
\(\{w_-=0\}\) above the centre \(a\) along \(x^7\), and, up to normalization,
\(|w_\pm|\simeq e^{\pm x^7/R}|z|^{K/2}\) at large distance from the centres,
with \(R\) the asymptotic radius of the fibre, so that \(|\nu|\) controls the
asymptotic position of the D5-brane along \(x^7\). If this position lies below
a centre \(a\in S\), where the curve meets the cigar above the centre, near
\(z=z_a\) the curve runs up along the fibre: locally it is a tube
\(S^1_{\tilde 4}\times I\), with \(I\) an interval along \(x^7\), i.e.\ the
T-dual of a D4\('\)-brane stretched between the boundary D6-brane and the bulk
NS5-brane \(a\). Moving the D6-brane across a bulk NS5-brane is thus a
continuous deformation of the curve, and the datum \(S\) does not change.

The limit $\nu\to0$ corresponds to sending the boundary D6-brane
to $x^7\to-\infty$. The limiting ideal is
\[
\left(w_+,\prod_{a\in S}(z-z_a)\right),
\]
so the part remaining at finite distance is the union of the
lower cigars over the centres $a\in S$. Under T-duality these
are the semi-infinite D4\('\)-branes left behind by the D6-brane.
Their number is
$m=|S|=L_{\rm D6}+K/2$, consistently with
\eqref{e:linkingD6} and with the s-rule, since $S\subset\{1,\ldots,K\}$.

As anticipated, this geometric interpretation of brane creation is analogous
to the mechanism described in~\cite[Section~6.1]{Collinucci:2026kom}
for D5-branes ending on D7-branes in the presence of an NS5-brane.
In the dual geometric description, the Hanany--Witten transition
is represented by a deformation of a holomorphic curve, with
an elongated tube accounting for the stretched brane.
At the level of this mechanism, the D7-brane, D5-branes and
NS5-brane of~\cite{Collinucci:2026kom} play the roles of our
boundary D6-brane, D4\('\)-branes and bulk NS5-branes,
respectively; the T-dual D6-brane in that construction
corresponds to our D5-brane.
This comparison supports the geometric interpretation of the
transition.

When the centres coincide in all of $789$, this limit is
described in the affine surface $w_+w_-=z^K$ by the ideal
$(w_+,z^m)$. For $m\ge1$ it has a single reduced support,
the lower cigar, with multiplicity $m$; for $m=0$ no
component remains at finite distance.

\paragraph{The D4\('\)-type configuration.}
For centres coinciding in all of \(789\), the geometry near the centres is the
orbifold \(\mathbb C^2/\mathbb Z_K\), with
covering coordinates \((\alpha,\beta)\), \(\mathbb Z_K\) acting as
\((\alpha,\beta)\mapsto(\omega\alpha,\omega^{-1}\beta)\), and invariants
\begin{equation}\label{e:TNcomplex}
w_+=\alpha^K\,,\qquad w_-=\beta^K\,,\qquad z=\alpha\beta\,,\qquad w_+w_-=z^K\,.
\end{equation}
The fibre acts as \((\alpha,\beta)\mapsto(e^{i\psi}\alpha,e^{-i\psi}\beta)\), and
\(x^7\) is proportional to its moment map \(|\alpha|^2-|\beta|^2\); the two
cigars over \(z=0\), \(\{w_+=0\}\) and \(\{w_-=0\}\), lie at \(x^7\le0\) and
\(x^7\ge0\) respectively. On the regular-brane locus of
Section~\ref{srelationtostringtheory}, the \(i\)-th recombined D3-brane sits at
\begin{equation}\label{e:branecoords}
z_i=\alpha_i\beta_i\,,\qquad w_{+,i}=\alpha_i^K\,,\qquad w_{-,i}=\beta_i^K\,,
\end{equation}
i.e.\ at the eigenvalues of \(X_aY_a\), of \(X_aX_{a+1}\cdots X_{a+K-1}\) and of
\(Y_{a+K-1}\cdots Y_a\)~\cite{Douglas:1996sw}. In particular, \(X_a=0\) gives
\(w_{+,i}=0\) for every \(i\): the condition \eqref{e:D4pbc_hyper} restricts the
recombined branes to the cigar \(\{w_+=0\}\).

For $m=1$, the remaining boundary brane is a single D5-brane
wrapping $0126$ times the lower cigar. On the regular-brane
locus, this motivates imposing $X_a=0$ for every link. In the covering space the
cigar lifts to the line \(\{\alpha=0\}\), so that the configuration is locally the \(\mathbb Z_K\) projection of D3-branes
ending on a D5-brane extended along
\(x^6\) and the \(\beta\)-plane, the D3--D5
system of~\cite{Gaiotto:2008sa,Gaiotto:2008ak}. The directions transverse to the
D3-branes and tangent to the D5-brane, \(x^6\) and the \(\beta\)-plane, give the
Nahm triple, i.e.\ \(\phi_2\) and \(q_{\tilde I2}\); the directions transverse to
both, \(x^5\) and the \(\alpha\)-plane, give the fixed fields \(\phi_1\) and
\(q_{\tilde I1}\), in agreement with the boundary-field assignments of
Section~\ref{srelationtostringtheory}. For $K=1$ the orbifold is absent, and locally near the tip
the configuration reduces to the standard D3--D5 system.
The precise form of the resulting BPS equations, which for \(K>1\) is not
simply the orbifold projection of Nahm's equations, is derived from field
theory in Section~\ref{sendingond6}.

The condition \(X_a=0\) fixes neither the number of boundary branes nor the
pole residues: unlike in the D3--D5 system~\cite{Gaiotto:2008sa}, the
dictionary between the two requires data, such as Chan--Paton structure and
fractional-brane charges, that are not determined by where the branes
sit~\cite{Witten:2009xu}.

%----------------------------------------------------------
\subsubsection{The image of boundary NS5-branes}
\label{secDualNS5}
%----------------------------------------------------------

Starting from the boundary NS5-brane of Table~\ref{tIIAbr}, the first T-duality
is along a longitudinal direction and again gives an NS5-brane, \(S\)-duality
maps it to a D5-brane, and the second T-duality gives a D4-brane:
\begin{equation}\label{e:NS5image}
\text{NS5}_{\partial}(012457)
\xrightarrow{\;T_4\;}\text{NS5}_{\partial}(012\tilde457)
\xrightarrow{\;S\;}\text{D5}_{\partial}(012\tilde457)
\xrightarrow{\;T_4\;}\text{D4}_{\partial}(01257)\,.
\end{equation}
The resulting configuration is supersymmetric, being the image of a
supersymmetric one. Up to a rotation by \(\pi/2\) of the \(56\)-plane, which
acts on the Coulomb-branch scalar as the \(U(1)_r\) symmetry of the bulk
superconformal theory, the image \eqref{e:NS5image} is a D4\('\)-brane,
extended along \(01267\). Since \(U(1)_r\) is a symmetry of the bulk theory,
this rotation relates equivalent boundary conditions.

The image, however, extends along the whole \(x^7\) axis: it can be viewed as
a pair of semi-infinite D4\('\)-branes, one along \(x^7\le0\) and one along
\(x^7\ge0\).
Since the NS5-type terminations realize Neumann-type boundary conditions for
the gauge field (Appendix~\ref{app:openstrings}), these observations suggest
that the D4\('\)-type class, and in particular the configuration with a single
D4\('\)-brane, may be a good place to look for the S-dual of pure Neumann
boundary conditions.

\section{A class of boundary conditions}\label{sboundaryconditions}

We now specialize the general discussion of Section~\ref{srelationtostringtheory} to a concrete and string-motivated family of $\tfrac12$-BPS boundary conditions for the $A_{K-1}$ circular quiver. We will restrict to boundary setups that admit a brane engineering description in Type~IIA and whose behavior under the bulk duality can be followed, at the level of the brane picture, through the $T_4ST_4$ chain.

We take as motivation the two pure configurations of
Section~\ref{sboundary:setup}: the bulk D4 segments terminating on a
\emph{single boundary NS5-brane}, and on a \emph{single boundary D6-brane} sent
to infinity along \(x^7\), leaving a single D4\('\)-brane behind
(Section~\ref{sboundary:setup}).
More elaborate boundary webs (mixtures of NS5- and D6-branes producing
additional boundary matter, or interfaces with extra 3d degrees of freedom) are
left for future work, and we introduce no additional boundary matter in the
field-theory problems considered below. The boundary branes indicate which of
the 3d $\mathcal N=2$ multiplets obtained from the restriction of the bulk
fields are fixed and which fluctuate, and hence motivate the boundary
conditions for the bulk vector and hypermultiplets. For the D4\('\)-type
condition the BPS equations form a generalization of Nahm's equations to the
circular quiver, whose single-pole sector we analyze in detail.

\subsection{D4s on a single NS5}\label{sd4onns5}

We start from the simplest boundary realization in Table~\ref{tIIAbr}, in which the bulk D4 segments end on a \emph{single} boundary NS5. In this configuration no additional boundary degrees of freedom are expected at low energy, and the boundary condition is entirely captured by fixing the appropriate 3d $\mathcal N=2$ multiplets obtained from the restriction of the bulk fields to the boundary.

For simplicity in this subsection we set $\theta=0$, so that $\gamma=0$ in \eqref{Eq: definition of gamma}. As reviewed in Section~\ref{srelationtostringtheory}, in \eqref{e:JNbg_brane_main}, D4s ending on a boundary NS5 realize Neumann-type boundary conditions for the bulk gauge field. In the 3d $\mathcal N=2$ covariant language this amounts to fixing the linear multiplet $\mathcal{J}_{\mathrm N}$ to a background value.

The bifundamental hypermultiplets require a separate comment. They are not
captured by the open-string endpoint analysis of
Appendix~\ref{app:openstrings}, and the geometric picture of
Section~\ref{srelationtostringtheory} is compatible with fixing either of the
two 3d \(\mathcal N=2\) chirals in \eqref{eq:bchyp} to a background, as in
\eqref{ephibg}. Since the NS5-type condition is not analyzed further in this
paper, we do not need to specify which.

Assuming that a single boundary NS5-brane induces no further boundary
interactions, such as the superpotential couplings \(W_{\mathrm{3d}}\)
discussed in Section~\ref{sec4dBC}, the boundary condition for the vector
multiplet reads
\begin{equation}\label{ens5bc}
\Big\{2\phi_2,\lambda_2^-,-\lambda_1^+,F_{\perp i},2D_\perp\phi_1\Big\}
=\{a,0,0,0,0\}\,,
\end{equation}
while for the hypermultiplets one of the two chirals is fixed to a background.

The brane interpretation of the boundary data is more direct for the vector
multiplet than for the hypermultiplets. Non-zero bifundamental data control
the recombination of adjacent D4 segments at the boundary, and become
positions only for regular branes (Sections~\ref{srelationtostringtheory}
and~\ref{secTNframe}). The background value \(a\) instead prescribes the
boundary value of \(\phi_2\), which corresponds, heuristically, to the
position of the D4 endpoints along \(x^6\), the direction transverse to the
boundary NS5-brane. These interpretations play no role in what follows.

\subsection{D4\('\)-type boundary conditions and single-pole solutions}\label{sendingond6}

The boundary conditions of this subsection are motivated by D4 segments ending
on a single boundary D6-brane. Ideally, one would like to study the boundary condition realized by
the boundary D6-brane at \(x^7=0\), through the colour branes, i.e.\ the
D6-type termination of Section~\ref{srelationtostringtheory}. For the vector
multiplet this is a Dirichlet-type boundary condition for the gauge field, see
equation \eqref{e:JDbg_brane_main}. For the hypermultiplets, as discussed in
Section~\ref{srelationtostringtheory}, the D6-brane extends along \(x^8,x^9\),
so that the recombined D4-branes can move along it and
\(X_aY_a\) need not vanish at the boundary. We expect the corresponding boundary condition to be a deformation of \(q_{\tilde I1}=0\) which may introduce a scale, in which case the resulting BPS equations are considerably harder to solve.

For simplicity, we therefore focus on the limit in which the boundary D6-brane
is sent to infinity along \(x^7\), i.e.\ on the D4\('\)-type termination of
Sections~\ref{sboundary:setup} and~\ref{srelationtostringtheory}: the boundary
condition for the vector multiplet is unchanged, while for the hypermultiplets
it becomes \(q_{\tilde I1}=0\), eq.~\eqref{e:D4pbc_hyper}, leaving
\(q_{\tilde I2}\) fluctuating. This condition only specifies which fields are
fixed: it does not depend on the number of D4\('\)-branes left behind, nor on
the bulk NS5-branes they end on, and the analysis below does not determine
which of its solutions are realized by a given brane configuration.

\subsubsection{Boundary conditions and BPS equations}\label{sbpsequations}

We assume that no boundary interactions or boundary degrees of freedom are
present; this is an assumption of the field-theory problem, not a consequence
of the limit. Accordingly, the background \(b\) for \(q_{\tilde I1}\) in
\eqref{ephibg} vanishes. The background \(a\) for \(\phi_1\) in \eqref{eJbg}
is instead the position of the D6-brane along \(x^5\), which the limit does
not affect: for a single boundary D6-brane at the \(x^5\) position of the
colour branes, $a$ is proportional to the identity, and tracelessness sets
\(a=0\), so that the colour branes are not separated along \(x^5\).

With these specifications, the only nontrivial components of the BPS boundary conditions
\eqref{ebcvectors} and \eqref{ephibg}
are the top components\footnote{There remains a decoupled diagonal $U(1)$ factor whose equations are trivial.}
\begin{align}\label{enahm1}
\begin{split}
D_\perp \phi_2^{x_a}+\frac i2 D_{12}^{x_a}&=0\,,\\
D_\perp q_{\tilde I 2}+2(\phi_2)_{\tilde I}^{\ \tilde J}q_{\tilde J 2}&=0\,,
\end{split}
\end{align}
where $x_a$ denotes the adjoint index of the $a^{\text{th}}$ $SU(N)$ gauge node. As discussed below \eqref{ephi1phi2lamdapm}, the auxiliary field $D_{12}$ must be replaced by its on-shell value obtained from the bulk equations of motion. Using \eqref{eYMCSflat} and \eqref{ehyperaction4dflat}, one finds
\begin{align}
D_{12}^{x_a}
=\frac i2\,g_a^2\,K^{x_a y_a}
\bigg(
(q_{\tilde I 1})^*(T_{y_a})_{\tilde I}^{\ \tilde J}q_{\tilde J 1}
-(q_{\tilde I 2})^*(T_{y_a})_{\tilde I}^{\ \tilde J}q_{\tilde J 2}
\bigg)\,,
\end{align}
with $K_{x_a y_a}=\text{Tr} (T_{x_a}T_{y_a})$ and $K^{x_a y_a}=(K_{x_a y_a})^{-1}$.

In the gauge \(A_\perp=0\), with \(Q_a\) the \(N\times N\) matrix of the
components of \(q_{\tilde I2}\) between the nodes \(a\) and \(a+1\) and
\((\phi_2)_a\) the restriction of \(\phi_2\) to the node \(a\), the system
\eqref{enahm1} reads
\begin{equation}\label{egennahm}
\begin{split}
\frac{d(\phi_2)_a}{dx_\perp}&=-\frac{g_a^2}{4}\Big(Q_aQ_a^\dagger-Q_{a-1}^\dagger Q_{a-1}
-\frac1N\,\mathds 1_a\,\mathrm{Tr}\big(Q_aQ_a^\dagger-Q_{a-1}^\dagger Q_{a-1}\big)\Big)\,,\\
\frac{dQ_a}{dx_\perp}&=-2\big((\phi_2)_aQ_a-Q_a(\phi_2)_{a+1}\big)\,,
\end{split}
\end{equation}
with node labels understood modulo \(K\). The subtraction of the traces is the
projection appropriate to the \(SU(N)^K\) theory, in which the relative abelian
factors are frozen and the diagonal one
decouples~\cite{Witten:1997sc}: the equations are
derived directly for this theory, and are not a modification of the unitary
system. This is the generalization of Nahm's
equations to the circular quiver: for \(K=1\) it reduces, after a rescaling of
\(Q\), to Nahm's equations of \(\mathcal N=4\) super Yang--Mills theory written
in complex form. The appearance of Nahm's equations is motivated by the brane
picture: in the Type~IIB frame of Section~\ref{secTNframe} the configuration
with a single D4\('\)-brane is locally the \(\mathbb Z_K\) projection of
D3-branes ending on a D5-brane, whose Nahm triple is \((\phi_2,q_{\tilde I2})\).
The orbifold projection motivates the choice of fields and the Nahm structure,
while the precise form of \eqref{egennahm}, with the trace parts subtracted
node by node, follows from the field theory. 

The equations \eqref{egennahm} are closely related to those of~\cite{Heckman:2016xdl}, who studied the five-dimensional affine quiver on
a half-line to describe punctures of class \(\mathcal S_\Gamma\) theories with
\(\Gamma=\mathbb Z_K\); see also~\cite{Hassler:2017arf}. Their equations are
recovered from \eqref{egennahm} for gauge group \(U(N)^K\), where the trace
terms are absent, at equal couplings and after a rescaling of \(Q\). In this
case the pole solutions are \(\mathfrak{su}(2)\) representations on the
orbifold covering space, labelled by Young diagrams whose boxes carry
\(K\)-th roots of unity, and chains of eigenspaces along the affine quiver
play the role of the \(\hat\phi\)-families introduced below. The system
studied here differs in two respects: the gauge group is
\(\prod_aSU(N)_a\), so that each \(\phi_a\) is traceless node by node, and
the couplings are generic. As a consequence, the spectrum of the residues is
in general not that of \(\mathfrak{su}(2)\) representations, and the
solutions depend on the couplings.

Motivated by the brane picture of Sections~\ref{sboundary:setup}
and~\ref{secTNframe}, we look for Nahm-pole-like boundary conditions. For these
the fields have no finite restriction to \(x_\perp=0\), so the boundary
condition cannot be imposed by fixing boundary values to a constant background
as in \eqref{eJbg}. We therefore proceed as for the Nahm pole
in~\cite{Gaiotto:2008sa,Gaiotto:2008ak}: we require \eqref{enahm1} to hold at
every \(x_\perp>0\), and specify the boundary condition by the leading
behaviour of the fields as \(x_\perp\to0\), a \emph{simple pole} whose residues
label the boundary condition. Since no scale is present, we look for exact solutions of \eqref{enahm1}
invariant under rescalings of \(x_\perp\), i.e.\ a pure pole with no
subleading terms. For the Nahm pole of \(\mathcal N=4\) super Yang--Mills
theory, the perturbations of the pure pole compatible with the boundary
condition were determined in~\cite{Mazzeo:2013zga}, using the
\(\mathfrak{su}(2)\) representation carried by the residues. Here, as noted above, the residues do not in general form
\(\mathfrak{su}(2)\) representations, and the linearized problem depends on
the couplings and couples different nodes, so that this analysis does not
extend directly and we do not attempt it.

With $q_{\tilde I 1}=0$ as above, we choose the gauge
\begin{equation}\label{eaperpzero}
A_\perp=0\,.
\end{equation}
We adopt the \emph{single-pole ansatz}
\begin{equation}\label{epolefields}
\phi_2=\frac12\,\frac{\phi}{x_\perp}\,,
\qquad
q_{\tilde I 2}=\frac{q_{\tilde I}}{x_\perp}\,,
\end{equation}
where $\phi$ and $q_{\tilde I}$ are independent of $x_\perp$. By 3d Lorentz invariance along the boundary, we also take them to be constant along the boundary directions $x^i$.

With these choices, the differential system \eqref{enahm1} reduces to an algebraic one. Using the identity for the generators in the fundamental,
\begin{equation}
K^{xy}T_{x,i}^{\ \ j}T_{y,k}^{\ \ l}
=
\delta_i^l\delta_k^j-\frac1N\delta_i^j\delta_k^l\,,
\end{equation}
and defining
\begin{align}
\phi_a=\phi^{x_a}T_{x_a,i_a}^{\ \ \ j_a}\,,
\qquad
q_a=q_{i_a}^{\ \ i_{a+1}}\,,
\qquad
\bar q_a=(q_a)^\dagger\,,
\end{align}
we obtain
\begin{equation}\label{efundeqmain2}
\begin{split}
\phi_a
&=
\frac{g_a^2}{2}\big(q_a\bar q_a-\bar q_{a-1}q_{a-1}\big)
-\frac{g_a^2}{2N}\,\mathds{1}_a\,
\text{Tr}\big(q_a\bar q_a-\bar q_{a-1}q_{a-1}\big)\,,
\\
q_a&=\phi_a q_a-q_a\phi_{a+1}\,.
\end{split}
\end{equation}
Thus the single-pole ansatz reduces the BPS equations to the algebraic system
\eqref{efundeqmain2}. In the rest of this section we study this system and
characterize a broad class of its solutions; as we will see, some of them have
a discrete stabilizer (see Section~\ref{sdualneumann}).

%----------------------------------------------------------
\subsubsection{$\hat{\phi}$-family structure of the single-pole solution}\label{sstructural}
%----------------------------------------------------------

The solutions of \eqref{efundeqmain2} have a rigid structure, which follows from
the equations alone and which we now describe. For each property we give the
key step of the argument; the complete derivations are collected in
Appendix~\ref{astructureofsolution}.

\begin{itemize}

    \item \textbf{Simultaneous diagonalization.}
The Hermitian matrices \(\phi_a\), \(q_a\bar q_a\), and
\(\bar q_{a-1}q_{a-1}\) commute,
\begin{equation}
[\phi_a,q_a\bar q_a]=[\phi_a,\bar q_{a-1}q_{a-1}]
=[q_a\bar q_a,\bar q_{a-1}q_{a-1}]=0\,,
\end{equation}
and can therefore be simultaneously diagonalized. The second equation in
\eqref{efundeqmain2} gives \(\phi_aq_a=q_a+q_a\phi_{a+1}\), hence
\(\phi_aq_a\bar q_a=q_a\bar q_a+q_a\phi_{a+1}\bar q_a=q_a\bar q_a\phi_a\); the
second commutator follows in the same way, and the third because, by the first
equation in \eqref{efundeqmain2}, \(\phi_a\) is a linear combination of
\(q_a\bar q_a-\bar q_{a-1}q_{a-1}\) and the identity.

  \item \textbf{Unit jumps and the \(\hat{\phi}\)-family structure.}
Writing
\begin{equation}\label{ephiaeigenvalue}
\phi_a=\mathrm{diag}\{\phi_{a,1},\ldots,\phi_{a,N}\}\,,
\end{equation}
the second equation in \eqref{efundeqmain2} reads, in components,
\(\big(\phi_{a,i}-\phi_{a+1,j}-1\big)\,q_{a,i}^{\ \ j}=0\), so that
\begin{equation}\label{ephinonzeroqmain}
q_{a,i}^{\ \ j}\neq 0
\qquad\Longrightarrow\qquad
\phi_{a,i}-\phi_{a+1,j}=1\,.
\end{equation}
Thus a non-zero bifundamental entry can connect only eigenspaces whose \(\phi\)-eigenvalues differ by one unit. This gives a convenient way to organize the single-pole data. Starting from an eigenspace of \(\phi_a\), one can follow the non-zero blocks of \(q_a\) from node to node; at each step the eigenvalue decreases by one. We call the resulting chain of eigenspaces a \(\hat\phi\)-family.

More precisely, a \emph{\(\hat\phi\)-family} is specified by a starting node \(A_{\hat\phi}\)
and by an eigenvalue \(\hat\phi\) of \(\phi_{A_{\hat\phi}}\). Starting from this eigenspace and
following the non-zero blocks of \(q_a\) from node to node, the eigenvalue decreases by one unit
at each step (see~\eqref{ephinonzeroqmain}); we call \emph{characteristic length} \(L_{\hat\phi}\)
the number of steps until the chain terminates. We label the family by \(\hat\phi\), the maximal
eigenvalue: it sits at the head of the chain, where no block of the same family enters.
The eigenvalues are then distributed along the quiver as
\begin{equation}\label{emaxtomin}
\phi_{A_{\hat\phi}+J}\ni \hat\phi-J\,,
\qquad
0\le J\le L_{\hat\phi}\,.
\end{equation}
Here the node label is understood modulo \(K\), since the quiver is circular.
Schematically, the eigenvalues of a family take the form
\[ \hat\phi,\qquad \hat\phi-1,\qquad \hat\phi-2,\qquad \ldots \]
on consecutive nodes.
The circular identification of the nodes introduces a possibility that is absent for
linear quivers. If \(L_{\hat\phi}\ge K\), the chain returns to a node that it has already
visited. In that case several eigenspaces belonging to the same \(\hat\phi\)-family can
appear at a single gauge node, with eigenvalues separated by integer multiples of \(K\).
We refer to this phenomenon as \emph{winding}, and call a family \emph{non-winding}
when \(L_{\hat\phi}<K\).

\item \textbf{\(q_a\)-blocks of constant rank proportional to unitary matrices.}
Within a given \(\hat\phi\)-family, let \(M_1\) and \(M_2\) be the dimensions of
two joint eigenspaces of \((\phi_a,q_a\bar q_a)\) and of
\((\phi_{a+1},q_{a+1}\bar q_{a+1})\) whose \(\phi\)-eigenvalues differ by one
unit. Equation~\eqref{ephinonzeroqmain} then allows for a non-zero
\(M_1\times M_2\) block \(Q_a\subset q_a\) connecting them. Two consequences
follow from \eqref{efundeqmain2} and the simultaneous diagonalization; both are
derived in Appendix~\ref{astructureofsolution}. First, all such dimensions along
a given \(\hat\phi\)-family are equal, so every block of the family has the same
rank, which we denote by \(M_{\hat\phi}\). This multiplicity of the family can
differ from the multiplicity of the corresponding eigenvalue of \(\phi_a\), when
that eigenvalue is shared with slots outside the family. Second, when the family winds around the quiver, several
copies of these blocks can sit at the same node; we label them by a winding
index \(w_{\hat\phi}\) (see the discussion of winding above). Allowing also
for several \(\hat\phi\)-families, each block
\(Q_{a,w_{\hat\phi}}^{\hat\phi}\subset q_a\) then takes the form
\begin{align}\label{eqstructure}
Q_{a,w_{\hat\phi}}^{\hat\phi}
=
C_{a,w_{\hat\phi}}^{\hat\phi}\,
U_{a,w_{\hat\phi}}^{\hat\phi}\,,
\end{align}
where \(C_{a,w_{\hat\phi}}^{\hat\phi}\in \mathbb C\) and
\(U_{a,w_{\hat\phi}}^{\hat\phi}\in U(M_{\hat\phi})\).
Once this block structure is imposed, the second equation in
\eqref{efundeqmain2} is automatically satisfied. The remaining equations depend
only on the non-negative variables
\begin{equation}
x_{a,w_{\hat\phi}}^{\hat\phi}
:=
\left|C_{a,w_{\hat\phi}}^{\hat\phi}\right|^2
=
\frac{1}{M_{\hat\phi}}
\operatorname{Tr}\!\left(
Q_{a,w_{\hat\phi}}^{\hat\phi}
Q_{a,w_{\hat\phi}}^{\hat\phi\dagger}
\right).
\end{equation}

\end{itemize}

%----------------------------------------------------------
\subsubsection{Discrete data and reduced equations}\label{ssummaryofgeneralsolution}%----------------------------------------------------------
The structure of Section~\ref{sstructural} reduces the single-pole problem to a
small set of discrete and continuous unknowns.
A \(\hat\phi\)-family is specified by three discrete data:
the starting node \(A_{\hat\phi}\), the characteristic length \(L_{\hat\phi}\)
along the quiver, and the constant rank \(M_{\hat\phi}\) of the blocks
\(Q_a\subset q_a\). The continuous unknowns are the maximal eigenvalue
\(\hat\phi\) of \(\phi_{A_{\hat\phi}}\) in the family and the
gauge-invariant moduli
\begin{equation}
\bigl|C^{\hat\phi}_{a,w_{\hat\phi}}\bigr|^2
\;=\;
\frac{1}{M_{\hat\phi}}\,
\operatorname{Tr}\!\Bigl(
Q^{\hat\phi}_{a,w_{\hat\phi}}
Q^{\hat\phi\dagger}_{a,w_{\hat\phi}}
\Bigr)\,.
\end{equation}
When the stabilizer is continuous, gauge-invariant relative phases of the
blocks may also survive, at most as many as the rank of the stabilizer
(Appendix~\ref{astructureofsolution}); they are part of the boundary datum but
do not enter the reduced equations below. The second equation
in~\eqref{efundeqmain2} is solved automatically by the block
structure~\eqref{eqstructure}. It remains to impose the first equation
in~\eqref{efundeqmain2}: projecting it onto the joint eigenspace associated
with each position \(J\) along a given \(\hat\phi\)-family yields the algebraic
relation to be solved. To lighten the notation, in the equations below we
write \(a\equiv A_{\hat\phi}+J\) for the node at position \(J\) within the
family, and suppress the labels \((\hat\phi,w_{\hat\phi})\) on the moduli,
which are understood to refer to the family under consideration. The
relation reads
\begin{align}\label{efinalequationgeneral}
\hat{\phi}-J
\;=\;
\frac{g^2_{a}}{2}\bigl(|C_{a}|^2-|C_{a-1}|^2\bigr)
\;-\;
\frac{g^2_{a}}{2N}\Bigl(
\operatorname{Tr}(q_{a}\bar q_{a})
-\operatorname{Tr}(\bar q_{a-1}q_{a-1})
\Bigr)\,,
\end{align}
while the eigenvalues of \(\phi_{a}\) that do not belong to any
\(\hat\phi\)-family are fixed to
\begin{align}\label{eforreference}
\phi_{a,i}
\;=\;
-\frac{g^2_{a}}{2N}\Bigl(
\operatorname{Tr}(q_{a}\bar q_{a})
-\operatorname{Tr}(\bar q_{a-1}q_{a-1})
\Bigr)\,.
\end{align}
Here the traces collect all the blocks contributing at node \(a\):
\begin{equation}\label{etracesum}
\operatorname{Tr}(q_{a}\bar q_{a})
\;=\;
\sum_{\hat\phi'}\,M_{\hat\phi'}
\sum_{w'}\,
\bigl|C^{\hat\phi'}_{a,w'}\bigr|^2\,,
\end{equation}
with the sum running over every \(\hat\phi\)-family and every winding copy
that visits node \(a\).
Note that \eqref{eforreference} is \eqref{efinalequationgeneral} for a family of
length zero, with no blocks entering or leaving it; in
Section~\ref{sdualneumann} it will be convenient to treat such eigenvalues as
families of this kind.
Equation~\eqref{efinalequationgeneral} is therefore non-local along the
family: the trace terms couple \(|C_a|^2\) to the moduli of all the other
winding copies of the same family and of any other family contributing at
the same node, making a general closed-form solution difficult to obtain.

The system can nevertheless be solved explicitly in three sectors, which play
different roles below
(Appendices~\ref{asinglephihatandnowinding}--\ref{asymmetriccase}): a single
non-winding family, the simplest solution, which illustrates the general
structure and its brane interpretation (Section~\ref{secSinglPhiNoWind}); and
two configurations that saturate the bound on the total length of the families
in opposite ways, the maximal-winding family, consisting of a single family,
and the maximally symmetric configuration, with a family starting at every
node and invariant under cyclic relabelling (Section~\ref{sdualneumann}). In
Section~\ref{sdualneumann} we also use this structure to determine which
solutions have a discrete stabilizer.

A formal solution of the algebraic equations does not necessarily define a
boundary condition. The variables \(|C_a|^2\) are squared moduli and must
therefore be non-negative. To realize a prescribed set of families, every
block connecting consecutive slots of a family must in fact have strictly
positive modulus. These conditions can fail or restrict the allowed region
in the space of gauge couplings. If a modulus becomes negative, there is no
solution with those data; if it vanishes, the corresponding chain splits,
and the family decomposition and stabilizer must be determined anew.

\subsubsection{Brane interpretation of the single-pole data}
\label{secSinglPhiNoWind}

This subsection gives a physical picture of the single-pole data. The choice of
fixed and fluctuating fields is motivated by the branes, while the BPS
equations and their solutions follow from field theory; the identification of
individual residues with brane data is not established, and the picture below
is a motivation, not a reconstruction of each solution.

The single-pole ansatz does not specify a unique brane profile. In this
picture, different solutions correspond to different patterns by which the
suspended D4 segments recombine through the bulk NS5-branes before bending
along \(x^7\) towards the boundary D6-brane. The non-winding solution gives the
simplest local recombination pattern, while winding solutions involve profiles
that wrap around the circular quiver and can revisit the same gauge node. The
equations do not fix the boundary branes either: for the Nahm pole of
\(\mathcal N=4\) super Yang--Mills theory, the irreducible blocks of the pole
correspond to D3-branes ending on different
D5-branes~\cite{Gaiotto:2008sa,Gaiotto:2008ak}, and solutions with several
\(\hat\phi\)-families may similarly correspond to several D4\('\)-branes, left
behind by one or several boundary D6-branes.

We illustrate this picture using the simplest explicit solution, namely the
single \(\hat\phi\)-family without winding solved analytically in
Appendix~\ref{asinglephihatandnowinding}. For a single family we drop the
\(\hat\phi\) labels and write the discrete data as \((A,L,M)\). The solution
for the maximal eigenvalue and the moduli reads
\begin{equation}\label{egeneralhatphi}
\begin{aligned}
\hat\phi
&=
\frac{\displaystyle\sum_{J=0}^{L}
\frac{J}{g_{A+J}^2}}
{\displaystyle\sum_{J=0}^{L}
\frac{1}{g_{A+J}^2}}\,,
 \qquad\qquad
|C_{A+J}|^2=\frac{2N}{N-M}
\frac{\displaystyle\sum_{i=0}^{J}\sum_{j=J+1}^{L}
\frac{j-i}{g_{A+i}^2g_{A+j}^2}}
{\displaystyle\sum_{i=0}^{L}\frac{1}{g_{A+i}^2}}\,,
\quad
0\le J\le L-1\,,
\end{aligned}\end{equation}
together with
\begin{equation}\label{egeneralCsq}
|C_{A+L}|^2=0\,.
\end{equation}
Since every term in the double sum in \eqref{egeneralhatphi} has \(j>i\), all
moduli are automatically non-negative for positive couplings \(g_a^2>0\).

These formulas show that every choice of discrete data \((A,L,M)\), with
\(1\le L\le K-1\) (the non-winding condition) and \(M\le N-1\), admits a
solution. The bound on \(M\) is required by the equations of the chain
(Appendix~\ref{asinglephihatandnowinding}), while \(L=0\) gives the trivial
datum. In particular, the continuous moduli are not arbitrary: once the family
data and the couplings are fixed, the quantities \(|C_{A+J}|^2\) are
determined. Since the length of such a family is below the bound of
Section~\ref{sdualneumann}, its stabilizer is continuous.

\begin{figure}[h!]
\begin{center}
\includegraphics[width=40mm]{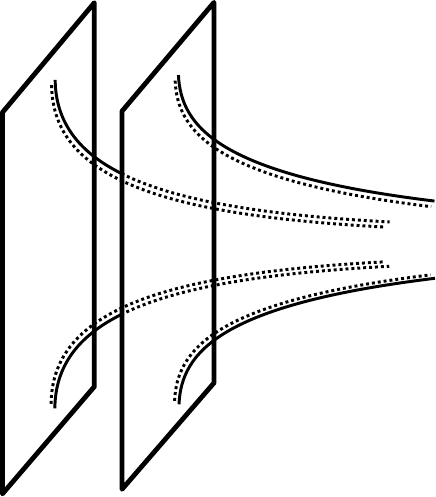}
\caption{In this figure we present the original Nahm pole configuration
from~\cite{Gaiotto:2008ak}, with \(N=4\) and \(p_1=p_2=2\).}\label{foriginalnahm}
\end{center}
\end{figure}

This block-diagonal solution is reminiscent of Nahm-pole solutions for the
D3--D5 system discussed in~\cite{Gaiotto:2008sa, Gaiotto:2008ak}, see
Fig.~\ref{foriginalnahm}, and suggests the following physical interpretation.
In that setting, a pole \(X^i \sim \tfrac{t^i}{x_\perp}\) describes how groups
of D3-branes end on D5-branes through a fuzzy-funnel profile, with the
corresponding \(SU(2)\) embedding determining the boundary data.\footnote{The
analogy is at the level of the pole structure and of the brane recombination
pattern, rather than of an underlying \(SU(2)\) embedding.}

\begin{figure}[htbp]
  \centering

  %------------------ 1st row ------------------%
  \begin{subfigure}[b]{0.4\textwidth}
    \centering
    \includegraphics[width=\linewidth]{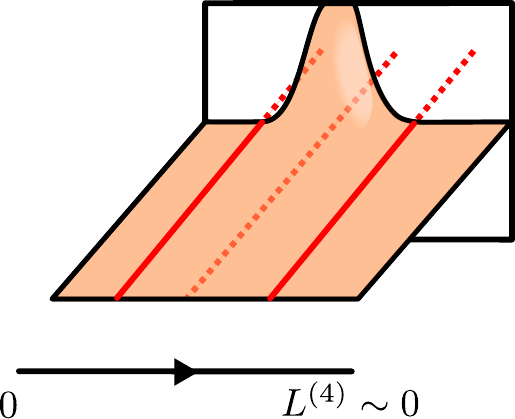}
 \caption{}\label{fournahm1}
  \end{subfigure}\hspace{0.5 cm}
  \begin{subfigure}[b]{0.4\textwidth}
    \centering
    \includegraphics[width=\linewidth]{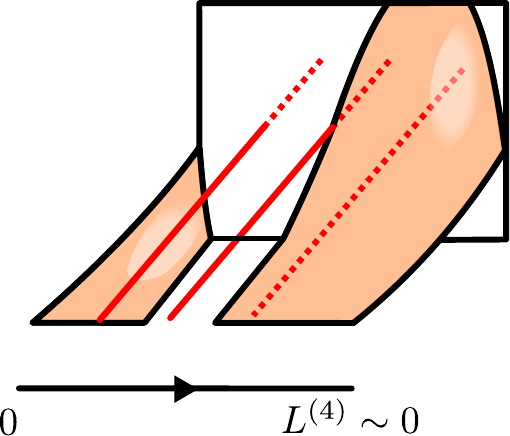}
 \caption{}\label{fournahm2}
  \end{subfigure}

 \caption{Red lines: NS5-branes; orange surfaces: D4-branes; white plane:
the boundary D6-brane, drawn at finite distance for illustration.
The horizontal direction is periodic, and the vertical direction
is $x^7$. In (a), two adjacent D4-branes recombine above an
 NS5-brane and, due to the pole behaviour of the boundary condition, move off
 vertically towards the boundary D6-brane. In (b) the D4-branes move off
 vertically without recombining at the central bulk NS5-brane. The \(x^6\)
 direction, along which the profile also extends, is not shown.}
  \label{fournahm}
\end{figure}

Our block-diagonal pattern admits an analogous brane interpretation.
Non-vanishing entries of \(q_a\) trigger the recombination of adjacent D4
segments across the bulk NS5-branes (Section~\ref{srelationtostringtheory}).
Around such recombination loci, we interpret the single-pole behaviour
\eqref{epolefields} as the recombined D4-branes bending along \(x^7\) with a
pole-like profile, in the sense specified below, running off towards the
boundary D6-brane at infinity along the direction of the D4\('\)-branes, as
schematically depicted in Fig.~\ref{fournahm}(a). The unit-jump condition
\eqref{ephinonzeroqmain} then organizes the recombined segments into
\(\hat\phi\)-families: whenever \(q_{a,i_a}^{\ \ i_{a+1}}\neq 0\), the
corresponding eigenvalues satisfy \(\phi_{a,i_a}-\phi_{a+1,i_{a+1}}=1\).

The single-pole ansatz also determines how this profile scales, and it is
worth separating the exact consequences of the ansatz from the geometric
picture. In the Type~IIB frame, the Nahm triple \((\phi_2,q_{\tilde I2})\) corresponds
to the position along \(x^6\) and to the coordinate \(\beta\) of the covering
space near the tip of the cigar (Section~\ref{secTNframe}), and both grow as
\(1/x_\perp\). Reading the pole as a fuzzy
funnel that opens into the D5-brane is the same semiclassical picture as for
the D3--D5 system~\cite{Gaiotto:2008sa,Gaiotto:2008ak}: it treats
non-commuting matrices as positions. For \(K=1\) the residues, after the
appropriate rescaling, form a representation of \(\mathfrak{su}(2)\), and each
non-trivial irreducible block has the usual interpretation as a fuzzy sphere in
\((x^6,\mathrm{Re}\,\beta,\mathrm{Im}\,\beta)\). For \(K>1\) this is the case
only for special solutions, such as the maximally symmetric one of
Appendix~\ref{asymmetriccase}, at equal couplings with \(u_a=0\) and all
non-zero blocks chosen real and positive. Its moduli are then given by
\eqref{esymmsympoint_app}, and, after rescaling the \(q_a\), the residues at
every node have the form of the principal \(\mathfrak{su}(2)\) embedding. The single-pole scaling remains an exact consequence of the ansatz.

In the Type~IIA frame the picture requires one further step. The position along
\(x^7\) is not a field of the triple but, in the sector of regular branes, the
link combination \(X_aX_a^\dagger-Y_a^\dagger Y_a\) of \eqref{e:linkpositions},
quadratic in the hypermultiplets, which reduces to \(-Y_a^\dagger Y_a\) for
\(q_{\tilde I1}=0\). Two facts are exact consequences of the ansatz: with
\(q_{\tilde I1}=0\) the matrix \(-Y_a^\dagger Y_a\) is negative semidefinite,
and it scales as \(-1/x_\perp^2\). A third one constrains the interpretation.
By the second equation in \eqref{efundeqmain2}, the product
\(P_a\equiv q_aq_{a+1}\cdots q_{a+K-1}\) around the quiver satisfies
\([\phi_a,P_a]=K\,P_a\), hence \([\phi_a,P_a^r]=rK\,P_a^r\) and
\(\mathrm{Tr}\,P_a^r=0\) for every \(r\ge1\); by Newton's identities and the
Cayley--Hamilton theorem, \(P_a^N=0\): \(P_a\) is nilpotent. The residues
therefore do not describe commuting regular branes displaced along the cigar,
which would give non-zero eigenvalues of the products around the quiver, i.e.\
of \(w_-\). This does not contradict the fuzzy-funnel interpretation: in the
ordinary Nahm pole the complex residue \(q\) is likewise nilpotent, being
proportional to the raising operator of an \(\mathfrak{su}(2)\) representation,
while the Hermitian residue \(\phi\) is not nilpotent in general. It shows that
the pole is not a configuration of regular branes, and that the geometric
dictionary applies to it only in the non-commutative sense of the funnel. The
identification of \(-Y_a^\dagger Y_a\) with the position along \(x^7\), which
holds for regular branes on the Higgs branch~\cite{Douglas:1996sw}, is
therefore the interpretive step, and it goes beyond the D3--D5 case. Moreover,
the orbifold description is local near the centre, and its extension to the
region \(x^7\to-\infty\) of the Taub--NUT geometry, where the boundary D6-brane
sits, is not automatically uniform. Within this interpretation, and with these
caveats, the profile extends along \(x^7\) as \(-1/x_\perp^2\), up to factors
involving \(\alpha'\) and the Taub--NUT radius, and only towards \(x^7<0\): for
\(K=1\) the fuzzy sphere is mapped to a cap in the \((x^6,x^7)\)-plane, bulging
towards \(x^7<0\), which widens and deepens as \(x_\perp\to0\). In this
picture, this is how the colour branes reach the boundary D6-brane at infinity.

There can also be blocks for which \(q_{a,i_a}^{\ \ i_{a+1}}=0\) while both
\(\phi_a\) and \(\phi_{a+1}\) are non-vanishing. In the brane picture this
corresponds to D4 segments that bend towards the boundary without recombining
at the intermediate NS5-brane, as illustrated in Fig.~\ref{fournahm}(b).

When the pole is absent altogether, the only solution within the pure
single-pole ansatz is the trivial one, with vanishing fields, in analogy with
the trivial \(SU(2)\) embedding in~\cite{Gaiotto:2008ak}. In the Type~IIB frame of Section~\ref{secTNframe}, the same discussion applies
to fractional D3-branes recombining before ending on the D5-branes obtained
from the D4\('\)-branes, which wrap a cigar.

%==========================================================
\subsubsection{Boundary data with discrete stabilizer}\label{sdualneumann}
%==========================================================

The duality chain of Section~\ref{secDualNS5} suggests that the D4\('\)-type
class may be a good place to look for the S-dual of pure Neumann boundary
conditions, but not which datum within it. In this subsection we address a
purely field-theoretic question, suggested by the analogy with
\(\mathcal N=4\) super Yang--Mills theory: which single-pole data leave a
residual symmetry that is as small as possible?

\paragraph{The stabilizer.}
Throughout this subsection, by boundary datum we mean the residues
\((\phi_a,q_a)\) of the pole \eqref{epolefields}. The stabilizer of a boundary
datum is the subgroup of constant gauge transformations \((g_1,\ldots,g_K)\in SU(N)^K\) (which
preserve the gauge choice \(A_\perp=0\)) satisfying
\begin{equation}\label{estabcondmain}
g_a\,\phi_a\,g_a^{-1}=\phi_a\,,\qquad g_a\,q_a\,g_{a+1}^{-1}=q_a\,,\qquad a=1,\ldots,K\,.
\end{equation}
For a Dirichlet-type condition, the constant gauge transformations at the
boundary act as global symmetries of the boundary condition rather than as
redundancies (Section~\ref{srelationtostringtheory}), and the stabilizer is the
part of this global symmetry left unbroken by the pole, the analogue of the
centralizer of the Nahm pole in~\cite{Gaiotto:2008ak}. There, a further datum
is the subgroup of the centralizer that is gauged at the boundary, possibly
with additional boundary degrees of freedom. Here we retain Dirichlet-type
conditions and do not gauge any subgroup of the centralizer. The diagonal centre \(\mathbb Z_N\subset SU(N)^K\)
acts trivially on adjoint and bifundamental fields, so it stabilizes every
boundary datum: no boundary datum can break the symmetry further. The reason to
look for data attaining this floor is the matching of global symmetries under
the duality. Pure Neumann boundary conditions introduce no boundary degrees of
freedom and leave the gauge field dynamical at the boundary, so they have no
continuous boundary global symmetry coming from the gauge group; a
Dirichlet-type dual should then have none either, i.e.\ a discrete stabilizer.
In the \(\mathcal N=4\) setting this argument selects the regular Nahm pole as
the dual of pure Neumann~\cite{Gaiotto:2008sa,Gaiotto:2008ak}.\footnote{The
bulk theory also has the \(U(1)^K\) flavour symmetry rotating the bifundamental
hypermultiplets, preserved by pure Neumann boundary conditions. For \(N\ge3\),
a datum with discrete stabilizer preserves a combination of it with gauge
transformations of the same rank, as follows from counting the phases of the
slots, so the criterion concerns only symmetries beyond the bulk flavour
symmetry.} In the \(\mathcal N=2\) quiver the criterion is necessary only if
boundary global symmetries are matched across the duality as in
\(\mathcal N=4\), which we do not show, and it is not a characterization of the
dual.

\paragraph{Families and the matrix \(B\).}
We regard every eigenvalue of \(\phi_a\) that does not belong to a
\(\hat\phi\)-family as a family of length zero, so that the \(NK\) diagonal
slots are partitioned into families, and record their distribution along the
quiver in the integer matrix
\begin{equation}\label{eBmatrix}
B_{a\hat\phi}\equiv\#\{\text{slots of the family }\hat\phi\text{ at the node }a\}\,,
\qquad \sum_{\hat\phi}B_{a\hat\phi}=N\,.
\end{equation}
If some family has rank \(M_{\hat\phi}\ge2\), the stabilizer is continuous: the
unitary factors in \eqref{eqstructure} can be gauged to the identity along the
family, which has no closed loops, and an \(SU(M_{\hat\phi})\) rotation acting
identically on all its slots then commutes with the whole datum. When all the
ranks are one, the number of families is \(c=NK-\sum_{\hat\phi}L_{\hat\phi}\).
A phase \(e^{i\theta_{\hat\phi}}\) acting on all the slots of each family
preserves the datum, and \(\det g_a=1\) requires
\(\sum_{\hat\phi}B_{a\hat\phi}\theta_{\hat\phi}\in2\pi\mathbb Z\), so the
stabilizer contains a torus of rank \(c-\operatorname{rank}B\). Nothing else
survives: every \(g_a\) commutes with \(\phi_a\) and \(q_a\bar q_a\), hence
preserves their joint eigenspaces, which are one-dimensional
(Appendix~\ref{astructureofsolution}), and each non-zero block identifies the
phases of the two slots it joins.

A discrete stabilizer therefore requires all ranks to be one and
\(\operatorname{rank}B=c\le K\), that is,
\begin{equation}\label{ebudget}
\sum_{\hat\phi}L_{\hat\phi}\ \ge\ (N-1)K\,.
\end{equation}
For \(c=K\) the stabilizer is the kernel of \(B:T^K\to T^K\), finite if and only
if \(\det B\neq0\), and then of order \(|\det B|\).

\paragraph{Saturation of the bound.}
Tracelessness also bounds the total length of the families from above. By the
second equation in \eqref{efundeqmain2}, \(\phi_aq_a=q_a(\phi_{a+1}+1)\). If
\(q_a\) were invertible, the trace of \(\phi_a=q_a(\phi_{a+1}+1)q_a^{-1}\) would
give \(0=N\). Hence
\begin{equation}\label{erankbound}
\operatorname{rank}q_a\le N-1\,,\qquad a=1,\ldots,K\,.
\end{equation}
The matrix \(q_a\) is the direct sum of the non-zero blocks \eqref{eqstructure}
it contains, which join distinct slots, so its rank is the number of these
blocks, each counted with its rank \(M_{\hat\phi}\). Summing
\eqref{erankbound} over the nodes gives
\begin{equation}\label{ebudgetup}
\sum_{\hat\phi}M_{\hat\phi}L_{\hat\phi}\ \le\ (N-1)K
\end{equation}
for every single-pole datum. Together with \eqref{ebudget}, a discrete
stabilizer therefore requires \(c=K\), i.e.\ \eqref{ebudget} saturated, and
\(\operatorname{rank}B=K\), i.e.\ \(\det B\neq0\); every \(q_a\) then has rank
exactly \(N-1\).

This also fixes the stabilizer. Since every \(q_a\) has rank \(N-1\), each
node carries exactly one head and one tail of a family, and this implies
\(\det B\in\{0,N\}\) (Appendix~\ref{adetB}). A discrete stabilizer requires
saturation and \(\det B\neq0\), and is then the kernel of \(B\), of order \(N\),
which contains the diagonal centre. Every single-pole datum with discrete
stabilizer therefore has stabilizer exactly
\begin{equation}\label{estabZN}
\mathbb Z_N\subset SU(N)^K\,.
\end{equation}

\paragraph{Rigidity of the spectrum.}
Summing the equations \eqref{efinalequationgeneral} over the slots of a given
node, the terms in the moduli cancel and only the tracelessness of \(\phi_a\)
survives. In terms of \eqref{eBmatrix} it reads
\begin{equation}\label{etracesystem}
\sum_{\hat\phi}B_{a\hat\phi}\,\hat\phi=b_a\,,\qquad
b_a\equiv\sum_{\text{slots at }a}J\,,\qquad a=1,\ldots,K\,,
\end{equation}
where \(J\) is the position of each slot within its own family. These are \(K\)
linear equations for the \(c\) unknowns \(\hat\phi\), with coefficients that do
not involve the couplings. When \(c=K\) and \(\det B\neq0\) they determine every
\(\hat\phi\) uniquely: the whole spectrum of the residues \(\phi_a\) is then
fixed to rational values independent of the gauge couplings. This explains the rigidity
of the maximal-winding solution found in Appendix~\ref{asinglephihatwinding}, and
shows that it is shared by every single-pole datum with finite stabilizer, since
all such data have \(c=K\) and \(\det B\neq0\).

\paragraph{Solutions.}
The maximal-winding and the maximally symmetric solutions of
Appendix~\ref{ad6bc} saturate \eqref{ebudget} in opposite ways.
\begin{itemize}
\item The \emph{maximal-winding} solution (Appendix~\ref{asinglephihatwinding})
consists of one family of length \((N-1)K\), which winds \(N-1\) times around
the quiver and completely fills one node, and of \(K-1\) free eigenvalues, one
at each of the other nodes. The column of \(B\) belonging to the family has the
entry \(N\) in a row that no other column reaches, so \(\det B=N\) and the
stabilizer is \(\mathbb Z_N\) for every \((N,K)\)
(Appendix~\ref{awindstabilizer}). Taking the filled node to be node \(1\), the
system \eqref{etracesystem} gives \(\hat\phi=K(N-1)/2\), the free eigenvalue at
node \(1+j\), \(j\ge1\), is \((N-1)(j-K/2)\), and the moduli, linear in the
inverse couplings, are given in closed form in \eqref{ewindingmoduli_app}.
Of the \((N-1)K\) positivity conditions only two, those of the two outermost
links, are ever active, so the solution exists in a chamber cut out by two linear
inequalities,
which is the whole conformal manifold for \(K\le2\); for \(K\ge3\) the
chambers of the \(K\) cyclic copies together cover the whole conformal
manifold (Appendix~\ref{awindpositivity}). An explicit instance is worked out
in Appendix~\ref{awindexample}. For \(K=1\) there are no free eigenvalues and \(\hat\phi=(N-1)/2\): the spectrum
is that of the principal \(\mathfrak{su}(2)\) embedding, and the solution
reduces to the regular Nahm pole of \(\mathcal N=4\) super Yang--Mills theory
(Appendix~\ref{awindrecursion}).
\item The \emph{maximally symmetric} solution (Appendix~\ref{asymmetriccase})
consists of \(K\) families of length \(N-1\), one starting at each node, all with
\(\hat\phi=(N-1)/2\). Here \(B\) is circulant, and \(\det B=N\) for
\(\gcd(N,K)=1\) and \(\det B=0\) otherwise, in agreement with the stabilizers
\eqref{esymmstabres_app} and \eqref{esymmstabgcd_app}. For \(\gcd(N,K)=d>1\) the
solution is moreover not rigid, and exists only on the locus
\eqref{esymmgcdlocus_app} of codimension \(d-1\) in the conformal manifold. For
\(\gcd(N,K)=1\) it is rigid, but, being invariant under cyclic relabelling, it
has no other copies to cover the couplings where its moduli fail to be positive:
at \((N,K)=(4,3)\) with \((g_1^2,g_2^2,g_3^2)=(1,2,10)\) one of them equals
\(-1/5\), while the maximal-winding solution with filled node \(1\) exists there.
\end{itemize}
These are not the only saturating data with finite stabilizer. For example, at
\((N,K)=(5,2)\), one family of length \(2\) starting at node \(1\) and one of
length \(6\) starting at node \(2\) give, from \eqref{etracesystem},
\(\hat\phi=1\) and \(\hat\phi=3\) respectively, hence
\begin{equation}\label{e52spectrum}
\phi_1=\mathrm{diag}(2,1,0,-1,-2)\,,\qquad \phi_2=\mathrm{diag}(3,1,0,-1,-3)\,,
\end{equation}
with distinct eigenvalues at both nodes and \(\det B=5\). Writing
\(t_a\equiv1/g_a^2\), the moduli along the two families are
\begin{equation}\label{e52moduli}
\begin{split}
&\big(2t_1,\ 2t_1\big)\,,\\
&\big(6t_2,\ 2(2t_1+3t_2),\ 4(t_1+2t_2),\ 4(t_1+2t_2),\ 2(2t_1+3t_2),\ 6t_2\big)\,,
\end{split}
\end{equation}
positive for all couplings, and the stabilizer is \(\mathbb Z_5\).

To see how common such data are, we enumerated all single-pole data with
\(M_{\hat\phi}=1\) for small \((N,K)\), solved the BPS equations symbolically in
the couplings and determined where the moduli are positive. The results are
summarized in Table~\ref{tdiscretestab}.

\begin{table}[h]
\centering
\begin{tabular}{c|c|c|c|c}
\((N,K)\) & maximal winding & maximally symmetric & other data & existence \\
\hline
\((3,2)\) & \(\checkmark\) & \(\checkmark\) & --- & all couplings \\
\((4,2)\) & \(\checkmark\) & --- & 1 & all couplings \\
\((5,2)\) & \(\checkmark\) & \(\checkmark\) & 1 & all couplings \\
\((6,2)\) & \(\checkmark\) & --- & 2 & all couplings \\
\((2,3)\) & \(\checkmark\) & \(\checkmark\) & --- & chambers \\
\((3,3)\) & \(\checkmark\) & --- & --- & chambers \\
\((4,3)\) & \(\checkmark\) & \(\checkmark\) & --- & chambers \\
\((3,4)\) & \(\checkmark\) & \(\checkmark\) & 2 & chambers
\end{tabular}
\caption{Single-pole data with finite stabilizer and positive moduli somewhere
on the conformal manifold, up to cyclic relabelling of the nodes. ``Other data''
counts the saturating data that are neither maximal winding nor maximally
symmetric. For \(K=2\) every such datum exists at all couplings; for \(K\ge3\),
in the cases examined, only the cyclic copies of the maximal-winding solution
cover the whole conformal manifold.}\label{tdiscretestab}
\end{table}

\paragraph{What the stabilizer does not determine.}
A discrete stabilizer therefore does not single out a boundary datum within
the single-pole class. By the results above, all the data of Table~\ref{tdiscretestab} have a rigid spectrum and stabilizer \(\mathbb Z_N\). For \(K=2\) every datum in the table moreover exists
at all couplings, so that none of these properties distinguishes among them.
For \(K\ge3\), in the cases examined, the data differ in where they exist: only
the maximal-winding solution has \(K\) cyclic copies that together cover the
whole conformal manifold (Appendix~\ref{awindpositivity}), while the other data,
including the maximally symmetric one, exist only in part of it. The two
solutions of Appendix~\ref{ad6bc} also behave differently on the equal-coupling
locus, where the bulk theory acquires an additional \(\mathbb Z_K\) symmetry. For \(\gcd(N,K)=1\) the maximally symmetric solution is invariant under it, while the \(K\) copies of the maximal-winding solution all exist there and are permuted
among themselves.

The brane picture suggests a further criterion, which we do not use. If only
the families of non-zero length corresponded to boundary branes, a single
family would be natural for the image of a single boundary NS5-brane
(Section~\ref{secDualNS5}), and among the data with discrete stabilizer only
the maximal-winding solution has this property: the maximally symmetric
solution has \(K\) families of non-zero length, and every other saturating
datum has at least two. Nothing in the analogy
with~\cite{Gaiotto:2008sa,Gaiotto:2008ak}, however, allows us to ignore the
\(K-1\) families of length zero of the maximal-winding solution, and deciding
the counting would require the dictionary of the orbifold discussed in
Section~\ref{secTNframe}. The fact that the maximal-winding solution has a
single family of non-zero length is therefore not, by itself, evidence that it
is the S-dual of pure Neumann.

Whether any of these data is the S-dual of pure Neumann boundary conditions
cannot be decided on the basis of symmetry
and existence alone. Answering it would require determining the action of the
duality on the fields, the couplings and the fractional charges, matching protected observables, such as hemisphere partition
functions or half-indices, and incorporating the $\theta$-angles.
These tasks lie beyond the present analysis.

\section{Conclusions}\label{sconclusions}

In this paper we studied a concrete class of \(\tfrac12\)-BPS boundary
conditions for four-dimensional \(\mathcal N=2\) \(A_{K-1}\) circular quivers,
combining brane engineering with a direct analysis of the BPS equations. Our
goal was not to classify all supersymmetric boundaries of the theory, but rather
to isolate a natural and tractable string-motivated sector and analyze it
systematically.

The brane construction provides the physical origin of the class considered
here. Starting from the Type~IIA D4--NS5 realization of the circular quiver, we
introduced boundary NS5- and D6-branes, which preserve a 3d \(\mathcal N=2\)
subalgebra and motivate distinct field-theoretic boundary data. Sending the
boundary D6-brane to infinity along \(x^7\), which leaves D4\('\)-branes
behind, motivates singular D4\('\)-type boundary conditions governed by a
generalization of Nahm's equations to the circular quiver. Under the duality
chain of Section~\ref{secDualBoundaryConfigurations}, the boundary NS5-brane,
which realizes a Neumann-type condition, is mapped, up to a \(U(1)_r\)
rotation, to a pair of semi-infinite D4\('\)-branes. This makes the
D4\('\)-type class a natural place to look for the S-dual of pure Neumann
boundary conditions, although the argument does not select a datum within it.

On the field-theory side, our main technical result is the rigid structure
imposed by the BPS equations in the single-pole sector. The residues organize
into \(\hat\phi\)-families, chains of joint eigenspaces along which the
eigenvalue decreases by one unit from node to node and the multiplicity stays
constant. For circular quivers these families can wind around the quiver and
contribute more than once to a given node, a phenomenon with no analogue in
linear quivers. We solved in closed form a non-winding sector, a
maximal-winding sector and the maximally symmetric sector, and determined
which single-pole data have a discrete stabilizer, a property suggested by the
matching of global symmetries under the duality, as for the regular Nahm pole
dual to pure Neumann in \(\mathcal N=4\) super Yang--Mills theory. A discrete
stabilizer requires all family ranks to be one. In that case it is controlled
by the integer matrix \(B\) recording how the families are distributed over the
nodes, and discreteness requires \(\sum_{\hat\phi}L_{\hat\phi}\ge(N-1)K\).
Since tracelessness implies the opposite inequality, this bound must be
saturated and \(\det B\neq0\). The stabilizer is then always the diagonal
centre \(\mathbb Z_N\), and the spectra of the Hermitian residues \(\phi_a\)
are fixed independently of the couplings. The maximal-winding solution has this property
for every \(N\) and \(K\), the maximally symmetric one only for
\(\gcd(N,K)=1\), and for small \(N\) and \(K\) we found further such data.
These properties therefore do not single out a boundary datum: for \(K=2\) all
the data we found exist at all couplings, while for \(K\ge3\), in the cases
examined, only the cyclic copies of the maximal-winding solution together cover
the whole conformal manifold.

Several directions remain open. Within the single-pole class, one would like to
extend the list of data with discrete stabilizer beyond the small values of
\(N\) and \(K\) examined here, and to understand the positivity chambers of the
multi-family solutions. Our statements concern the residues of the pole: the
perturbations of the pure pole compatible with the boundary condition,
determined for the Nahm pole in~\cite{Mazzeo:2013zga}, remain to be found for
the present system. On the brane side, the boundary condition realized by a
boundary D6-brane at finite position along \(x^7\) remains to be determined, as
does the dictionary between the residues and the brane data, including the
Chan--Paton structure and the fractional charges of the D4\('\)-branes in the
orbifold (Section~\ref{secTNframe}). One should also analyze more general
boundary brane systems, including Dirichlet-type conditions with boundary
degrees of freedom and their possible role in a duality-wall description, in
analogy with \(T(G)\) in the Gaiotto--Witten analysis, as well as fundamental
matter from bulk D6-branes~\cite{Witten:1997sc}.

Deciding whether any of the solutions with discrete stabilizer is the S-dual of
pure Neumann boundary conditions requires ingredients that are absent from the
present analysis: the action of the duality on the fields and on the brane
data, a comparison of protected quantities, such as hemisphere partition
functions, supersymmetric indices or boundary anomalies, and a treatment of the
\(\theta\)-angles. The single-pole solutions depend only on the gauge couplings
\(g_a\), but the duality does not preserve \(\theta=0\): as shown in
Section~\ref{stypeiiamth}, at \(\theta=0\) the transformed frame has \(K-1\)
couplings at the boundary \(\mathrm{Im}\,\tau_i'=0\) of the space of
couplings, while the last node acquires a non-zero \(\theta\)-angle. A
comparison across the duality therefore requires control of supersymmetric
Neumann-type boundary conditions at non-zero \(\theta\), including their
boundary Chern--Simons couplings~\cite{Gaiotto:2008sd}, in a regime of
couplings different from the one considered here.

Finally, it would be interesting to understand whether the algebraic
constraints derived here can be reformulated in terms of moduli spaces or
algebroid-type structures, along the lines of~\cite{Bertle:2024djm}, and
whether an M-theoretic perspective could shed further light on them.

\section*{Acknowledgements}

We thank B.~S.~Acharya, D.~Baldwin, M.~Bertolini, G.~Bonelli, F.~Marino, M.~Moleti, and A.~Sangiovanni for interesting and stimulating discussions. We are especially grateful to L.~Di Pietro for carefully reading the draft and for many valuable comments that helped improve the paper, and to A.~Shri for sustained support and valuable feedback. We also acknowledge the use of Claude (Anthropic) and ChatGPT (OpenAI)  for assistance in checking intermediate calculations, improving the presentation, and preparing the manuscript; all AI-assisted output was independently reviewed and verified by the authors, who take full responsibility for the content of this work. R.V.\ acknowledges support by INFN Iniziativa Specifica ST\&FI.

\appendix

\vspace{1cm}

\section{Tensor conventions}\label{atensorconv}

\begin{enumerate}
\item Unless otherwise specified, the following is the general index convention.
\begin{equation}
\begin{aligned}[t]
&\mu,\, \nu,\, \hdots\in\{0,1,2,3\}\ \text{for 4d space-time indices},\notag\\
&i,\, j,\, \hdots\in\{0,1,2\}\ \text{for 3d space-time indices},\notag\\
&\alpha,\, \beta,\, \hdots\ \text{for }\psi_\alpha\in(\textbf{2},\textbf{1})_{\text{Spin}(1,3)}\text{ or \textbf{2}}_{\text{Spin}(1,2)},\\
&\dot{\alpha},\, \dot{\beta},\, \hdots\ \text{for }\bar{\psi}^{\dot{\alpha}}\in(\textbf{1},\textbf{2})_{\text{Spin}(1,3)},\\
&A,\, B,\, \hdots\ \text{for }\lambda_A\in\textbf{2}_{SU(2)_{\mathcal{R}}},\\
&I,\,J,\,\hdots\ \text{for }USp(2N_F),\\
&\tilde{I},\,\tilde{J}\,\hdots\subset\ I,\,J,\,\hdots\text{once a Lagrangian sub-manifold is specified},\\
&a,\, b,\,\hdots\in\{1,\hdots,\,K\}\ \text{list the $K$ gauge nodes},\\
&x_a,\, y_a\,\hdots\ \text{for adjoint indices of $a^{\text{th}}$ }SU(N)_a\text{ gauge node},\\
&i_a,\, j_a\,\hdots\ \text{for }q_{i_a}\in\mathbf{N}_{SU(N)_a}\ \text{and }q^{i_a}\in\mathbf{\bar{N}}_{SU(N)_a}.
\end{aligned}
\end{equation}
Symmetrization and antisymmetrization of indices are denoted by
$A_{(\ldots)}$ and $A_{[\ldots]}$, respectively, both with weight
$1/n!$.
\item The Minkowski metric, both in 3d and 4d, is mostly plus and we choose $\varepsilon^{0123}=\varepsilon^{012}=1$ for the Levi-Civita tensor.
\item The Pauli-matrices are taken to be
\begin{align}
\tau^1=\begin{pmatrix}
0 & 1\\
1 & 0
\end{pmatrix}\,\ \ \ \tau^2=\begin{pmatrix}
0 & -i\\
i & 0
\end{pmatrix}\,\ \ \ \tau^3=\begin{pmatrix}
1 & 0\\
0 & -1
\end{pmatrix}\,.
\end{align}
\item We define the chiral sub-matrices of the 4d gamma matrices, the 3d gamma matrices, the $SU(2)_{\mathcal{R}}$ matrices and the anti-symmetric tensor matrices to be
\begin{align}\label{e:gammamatrices}
\begin{split}
(\sigma^a)_{\alpha\dot{\beta}}=(-&\tau_1,-i\tau_2,-i\tau_3,\mathds{1})\,,\ \ \ \ (\bar{\sigma}^a)^{\dot{\alpha}\beta}=(\tau_1,i\tau_2,i\tau_3,\mathds{1})\,,\\
&(\gamma^{a'})_\alpha^{\ \beta}=(-i\tau_1,\tau_2,\tau_3)\,,\ \ \ \ \vec{\tau}_A^{\ B}=\vec{\tau}\,,\\
\sigma^{ab}=&\sigma^{[a}\bar{\sigma}^{b]}\,,\ \ \ \bar{\sigma}^{ab}=\bar{\sigma}^{[a}\sigma^{b]}\,,\ \ \ \gamma^{a'b'}=\gamma^{[a'}\gamma^{b']}\,.
\end{split}
\end{align}
This peculiar choice of the $\sigma^a$ and $\bar{\sigma}^a$ matrices is due to the fact that it eases the reduction to 3d of the multiplets, while preserving simple complex conjugation properties in 4d and 3d.
\item All spinorial and $SU(2)_{\mathcal{R}}$-symmetry indices are raised and lowered with the $\varepsilon$-tensors as in the Wess \& Bagger notation~\cite{Wess:1992cp}, with $\varepsilon^{12}=-\varepsilon_{12}=1$. We employ the same notation for spinor bilinears as well as for the conjugation of Grassmann odd quantities $\theta_i$:
\begin{align}
(\theta_1\theta_2)^\dag=\theta_2^\dag\theta_1^\dag\,.
\end{align}

\item The Killing spinors are taken to be Grassmann even.

\item The gauge-covariant derivative is fixed to be
\begin{align}
D_\mu=\partial_\mu -iA_\mu\,.  
\end{align}
In particular for adjoint fields we fix $D_\mu=\partial_\mu-i[A_\mu,\cdot]$.

\item In the presence of a boundary we often interchange the index 3 with $\perp$. We also orient the manifold such that $\int \partial\cdot V=\int_\partial V^\perp$.
\end{enumerate}

\vspace{1cm}

\section{Supersymmetry transformations and reduction to boundary supersymmetry}\label{asusyandred}

The 4d $\mathcal{N}=2$ supersymmetry is controlled by two parameters $\epsilon_{A}$ and $\bar{\epsilon}_A$ that generate
\begin{align}
\mathcal{Q}=\epsilon_{A\alpha}Q^{A\alpha}+\bar{\epsilon}^{A\dot{\alpha}}\bar{Q}_{A\dot{\alpha}}\,.
\end{align}
In presence of a boundary we can preserve at most half of the supersymmetry. This is achieved by imposing
\begin{align}
\bar{\epsilon}_A^{\dot{\alpha}}=-\tau_{3,A}^{\ \ B}(\bar{\sigma}_3\epsilon_B)^{\dot{\alpha}}
\end{align}
and then relating $\epsilon_A$ in the following way to the 3d spinors
\begin{align}\label{e4dto3d}
\epsilon_{A=1}=\frac{1}{\sqrt{2}}\zeta\,,\ \ \ \ \epsilon_{A=2}=\frac{1}{\sqrt{2}}\bar{\zeta}\,,\ \ \ \gamma^i=i\sigma^i\bar{\sigma}^\perp=-i\sigma^\perp\bar{\sigma}^i\,.
\end{align}

\subsubsection*{4d supersymmetry transformations}\label{s4dsusytransf}
These transformations are the ones in~\cite{Hama:2012bg} adapted for our convention of gauge covariant derivatives and for flat 4d Minkowski space.
{\allowdisplaybreaks
\begin{align}
&\text{\textbf{Vector multiplet}}\notag\\*
&\delta\phi=-i\epsilon^A\lambda_A\,,\notag\\
&\delta\bar{\phi}=i\bar{\epsilon}^A\bar{\lambda}_A\,,\notag\\
&\delta\lambda_A=\frac{1}{2}\sigma^{\mu\nu}\epsilon_A F_{\mu\nu}+2\sigma^\mu\bar{\epsilon}_A D_\mu\phi+2i \epsilon_A[\phi,\bar{\phi}]+D_{AB}\epsilon^B\,,\label{e4dvectorflat}\\
&\delta\bar{\lambda}_A=\frac{1}{2}\bar{\sigma}^{\mu\nu}\bar{\epsilon}_A F_{\mu\nu}+2\bar{\sigma}^\mu \epsilon_A D_\mu\bar{\phi}-2i\bar{\epsilon}_A[\phi,\bar{\phi}]+D_{AB}\bar{\epsilon}^B\,,\notag\\
&\delta A_\mu=i\epsilon^A\sigma_\mu \bar{\lambda}_A-i\bar{\epsilon}^A\bar{\sigma}_\mu\lambda_A\,,\notag\\
&\delta D_{AB}=-2i\bar{\epsilon}_{(A}\bar{\sigma}^\mu D_\mu\lambda_{B)}+2i\epsilon_{(A}\sigma^\mu D_\mu\bar{\lambda}_{B)}-4[\phi,\bar{\epsilon}_{(A}\bar{\lambda}_{B)}]+4[\bar{\phi},\epsilon_{(A}\lambda_{B)}]\,.\notag\\\notag\\
&\text{\textbf{Hypermultiplet}}\notag\\*
&\delta q_{IA}=-i\epsilon_A\psi_{I}+i\bar{\epsilon}_A\bar{\psi}_{I}\,,\notag\\
&\delta \psi_{I}=2\sigma^\mu\bar{\epsilon}_A D_\mu q^A_I-4i \epsilon_A\bar{\phi} q^A_I+2\epsilon_AF^{\hat{A}}_I\,,\notag\\
&\delta \bar{\psi}_I=2\bar{\sigma}^\mu\epsilon_A D_\mu q^A_I-4i \bar{\epsilon}_A\phi q^A_I+2\bar{\epsilon}_AF^{\hat{A}}_I\,,\label{ehyperssusyflat}\\
&\delta F_{I\hat{A}}=i \epsilon_A\sigma^\mu D_\mu\bar{\psi}_I-2\epsilon_A\phi\psi_I-2\epsilon_A\lambda_B q^B_I-i \bar{\epsilon}_A\bar{\sigma}^\mu D_\mu\psi_I+2\bar{\epsilon}_A\bar{\phi}\bar{\psi}_I+2\bar{\epsilon}_A\bar{\lambda}_B q^B_I\,.\notag
\end{align}}
We are able to write the hypermultiplet's transformations in the off-shell formalism because we preserve just four supercharges (namely it is possible to explicitly show that if we impose \eqref{e4dto3d} the above hypermultiplet transformations do close).

\subsubsection*{3d supersymmetry transformations}\label{s3dsusytransf}

These transformations are the ones in~\cite{Closset:2012ru} adapted for our convention of gauge covariant derivatives and for flat 3d Minkowski space (in practice these are the same ones written in~\cite{Bason:2023bin}, but with $\varepsilon^{ijk}\rightarrow -i\varepsilon^{ijk}$).
{\allowdisplaybreaks
\begin{align}
&\text{\textbf{Chiral multiplet}}\notag\\*
&\delta q=\sqrt{2}\zeta \psi\,,\notag\\
&\delta\psi=\sqrt{2}\zeta F-\sqrt{2}iz\tilde{\zeta} q-\sqrt{2}i \gamma^i \tilde{\zeta} D_i q+\sqrt{2}i \sigma \tilde{\zeta}q\,,\label{e3dchiralsusyflat}\\
&\delta F=\sqrt{2}iz\tilde{\zeta}\psi-\sqrt{2}i D_i(\tilde{\zeta}\gamma^i\psi)-\sqrt{2}i \sigma\tilde{\zeta}\psi+2i\tilde{\zeta}\tilde{\lambda}q~.\notag\\\notag\\
&\text{\textbf{Anti-chiral multiplet}}\notag\\*
&\delta \tilde{q}=-\sqrt{2}\tilde{\zeta}\tilde{\psi}\,,\notag\\
&\delta\tilde{\psi}=\sqrt{2}\tilde{\zeta}\tilde{F}+\sqrt{2}iz\zeta\tilde{q}+\sqrt{2}i \gamma^i \zeta D_i \tilde{q}-\sqrt{2}i \tilde{q}\sigma \zeta\,,\label{e3dantichiralsusyflat}\\
&\delta\tilde{F}=\sqrt{2}iz\zeta\tilde{\psi}-\sqrt{2}i D_i(\zeta\gamma^i\tilde{\psi})-\sqrt{2}i \zeta\tilde{\psi}\sigma+2i\tilde{q}\zeta\lambda~.\notag\\\notag\\
&\text{\textbf{Vector multiplet}}\notag\\*
&\delta\sigma=-\zeta\tilde{\lambda}+\tilde{\zeta}\lambda\,,\notag\\
&\delta\lambda=\bigg(iD-\frac{1}{2}\varepsilon^{ijk}\gamma_k F_{ij}-i\gamma^i D_i\sigma\bigg)\zeta\,,\notag\\
&\delta\tilde{\lambda}=\bigg(-iD-\frac{1}{2}\varepsilon^{ijk}\gamma_k F_{ij}+i\gamma^i D_i\sigma\bigg)\tilde{\zeta}\,,\label{e3dvectorsusyflat}\\
&\delta A_i=-i(\zeta \gamma_i\tilde{\lambda}+\tilde{\zeta}\gamma_i\lambda)\,,\notag\\
&\delta D=\zeta\gamma^i D_i\tilde{\lambda}-\tilde{\zeta}\gamma^i D_i \lambda-[\sigma,\zeta\tilde{\lambda}+\tilde{\zeta}\lambda]~.\notag\\\notag\\
&\text{\textbf{Linear multiplet}}\notag\\*
&\delta J=i\zeta j+i\tilde{\zeta}\tilde{j}\,,\notag\\
&\delta j=\tilde{\zeta} K+i\gamma^i\tilde{\zeta}(j_i+i D_i J)+\tilde{\zeta}[\sigma,J]\,\notag\\
&\delta\tilde{j}=\zeta K-i\gamma^i\zeta(j_i-i D_i J)-\zeta[\sigma,J]\,,\label{e3dlinearflat}\\
&\delta j_i=\varepsilon_{ijk}D^j(\zeta\gamma^k j-\tilde{\zeta}\gamma^k\tilde{j})+[\sigma,\zeta\gamma_i j+\tilde{\zeta}\gamma_i\tilde{j}]+i[J,\tilde{\zeta}\gamma_i\lambda-\zeta\gamma_i\tilde{\lambda}]\,,\notag\\
&\delta K=-i D_i(\zeta\gamma^i j+\tilde{\zeta}\gamma^i\tilde{j})+[\zeta\tilde{\lambda}+\tilde{\zeta}\lambda,J]~.\notag
\end{align}}
This definition of the linear multiplet matches the one of~\cite{Closset:2012ru} in the abelian case (upon Wick rotation). It is written here in the case of a linear multiplet in the adjoint representation but it can be generalized to any representation. The supersymmetry transformations in flat space close upon imposing
\begin{align}\label{enonabelianconstraint}
\begin{split}
&D_i j^i-i[D,J]-i[\sigma,K]+[\tilde{\lambda},\tilde{j}]-[\lambda,j]=0\,,\\
&[F_{ij},J]=0\,.
\end{split}
\end{align}
A vector multiplet can be dualized into a linear one in the following way
\begin{align}
\begin{split}\label{evectortolinear}
 &J=\sigma\,,\ \ \ j=i\tilde{\lambda}\,,\ \ \ \tilde{j}=-i\lambda\,,\\
&j_i=-\frac{1}{2}\varepsilon_{ijk}F^{jk}\,,\ \ \ K=D\,.   
\end{split}
\end{align}

\vspace{1cm}

\section{Review of open-string modes and branes ending on branes}
\label{app:openstrings}

In this appendix we review the minimal open-string analysis that underlies the brane interpretation of the elementary boundary terminations used in the main text. Our goal is to identify which components of the D4 worldvolume fields are fixed when a suspended D4 segment meets a boundary brane. This is sufficient for the discussion of Section~\ref{srelationtostringtheory}, where these fixed components are reorganized into 4d $\mathcal N=2$ multiplets, whose restrictions to the boundary are then organized into 3d $\mathcal N=2$ multiplets.

The scope of this appendix is limited. First, bifundamental hypermultiplets associated with D4 segments stretched across NS5-branes are not captured here as ordinary perturbative open strings; this is a standard limitation of the Type~IIA description.\footnote{See for instance the original discussion in~\cite{Hanany:1996ie}.} Second, the argument does not distinguish among the possible supersymmetric backgrounds to which fields may be fixed, such as regular Dirichlet data versus Nahm-pole singular profiles. Nor does it determine the additional boundary interactions that may arise in more general coupled boundary systems. In particular, extra boundary multiplets from strings stretched between bulk and boundary branes may modify the naive Neumann/Dirichlet picture through supersymmetric boundary couplings.

What the open-string analysis \emph{does} provide is a clean kinematical
identification of the fixed versus fluctuating components of the D4 fields
\((A_M,\Psi)\) at the endpoint, and hence of the corresponding 3d
\(\mathcal N=2\) multiplets. Together with the supersymmetry conditions of the
boundary branes collected below, this is the input from the present appendix
used in the main text.

\subsubsection*{Conventions and 10d notation}

Let $M,N,\ldots \in \{0,\ldots,9\}$ be 10d Minkowski indices, with
\begin{equation}
  \eta_{MN}=\mathrm{diag}(-1,1,1,1,1,1,1,1,1,1)\,,
  \qquad
  \{\Gamma_M,\Gamma_N\}=2\eta_{MN}\,,
  \qquad
  \bar\Gamma \equiv \Gamma_0\Gamma_1\cdots\Gamma_9\,.
\end{equation}
We use Type~IIA supersymmetry parameters $\epsilon$ and $\tilde\epsilon$ of opposite chirality,
\begin{equation}
  \epsilon=\bar\Gamma\,\epsilon\,,
  \qquad
  \tilde\epsilon=-\bar\Gamma\,\tilde\epsilon\,.
\end{equation}
For a D4-brane extended along $01234$, the preserved supersymmetry obeys
\begin{equation}
  \tilde\epsilon=\Gamma_0\Gamma_1\Gamma_2\Gamma_3\Gamma_4\,\epsilon\,.
  \label{D4susy}
\end{equation}

\subsubsection*{Theory on a D4 and its modes}

The low-energy theory on a stack of D4-branes along $01234$ is five-dimensional maximally supersymmetric Yang--Mills theory, obtained by dimensional reduction of 10d $\mathcal N=1$ SYM:
\begin{equation}\label{Eq:appC10dAM}
  \mathcal L
  =
  \frac{1}{2g^2}\,\text{Tr}\!\left(
    F_{MN}F^{MN}
    -
    i\bar\Psi \Gamma^M D_M\Psi
  \right),
  \qquad
  A_M=(A_{\mu=0,\ldots,4},\,\phi_{I=5,\ldots,9})\,.
\end{equation}
The supersymmetry transformations can be written as
\begin{equation}
  \delta A_M=\epsilon\Gamma_M\Psi\,,
  \qquad
  \delta\Psi=\frac12\,F_{MN}\Gamma^{MN}\epsilon\,.
  \label{D4susyvar}
\end{equation}

As usual, the scalars $\phi_{5,\ldots,9}$ encode transverse fluctuations of the D4-brane. A vacuum expectation value proportional to the identity corresponds to a center-of-mass displacement of the stack.

\subsubsection*{Suspended D4s and the 4d \texorpdfstring{$\mathcal N=2$}{N=2} splitting}

We now turn to the bulk configuration of Table~\ref{tIIAbr}, in which the D4-branes are suspended between NS5-branes. The NS5-branes impose the standard supersymmetry conditions
\begin{equation}
  \epsilon=\Gamma_0\Gamma_1\Gamma_2\Gamma_3\Gamma_5\Gamma_6\,\epsilon\,,
  \qquad
  \tilde\epsilon=\Gamma_0\Gamma_1\Gamma_2\Gamma_3\Gamma_5\Gamma_6\,\tilde\epsilon\,.
  \label{NS5bulkSUSY}
\end{equation}
Together with \eqref{D4susy}, this leaves eight supercharges, namely 4d $\mathcal N=2$ supersymmetry on each suspended D4 segment.

A convenient 10d-covariant way to display the splitting of the D4 modes into a 4d $\mathcal N=2$ vector multiplet plus a 4d $\mathcal N=2$ hypermultiplet is to introduce the projectors
\begin{equation}
  P^{\pm}_{\rm bulk}
  \equiv
  \frac{1\pm \Gamma_4\Gamma_7\Gamma_8\Gamma_9}{2}\,,
  \qquad
  \Psi_\pm \equiv P^{\pm}_{\rm bulk}\Psi\,,
  \label{Pbulk}
\end{equation}
and to separate indices as
\begin{equation}
  a \in \{0,1,2,3,5,6\}\,,
  \qquad
  \tilde a \in \{4,7,8,9\}\,.
\end{equation}
Then the supersymmetry variations take the schematic form
\begin{align}
  \text{4d $\mathcal N=2$ vector:}\qquad
  &\delta A_a = \epsilon \Gamma_a \Psi_+\,,
  &\delta \Psi_+ = \frac12\,F_{ab}\Gamma^{ab}\epsilon\,,
  \nonumber\\[2mm]
  \text{4d $\mathcal N=2$ hyper:}\qquad
  &\delta A_{\tilde a} = \epsilon \Gamma_{\tilde a} \Psi_-\,,
  &\delta \Psi_- = F_{a\tilde a}\Gamma^{a\tilde a}\epsilon\,.
  \label{4dN2split}
\end{align}

For D4-branes suspended between NS5-branes, the endpoint conditions along $x^4$ fix the $\tilde a$ sector in the infrared. The surviving low-energy theory on each segment is therefore the expected 4d $\mathcal N=2$ vector multiplet.

\subsubsection*{Adding boundary branes and the 3d \texorpdfstring{$\mathcal N=2$}{N=2} multiplets at the boundary}

We now consider the boundary setup of Table~\ref{tIIAbr}. The elementary boundary terminations relevant for the main text are a D4 ending on a boundary NS5 and a D4 ending on a boundary D6. In both cases the system preserves four supercharges, corresponding to 3d $\mathcal N=2$ supersymmetry at the boundary.

The supersymmetry constraints coming from the boundary NS5- and D6-branes are
redundant with one another once the bulk conditions have been imposed, for
compatible orientations (\(s_{\rm NS}=-1\) in \eqref{eq:extraprojectors}
below). It is convenient to impose the D6 condition
\begin{equation}
\tilde\epsilon=\Gamma_0\Gamma_1\Gamma_2\Gamma_4\Gamma_6\Gamma_8\Gamma_9\,\epsilon\,,
  \label{D6susy}
\end{equation}
together with \eqref{D4susy} and \eqref{NS5bulkSUSY}. This system preserves
four independent supercharges.

\paragraph{Further boundary branes.}
The statements of Section~\ref{sboundary} on the other branes of
Table~\ref{tIIAbr}, and on the images of Section~\ref{secDualNS5}, follow from
the same analysis. Besides \eqref{D4susy}, \eqref{NS5bulkSUSY} and
\eqref{D6susy}, we impose
\begin{align}\label{eq:extraprojectors}
\begin{split}
\text{boundary NS5 }(012457):&\qquad
\epsilon=s_{\rm NS}\,\Gamma_0\Gamma_1\Gamma_2\Gamma_4\Gamma_5\Gamma_7\,\epsilon\,,\qquad
\tilde\epsilon=s_{\rm NS}\,\Gamma_0\Gamma_1\Gamma_2\Gamma_4\Gamma_5\Gamma_7\,\tilde\epsilon\,,\\
\text{D4}'\ (01267):&\qquad
\tilde\epsilon=s'\,\Gamma_0\Gamma_1\Gamma_2\Gamma_6\Gamma_7\,\epsilon\,,\\
\text{D4}_\partial\ (01257):&\qquad
\tilde\epsilon=s_\partial\,\Gamma_0\Gamma_1\Gamma_2\Gamma_5\Gamma_7\,\epsilon\,,
\end{split}
\end{align}
where the signs \(s_{\rm NS},s',s_\partial=\pm1\) encode the orientations, and we
count the independent solutions with an explicit \(32\times32\) representation
of the gamma matrices. The results are the following.
\begin{itemize}
\item Each of the boundary NS5-, D4\('\)- and D4\(_\partial\)-branes, added to the
bulk system \eqref{D4susy}, \eqref{NS5bulkSUSY}, preserves four supercharges,
for either orientation.
\item The D4\('\)-brane preserves the same four supercharges as the boundary
D6-brane \eqref{D6susy} for \(s'=+1\), and none of them for \(s'=-1\).
\item The bulk system with the boundary D6-brane, the boundary NS5-brane and the
D4\('\)-brane preserves the same four supercharges for \(s_{\rm NS}=-1\),
\(s'=+1\), and none for the other choices. Reversing the orientation of the
boundary D6-brane reverses the compatible orientations of the other two.
\item Without the colour D4-branes, the bulk NS5-brane, the boundary D6-brane
and the D4\('\)-brane with \(s'=+1\) preserve eight supercharges, as the NS5-,
D6- and D4-branes of~\cite{Witten:1997sc}; for \(s'=-1\) they preserve none.
\end{itemize}

To identify the fixed versus fluctuating components of the fermion $\Psi_+$ at
the endpoint, it is useful to introduce the boundary projector
\begin{equation}
  \Gamma_{{\rm NS5},\,{\rm br}}
  \equiv
  s_{\rm NS}\,\Gamma_0\Gamma_1\Gamma_2\Gamma_4\Gamma_5\Gamma_7\,,
  \qquad
  P^\pm_{{\rm NS5},\,{\rm br}}
  \equiv
  \frac{1\pm \Gamma_{{\rm NS5},\,{\rm br}}\bar\Gamma}{2}\,,
  \label{Pboundary}
\end{equation}
with \(s_{\rm NS}=-1\), the orientation compatible with the boundary D6-brane
\eqref{D6susy}.

The endpoint constraints are imposed locally on the D4 endpoint. For the
scalars \(A_M\) with \(M\) transverse to the D4, describing motion in the
\(x^M\) direction, we use the standard rule: motion along a direction
contained in the boundary brane is allowed, while motion in a direction
transverse to it is fixed. For the gauge field along the boundary, the
condition depends on the type of boundary brane: it is of Neumann type for a
D4-brane ending on an NS5-brane and of Dirichlet type for a D4-brane ending on
a D-brane, as for D3-branes ending on NS5- and D5-branes~\cite{Gaiotto:2008sa}.
The fermionic conditions are then fixed by supersymmetry: the variations
\(\delta A_M=\epsilon\Gamma_M\Psi\) and \(\delta\Psi\) must preserve the split
between fixed and fluctuating bosonic modes. Equivalently, the allowed fermions
are selected by the boundary projector compatible with the preserved four
supercharges.

Applying these rules to the boundary NS5 and D6 terminations gives the following schematic pattern:
\begin{align}
  \text{NS5 boundary brane:}\qquad
  &\text{Fixed: } A_{3,6}\,,\ P^-_{{\rm NS5},\,{\rm br}}\Psi_+\,,
  &&
  \text{Fluctuating: } A_{0,1,2,5}\,,\ P^+_{{\rm NS5},\,{\rm br}}\Psi_+\,,
  \nonumber\\[2mm]
  \text{D6 boundary brane:}\qquad
  &\text{Fixed: } A_{0,1,2,5}\,,\ P^+_{{\rm NS5},\,{\rm br}}\Psi_+\,,
  &&
  \text{Fluctuating: } A_{3,6}\,,\ P^-_{{\rm NS5},\,{\rm br}}\Psi_+\,.
  \label{endpointpattern}
\end{align}

In 3d $\mathcal N=2$ language, the fluctuating set for the NS5-type termination matches the field content of a vector multiplet, while the fluctuating set for the D6-type termination matches that of a chiral multiplet; in both cases these are restrictions of bulk fields to the boundary. Equivalently, the fixed sets are naturally interpreted as the 3d $\mathcal N=2$ multiplets that are fixed to background values at the boundary. This is the origin of the Neumann-like and Dirichlet-like interpretations described in Section~\ref{srelationtostringtheory}.

\vspace{1cm}

\section{Single-pole solutions in the D4\('\)-type class}\label{ad6bc}

In this appendix we collect the derivations underlying the analysis of Section~\ref{sendingond6}. We work within the single-pole ansatz \eqref{epolefields} and in the gauge $A_\perp=0$ \eqref{eaperpzero}. In this setup the BPS equations reduce to the algebraic system
\begin{align}
\phi_a
&=
\frac{g_a^2}{2}\big(q_a\bar q_a-\bar q_{a-1}q_{a-1}\big)
-\frac{g_a^2}{2N}\,\mathds{1}_a\,
\text{Tr}\big(q_a\bar q_a-\bar q_{a-1}q_{a-1}\big)\,,
\label{efundeq1app}\\
q_a&=\phi_a q_a-q_a\phi_{a+1}\,,
\label{efundeq2app}
\end{align}
with
\begin{align}
\phi_a=\phi^{x_a}T_{x_a,i_a}^{\ \ \ j_a}\,,
\qquad
q_a=q_{i_a}^{\ \ i_{a+1}}\,,
\qquad
\bar q_a=(q_a)^\dagger\,.
\end{align}

The goal of this appendix is threefold. First, we justify the structural
picture used in Section~\ref{sstructural}. Second, we present solvable classes
of \eqref{efinalequationgeneral}. Third, we prove the result on
\(\det B\) used in Section~\ref{sdualneumann}.

\subsection{Structure of the solution}\label{astructureofsolution}

In this subsection we establish the structural claims of
Section~\ref{sstructural}: the mutual commutativity of
$\phi_a$, $q_a\bar q_a$ and $\bar q_{a-1}q_{a-1}$,
and the decomposition of each $q_a$ into blocks of constant rank,
proportional to unitary matrices, along each $\hat\phi$-family.

%----------------------------------------------------------
\subsubsection*{Commuting data}
%----------------------------------------------------------

We now prove
\begin{equation}\label{ecommutator_app}
[\phi_a,q_a\bar q_a]=[\phi_a,\bar q_{a-1}q_{a-1}]=[q_a\bar q_a,\bar q_{a-1}q_{a-1}]=0\,.
\end{equation}
We start by multiplying \eqref{efundeq2app} on the right by \(\bar q_a\):
\begin{equation}
q_a\bar q_a
= \phi_a q_a\bar q_a - q_a\phi_{a+1}\bar q_a .
\end{equation}
Taking the Hermitian conjugate of \eqref{efundeq2app} gives
\begin{equation}
\bar q_a = \bar q_a\phi_a - \phi_{a+1}\bar q_a ,
\end{equation}
and therefore
\begin{equation}
q_a\phi_{a+1}\bar q_a
= q_a\bar q_a\phi_a - q_a\bar q_a .
\end{equation}
Substituting this relation into the previous equation gives
\begin{equation}
q_a\bar q_a
= \phi_a q_a\bar q_a - q_a\bar q_a\phi_a + q_a\bar q_a ,
\end{equation}
hence
\begin{equation}
[\phi_a,q_a\bar q_a]=0 .
\end{equation}
Replacing $a\to a-1$ and repeating the same argument gives
\begin{equation}
[\phi_a,\bar q_{a-1}q_{a-1}]=0\,.
\end{equation}
Using these commutation relations and \eqref{efundeq1app} we can also write
\begin{align}
\begin{split}
q_a\bar{q}_a\bar{q}_{a-1}q_{a-1}&=q_a\bar{q_a}\left(q_a\bar{q_a}-\frac{2}{g_a^2}\phi_a-\frac 1N\,\mathds{1}_a(\hdots) \right)=\left(q_a\bar{q_a}-\frac{2}{g_a^2}\phi_a-\frac 1N\,\mathds{1}_a(\hdots) \right)q_a\bar{q_a}=\bar{q}_{a-1}q_{a-1}q_a\bar{q}_a\,,
\end{split}
\end{align}
namely
\begin{align}\label{eqqqqcommutation}
[q_a\bar{q}_a,\bar{q}_{a-1}q_{a-1}]=0\,.
\end{align}
This proves \eqref{ecommutator_app}.

It follows that, by a constant gauge transformation preserving the gauge choice $A_\perp=0$, one may work in a basis in which for each node $a$
\begin{equation}
\phi_a,\qquad q_a\bar q_a,\qquad \bar q_{a-1}q_{a-1}
\end{equation}
are all diagonal.

%----------------------------------------------------------
%----------------------------------------------------------
\subsubsection*{$q_a$-blocks of constant rank proportional to unitary matrices}
%----------------------------------------------------------

The families are best described in terms of joint eigenspaces rather than
eigenspaces of $\phi_a$ alone. An eigenvalue of $\phi_a$ may be shared by a slot
of a family and by slots that do not belong to it, for instance a free
eigenvalue, so that its multiplicity can change along a family. By
\eqref{efundeq1app}, $\bar q_{a-1}q_{a-1}$ and $q_a\bar q_a$ differ by a linear
combination of $\phi_a$ and the identity,
\begin{equation}\label{ejointeigen_app}
\bar q_{a-1}q_{a-1}=q_a\bar q_a-\frac{2}{g_a^2}\,\phi_a+c_a\,\mathds{1}_a\,,
\qquad c_a\in\mathbb R\,,
\end{equation}
so at each node the joint eigenspaces of any two of the commuting operators
\eqref{ecommutator_app} coincide with those of all three. We denote by
$V_a(\lambda,s)$ the joint eigenspace at node $a$ on which $\phi_a=\lambda$ and
$q_a\bar q_a=s$.

The equation \eqref{efundeq2app} and its conjugate read
\begin{equation}
\phi_a q_a=q_a\big(\phi_{a+1}+1\big)\,,\qquad
\bar q_a\phi_a=\big(\phi_{a+1}+1\big)\bar q_a\,.
\end{equation}
Hence $q_a$ maps the joint eigenspace of $\big(\phi_{a+1},\bar q_aq_a\big)$ with
eigenvalues $(\lambda-1,s)$ into $V_a(\lambda,s)$, and $\bar q_a$ maps
$V_a(\lambda,s)$ back. For $s\neq0$ one has $\bar q_aq_a=s$ and $q_a\bar q_a=s$
on these two spaces, so the two maps are inverse to each other up to the factor
$s$: the two spaces have the same dimension, and on them
$q_a=\sqrt{s}\,U$ with $U$ unitary. By \eqref{ejointeigen_app} at the node
$a+1$, the joint eigenspace of $\big(\phi_{a+1},\bar q_aq_a\big)$ is itself one
of the spaces $V_{a+1}(\lambda-1,s')$.

Starting from a joint eigenspace and following the non-zero blocks of the
$q_a$, one therefore obtains a chain of joint eigenspaces of equal dimension,
connected by blocks proportional to unitary maps. This chain is a
$\hat\phi$-family. It starts at a joint eigenspace with $\bar q_{a-1}q_{a-1}=0$,
where no block enters, and ends at one with $q_a\bar q_a=0$, where no block
leaves. Its common dimension $M_{\hat\phi}$ is the multiplicity of the family,
which in general differs from the multiplicity of the corresponding eigenvalues
of $\phi_a$. In bases adapted to the joint eigenspaces, each block of the family
takes the form
\begin{equation}\label{eQpolar_app}
Q_{A+J}=C_{A+J}\,U_{A+J}\,,
\qquad
U_{A+J}\in U(M_{\hat\phi})\,,
\qquad
C_{A+J}\in \mathbb C\,.
\end{equation}
When the family winds around the quiver, several such blocks sit in the same
$q_a$, one per winding copy,
\begin{equation}
Q_a=\bigoplus_w C_{a,w}\,U_{a,w}\,,
\qquad
U_{a,w}\in U(M_{\hat\phi})\,.
\end{equation}

The equations \eqref{efundeq1app} and \eqref{efundeq2app} depend on the blocks
only through the non-negative quantities $|C_{a,w}|^2$. The phases of the
$C_{a,w}$ and the unitaries $U_{a,w}$ can be removed by $U(N)^K$ gauge
transformations acting on the families. Within $SU(N)^K$ this is possible
whenever the stabilizer of the data is discrete (Section~\ref{sdualneumann});
otherwise, some gauge-invariant relative phases of the blocks can survive, at
most as many as the rank of the stabilizer.

\subsection{The no-winding case}\label{asinglephihatandnowinding}
%----------------------------------------------------------

In this and the following subsections we solve special cases of the BPS system
\eqref{efinalequationgeneral} in closed form. We first treat the non-winding
case as a warm-up; this solution is used in Section~\ref{secSinglPhiNoWind} to
illustrate the brane interpretation of the single-pole data. We then turn to
the two saturating configurations discussed in Section~\ref{sdualneumann}: the
maximal-winding case in Appendix~\ref{asinglephihatwinding}, and the maximally
symmetric case, with a \(\hat\phi\)-family starting at every node, in
Appendix~\ref{asymmetriccase}.

In the first two cases the couplings enter only through a few fixed
combinations, which we define once and for all. For a family starting at node
$A$, we introduce the partial sums
\begin{equation}\label{eGHdef_app}
G_j\equiv \sum_{j'=0}^{j}\frac{1}{g_{A+j'}^2}\,,
\qquad
H_j\equiv \sum_{j'=0}^{j}\frac{j'}{g_{A+j'}^2}\,,
\qquad
-1\le j\le K-1\,,
\end{equation}
with the conventions $G_{-1}=H_{-1}=0$, together with the full-cycle sums
\begin{equation}\label{eGHfull_app}
G\equiv G_{K-1}=\sum_{j=0}^{K-1}\frac{1}{g_{A+j}^2}\,,
\qquad
H\equiv H_{K-1}=\sum_{j=0}^{K-1}\frac{j}{g_{A+j}^2}\,.
\end{equation}
Since the sum in $G$ runs over all the nodes of the quiver, $G$ is independent
of the starting node $A$; $G_j$, $H_j$ and $H$ do depend on $A$, but we suppress this dependence for economy of notation.

Consider a single non-winding family with data $(A,L,M)$, suppressing the $\hat\phi$ label for simplicity. The family contributes the eigenvalues
\begin{equation}
\hat\phi,\ \hat\phi-1,\ \ldots,\ \hat\phi-L
\end{equation}
at the nodes
\begin{equation}
A,\ A+1,\ \ldots,\ A+L\,.
\end{equation}
Using \eqref{eQpolar_app}, define
\begin{equation}
x_J\equiv |C_{A+J}|^2\,,
\qquad
0\le J\le L-1\,,
\end{equation}
and for convenience set
\begin{equation}
x_{-1}=x_L=0\,.
\end{equation}

Projecting \eqref{efundeq1app} onto the relevant eigenspace of the family at node $A+J$ gives
\begin{equation}\label{echainscalar_app}
\hat\phi-J
=
\frac{g_{A+J}^2}{2}\Big(x_J-x_{J-1}\Big)
-\frac{g_{A+J}^2}{2N}\,\text{Tr}\big(q_{A+J}\bar q_{A+J}-\bar q_{A+J-1}q_{A+J-1}\big)\,.
\end{equation}
For a single family of multiplicity $M$, one has
\begin{equation}
\text{Tr}(q_{A+J}\bar q_{A+J})=M\,x_J\,,
\qquad
\text{Tr}(\bar q_{A+J-1}q_{A+J-1})=M\,x_{J-1}\,,
\end{equation}
so \eqref{echainscalar_app} becomes
\begin{equation}\label{echainscalar2_app}
\hat\phi-J
=
\frac{g_{A+J}^2}{2}\left(1-\frac{M}{N}\right)x_J
-\frac{g_{A+J}^2}{2}\left(1-\frac{M}{N}\right)x_{J-1}\,.
\end{equation}
For $M=N$ the right-hand side vanishes identically, and \eqref{echainscalar2_app}
would require $\hat\phi-J=0$ at every position, which is impossible for $L\ge1$.
We therefore take $M\le N-1$ and $L\ge1$, the case $L=0$ being the trivial
datum.
Equivalently,
\begin{equation}\label{edifferenceeq_app}
x_J-x_{J-1}
=
\frac{2}{g_{A+J}^2}\,\frac{\hat\phi-J}{1-M/N}\,.
\end{equation}

Rather than solve \eqref{edifferenceeq_app} recursively immediately, it is convenient to sum suitable combinations and use the endpoint conditions $x_{-1}=x_L=0$.

%==========================================================
\subsubsection{Derivation of \texorpdfstring{$\hat\phi$}{phi-hat}}\label{anowindhatphi}
%==========================================================

Summing \eqref{edifferenceeq_app} from $J=0$ to $L$ and using the telescopic relation
\begin{equation}
\sum_{J=0}^{L}(x_J-x_{J-1})=x_L-x_{-1}=0\,,
\end{equation}
we obtain
\begin{equation}
0=\sum_{J=0}^{L}\frac{\hat\phi-J}{g_{A+J}^2}=\hat\phi\,G_L-H_L\,,
\end{equation}
where we used the partial sums \eqref{eGHdef_app} evaluated at $j=L$ (recall
that $L\le K-1$ in the non-winding case). Solving for $\hat\phi$ gives
\begin{equation}\label{egeneralhatphiapp}
\hat\phi
=
\frac{H_L}{G_L}\,.
\end{equation}

%==========================================================
\subsubsection{Derivation of the moduli \texorpdfstring{$|C_{A+J}|^2$}{|C|^2}}\label{anowindmoduli}
%==========================================================

We now solve for the $x_J=|C_{A+J}|^2$. From \eqref{edifferenceeq_app}, using $1/(1-M/N)=N/(N-M)$ together with \eqref{egeneralhatphiapp} and the partial sums \eqref{eGHdef_app},
\begin{equation}
x_J
=
\sum_{i=0}^{J}(x_i-x_{i-1})
=
\frac{2N}{N-M}\sum_{i=0}^{J}\frac{1}{g_{A+i}^2}\left(\frac{H_L}{G_L}-i\right)
=
\frac{2N}{N-M}\left(\frac{H_L}{G_L}\,G_J-H_J\right).
\end{equation}
We therefore arrive at the closed-form expression for the moduli,
\begin{equation}\label{egeneralCsqapp}
|C_{A+J}|^2
=
\frac{2N}{N-M}\,
\frac{H_L\,G_J-H_J\,G_L}{G_L}\,,
\qquad
0\le J\le L-1\,,
\end{equation}
together with $|C_{A+L}|^2=0$. Positivity of the moduli is not manifest in this form. However, expanding the products in the numerator, the terms with $j\le J$ cancel pairwise by antisymmetry, leaving a double sum with manifestly non-negative summands:
\begin{equation}\label{ePident_app}
H_L\,G_J-H_J\,G_L
=
\sum_{i=0}^{J}\sum_{j=0}^{L}\frac{j-i}{g_{A+i}^2g_{A+j}^2}
=
\sum_{i=0}^{J}\sum_{j=J+1}^{L}
\frac{j-i}{g_{A+i}^2g_{A+j}^2}
\ge0\,,
\end{equation}
since every term in the last double sum has $j>i$. All the coefficients $|C_{A+J}|^2$ are therefore non-negative for positive couplings $g_a^2>0$: the non-winding solution exists for arbitrary positive couplings. Substituting \eqref{ePident_app} into \eqref{egeneralCsqapp} reproduces the form quoted in the main text, eq.~\eqref{egeneralhatphi}.

%==========================================================
\subsection{Maximal winding case}\label{asinglephihatwinding}
%==========================================================

In this subsection we exploit the circular topology of the quiver to construct a second solution within the same single-pole D4\('\)-type class, in which a single $\hat\phi$-family winds around the quiver as many times as $SU(N)$ tracelessness allows. The resulting configuration is completely rigid: the eigenvalues of every $\phi_a$ are fixed to (half-)integer values independent of the gauge couplings, exactly one node of the quiver is completely filled by the family, and the moduli $|C_{a,w}|^2$ are determined uniquely and in closed form. As for every single-pole datum with discrete stabilizer, the boundary data $(\phi_a,q_a)$ break the gauge symmetry $SU(N)^K$ down to its diagonal center $\mathbb{Z}_N$, the minimal stabilizer that any boundary datum can have. Its place among the single-pole data with discrete stabilizer is discussed in Section~\ref{sdualneumann}.

%==========================================================
\subsubsection{The winding ansatz and its maximality}\label{awindansatz}
%==========================================================

We take a single family of multiplicity $M=1$, starting at node $A=1$ (any other choice of starting node is obtained by a cyclic relabeling of the nodes and yields a distinct copy of the solution), and the maximal length $L$ allowed for the family,
\begin{equation}\label{ewindansatz_app}
L=(N-1)K\,.
\end{equation}
The maximality of $L$ follows from \eqref{ebudgetup}; it is crucial for reducing the stabilizer of the boundary data to $\mathbb{Z}_N$. The choice $M=1$ is dictated by the same goal. A family of multiplicity $M$ fills each of its slots with an $M$-fold degenerate eigenvalue of $\phi_a$, and the unitary factors of its blocks \eqref{eQpolar_app} can be gauged to the identity along the chain, which is a path without closed loops; an $SU(M)$ rotation acting identically on all the slots of the family then commutes with the whole boundary data, so for $M\ge2$ the stabilizer would contain a continuous $SU(M)$ factor. Only for $M=1$ can the eigenvalues of $\phi_a$ be non-degenerate and the stabilizer discrete.

We label the family positions $J\in\{0,1,\ldots,L\}$ by
\begin{equation}
J=wK+j\,,\qquad 0\le j\le K-1\,,
\end{equation}
so that $j$ fixes the node $a=1+j$ visited at position $J$, and $w$ counts the completed windings. The position $J$ contributes the eigenvalue $\hat\phi-J=\hat\phi-wK-j$ to $\phi_{1+j}$. Since $L=(N-1)K$, the winding number takes the values $w=0,\ldots,N-1$ at node $1$ and $w=0,\ldots,N-2$ at the nodes $1+j$ with $j\ge1$. The occupancy of the quiver is therefore asymmetric:
\begin{itemize}
\item node $1$ is \emph{completely filled} by the family: its $N$ slots carry the eigenvalues $\hat\phi-wK$, $w=0,\ldots,N-1$, and no eigenvalue is left over to be fixed by \eqref{eforreference};
\item each node $1+j$ with $1\le j\le K-1$ carries the $N-1$ family eigenvalues $\hat\phi-wK-j$, $w=0,\ldots,N-2$, plus one \emph{free} eigenvalue $\phi_{1+j,N}$, fixed below by tracelessness.
\end{itemize}
Correspondingly, each bifundamental $q_a$ consists of $1\times1$ blocks
connecting consecutive family positions. No block acts on a free slot.
At the filled node $1$, however, all $N$ slots belong to the family: the
terminal slot receives a non-zero block from the preceding family position,
while no block leaves it.

Tracelessness fixes the highest eigenvalue:
\begin{equation}\label{ewindinghatphi_app}
0=\sum_{w=0}^{N-1}\big(\hat\phi-wK\big)=N\hat\phi-K\,\frac{N(N-1)}{2}
\qquad\Longrightarrow\qquad
\boxed{\ \hat\phi=\frac{K(N-1)}{2}\ }\,.
\end{equation}
This leads the $\phi_a$ matrices to have a \textit{coupling-independent} fixed structure in terms of eigenvalues, with
\begin{align}
\phi_1=K\text{diag}\left\{\frac{N-1}{2},\hdots\,,-\frac{N-1}{2}\right\}\,,
\end{align}
and all the $\phi_a$ eigenvalues fixed to integer or half-integer values, with tracelessness fixing their last eigenvalue to be
\begin{align}\label{entheigenv}
\phi_{a,N}=(N-1)\left(a-1-\frac{K}{2}\right)=(N-1)\left(\hat{\phi}+a-1-\frac{KN}{2}\right)\,.
\end{align}
The length \eqref{ewindansatz_app} is the maximal one allowed by the rank bound
\eqref{ebudgetup}.

\paragraph{Degenerate $\phi_a$ eigenvalues.} Within the family, the eigenvalues at a given node are spaced by $K$ and never collide. The $N^{\text{th}}$ eigenvalue \eqref{entheigenv} coincides with a family eigenvalue $\hat\phi-wK-j$ if and only if
\begin{equation}\label{ecollision_app}
Nj=K(N-1-w)\,,
\end{equation}
which admits solutions with $1\le j\le K-1$, $0\le w\le N-2$ if and only if $d\equiv\gcd(N,K)>1$. In that case the collisions occur exactly at the $d-1$ nodes $j=tK/d$, $t=1,\ldots,d-1$ (with $w=N-1-tN/d$), each carrying a single doubly degenerate eigenvalue. Because the spectrum is coupling-independent, these degeneracies can never be lifted by moving on the conformal manifold; as shown in Appendix~\ref{awindstabilizer}, however, they do not enhance the stabilizer in this case. For $\gcd(N,K)=1$, the $N$ eigenvalues of $\phi_a$ are distinct at every node.

%==========================================================
\subsubsection{BPS equations}\label{awindbps}
%==========================================================

Following \eqref{eQpolar_app}, let $C_{1+j,w}$ denote the $1\times1$ block of $q_{1+j}$ connecting the family position $J=wK+j$ to the position $J+1$, and set
\begin{equation}
x_J\equiv |C_{1+j,w}|^2\,,\qquad J=wK+j\,,\qquad 0\le J\le L-1\,,
\end{equation}
with the endpoint convention
\begin{equation}\label{eendpoint_app}
x_{-1}=x_L=0
\end{equation}
(no block enters the top position $J=0$, and no block leaves the terminal position $J=L$). The node traces collect the winding copies stacked at each node,
\begin{equation}\label{enodetrace_app}
T_j\equiv \mathrm{Tr}\big(q_{1+j}\bar q_{1+j}\big)=\sum_{w=0}^{N-2}x_{wK+j}\,,
\qquad
T_{-1}\equiv T_{K-1}\quad\text{(by circular identification of nodes)}\,,
\end{equation}
where for $j=0$ the would-be $w=N-1$ term is the terminal position and vanishes by \eqref{eendpoint_app}.

Projecting \eqref{efundeq1app} onto the family position $J=wK+j$ gives the analogue of \eqref{echainscalar_app},
\begin{equation}\label{echainscalarwind_app}
\hat\phi-J
=
\frac{g_{1+j}^2}{2}\big(x_J-x_{J-1}\big)
-\frac{g_{1+j}^2}{2N}\big(T_j-T_{j-1}\big)\,,
\qquad 0\le J\le L\,,
\end{equation}
a set of $(N-1)K+1$ equations, one per position, including the two endpoint positions where \eqref{eendpoint_app} is used. Projecting \eqref{efundeq1app} onto the free slot of node $1+j$, for $j\ge1$, only the trace term survives, since the free slot carries no $q$-entry:
\begin{equation}\label{efreebps_app}
\phi_{1+j,N}
=
-\frac{g_{1+j}^2}{2N}\big(T_j-T_{j-1}\big)\,,
\qquad 1\le j\le K-1\,,
\end{equation}
a set of $K-1$ further equations. The filled node $1$ provides no equation of this type.

The free-slot equations \eqref{efreebps_app}, combined with the tracelessness values \eqref{entheigenv}, fix the trace differences at the nodes carrying a free slot, while the remaining difference $T_0-T_{K-1}$ follows from the telescoping identity $\sum_{j=0}^{K-1}(T_j-T_{j-1})=0$, valid for arbitrary $x_J$ by the circular identification $T_{-1}=T_{K-1}$:
\begin{equation}\label{etracefixed_app}
T_j-T_{j-1}=\frac{2N(N-1)}{g_{1+j}^2}\left(\frac{K}{2}-j\right)\quad (1\le j\le K-1)\,,
\qquad
T_0-T_{K-1}=-2N(N-1)\,S\,,
\end{equation}
where we introduced the combination
\begin{equation}\label{eSdef_app}
S\equiv\sum_{j=1}^{K-1}\frac{1}{g_{1+j}^2}\left(\frac{K}{2}-j\right)
=\frac{K}{2}\big(G-G_0\big)-H\,.
\end{equation}

With the trace differences fixed, the chain projections \eqref{echainscalarwind_app} are $(N-1)K+1$ equations for the $(N-1)K$ unknowns $x_J$: apparently one too many. The excess is only apparent, and it is consistent with the fact that $\hat\phi$ has already been determined. Indeed, first solving \eqref{echainscalarwind_app} for the increment gives
\[
x_J-x_{J-1}
=\frac{2}{g_{1+j}^2}(\hat\phi-J)+\frac{T_j-T_{j-1}}{N},
\qquad J=wK+j.
\]
Summing over all family positions, the left-hand side telescopes to
$x_L-x_{-1}=0$ by \eqref{eendpoint_app}. The right-hand side also vanishes
after inserting \eqref{ewindinghatphi_app} and \eqref{etracefixed_app}.
More generally, if $\hat\phi$ is left undetermined while the free-slot
eigenvalues and trace differences are expressed using tracelessness at the
other nodes, the sum reduces to
$\tfrac{2N}{g_1^2}\big[\hat\phi-\tfrac{K(N-1)}{2}\big]$. The sum of the chain equations is therefore nothing but the tracelessness condition at the filled node, \eqref{ewindinghatphi_app}: closure of the maximally wound chain and tracelessness at the filled node are one and the same condition, which is the mechanism behind the rigidity of the spectrum. Likewise, the $K-1$ free-slot equations used above are implied by the chain equations: summing \eqref{echainscalarwind_app} at fixed residue $j\ge1$ over $w$ reproduces them. The system is thus exactly determined, and the solution constructed below is unique within the ansatz \eqref{ewindansatz_app}, with no continuous moduli.

%==========================================================
\subsubsection{Solving the recursion}\label{awindrecursion}
%==========================================================

Substituting \eqref{etracefixed_app} into \eqref{echainscalarwind_app} and solving for the increment gives, for $J=wK+j$,
\begin{equation}\label{ediffwind_app}
x_J-x_{J-1}
=
\frac{2}{g_{1+j}^2}\left(\frac{K(N-1)}{2}-wK-j\right)+2(N-1)\,v_j\,,
\qquad
v_j\equiv
\begin{cases}
\dfrac{1}{g_{1+j}^2}\left(\dfrac{K}{2}-j\right)\,, & 1\le j\le K-1\,,\\[2ex]
-S\,, & j=0\,,
\end{cases}
\end{equation}
so that $\sum_{j=0}^{K-1}v_j=0$ by construction. Iterating from $x_{-1}=0$, we obtain $x_J$ by summing \eqref{ediffwind_app} over the positions $J'=0,\ldots,J$, splitting the sum into $w$ complete winding cycles (levels $w'=0,\ldots,w-1$, each running over $j'=0,\ldots,K-1$) and a partial cycle at level $w$ running over $j'=0,\ldots,j$.

\medskip
\textit{One complete cycle at winding level $w'$.} Fixing $w'$ and summing \eqref{ediffwind_app} over $j'=0,\ldots,K-1$, the $v$-terms drop out and, in terms of the full-cycle sums \eqref{eGHfull_app},
\begin{equation}
\sum_{j'=0}^{K-1}\frac{2}{g_{1+j'}^2}\left(\frac{K(N-1)}{2}-w'K-j'\right)
=K(N-1-2w')\,G-2H\,.
\end{equation}

\medskip
\textit{Sum over $w$ complete cycles.} Summing over $w'=0,\ldots,w-1$ and using $\sum_{w'=0}^{w-1}(N-1-2w')=w(N-1)-w(w-1)=w(N-w)$,
\begin{equation}
\sum_{w'=0}^{w-1}\Big[K(N-1-2w')\,G-2H\Big]
=K\,w(N-w)\,G-2wH\,.
\end{equation}

\medskip
\textit{Partial cycle at winding level $w$.} Summing \eqref{ediffwind_app} at fixed level $w$ over $j'=0,\ldots,j$, the first term produces the partial sums \eqref{eGHdef_app},
\begin{equation}
\sum_{j'=0}^{j}\frac{2}{g_{1+j'}^2}\left(\frac{K(N-1)}{2}-wK-j'\right)
=K(N-1-2w)\,G_j-2H_j\,,
\end{equation}
while the $v$-terms give, using $v_0=-\sum_{j'=1}^{K-1}v_{j'}$ to convert the partial sum into the complementary tail,
\begin{equation}
2(N-1)\sum_{j'=0}^{j}v_{j'}
=-2(N-1)\!\!\sum_{j'=j+1}^{K-1}\!\frac{1}{g_{1+j'}^2}\left(\frac{K}{2}-j'\right)
=2(N-1)\left[\big(H-H_j\big)-\frac{K}{2}\big(G-G_j\big)\right].
\end{equation}

\medskip
\textit{Total.} Adding the three contributions and collecting the coefficients of $G$, $H$, $G_j$ and $H_j$,
\begin{align}
\begin{split}
x_{wK+j}
&=K\big[w(N-w)-(N-1)\big]G
+2\big[(N-1)-w\big]H\\
&\quad+K\big[(N-1-2w)+(N-1)\big]G_j
-2\big[1+(N-1)\big]H_j\\[1ex]
&=K(w-1)(N-1-w)\,G+2(N-1-w)\,H+2K(N-1-w)\,G_j-2N\,H_j\,,
\end{split}
\end{align}
where in the first term we used the factorization $w(N-w)-(N-1)=(w-1)(N-1-w)$. We therefore arrive at the closed-form expression for the moduli:
\begin{equation}\label{ewindingmoduli_app}
\boxed{\ |C_{1+j,w}|^2
=(N-1-w)\Big[K(w-1)\,G+2H+2K\,G_j\Big]-2N\,H_j\,,\ }
\end{equation}
valid for $0\le j\le K-1$ and $0\le w\le N-2$.

\medskip
\textit{Checks.} Formally extending \eqref{ewindingmoduli_app} to the terminal position $(w,j)=(N-1,0)$ gives $0\cdot[\ldots]-2N\,H_0=0$ (recall $H_0=0$), i.e.\ $x_L=0$: the chain closes, as required by \eqref{eendpoint_app}. The outermost links are
\begin{equation}\label{eouterlinks_app}
x_0=(N-1)\big(K\,G_0-2S\big)\,,
\qquad
x_{L-1}=(N-1)\big(K\,G-2H\big)=(N-1)\big(K\,G_0+2S\big)\,,
\end{equation}
where we used \eqref{eSdef_app}. 

Finally, at $K=1$ one should expect the result to reduce to the regular Nahm pole of $\mathcal{N}=4$ SYM. This is precisely what happens: $\phi=\phi_1$ becomes the Cartan generator in the $N$-dimensional irreducible representation of $\mathfrak{su}(2)$, while the hypermultiplet matrices become, upon rescaling, the corresponding ladder operators.

%==========================================================
\subsubsection{The stabilizer is the diagonal \texorpdfstring{$\mathbb{Z}_N$}{Z\_N}}\label{awindstabilizer}
%==========================================================

We now compute the stabilizer of the boundary data, i.e.\ the subgroup of constant gauge transformations $(g_1,\ldots,g_K)\in SU(N)^K$ (which preserve the gauge choice $A_\perp=0$) such that
\begin{equation}\label{estabcond_app}
g_a\,\phi_a\,g_a^{-1}=\phi_a\,,
\qquad
g_a\,q_a\,g_{a+1}^{-1}=q_a\,,
\qquad a=1,\ldots,K\,.
\end{equation}
\medskip
\textit{Step 1: the $g_a$ are diagonal.} The first condition in \eqref{estabcond_app} says that $g_a$ preserves every eigenspace of $\phi_a$. Wherever the eigenvalues are distinct this makes $g_a$ diagonal in the slot basis, $g_a=\mathrm{diag}\big(e^{i\theta_{a,s}}\big)$, with one phase per slot. The only exceptions are the $d-1$ resonant nodes of \eqref{ecollision_app}, present when $d\equiv\gcd(N,K)>1$, where the free slot and one family slot share an eigenvalue and $g_a$ is a priori block-diagonal with a $U(2)$ factor. Even there $g_a$ is diagonal: that family slot carries an incoming non-zero link, whose invariance under \eqref{estabcond_app} forces the family direction inside the degenerate $2$-plane to be an eigenvector of $g_a$, and unitarity then forces its orthogonal complement, the free slot, to be one as well. The steps below therefore apply for all $(N,K)$.

\medskip
\textit{Step 2: one common phase along the chain.} Written out in the slot basis, the second condition in \eqref{estabcond_app} imposes, for each non-zero block $C_{1+j,w}$ joining the slot at position $J$ (at node $1+j$) to the slot at position $J+1$ (at the next node),
\begin{equation}
e^{i(\theta_{J}-\theta_{J+1})}=1\,,
\end{equation}
where $\theta_J$ denotes the phase carried by the slot at position $J$. Since all the $(N-1)K$ links are non-zero in the interior of the existence chamber (Appendix~\ref{awindpositivity}), the phases of \emph{all} the $(N-1)K+1$ chain slots are equal to a single common phase $\alpha$. Crucially, this includes the links at $j=K-1$, which connect the winding level $w$ at node $K$ to the level $w+1$ back at node $1$, chaining all the winding levels together. The free slots, which carry no $q$-entry, have a priori unconstrained phases $\beta_{1+j}$, $1\le j\le K-1$.

\medskip
\textit{Step 3: determinants.} It remains to impose $\det g_a=1$ at each node --- the group-level counterpart of the tracelessness conditions used above, since the gauge group is $SU(N)^K$. At a node $1+j$ with $j\ge1$, the condition
\begin{equation}
\det g_{1+j}=e^{i(N-1)\alpha}\,e^{i\beta_{1+j}}=1
\end{equation}
merely \emph{fixes} the free phase, $\beta_{1+j}=-(N-1)\alpha \bmod 2\pi$, and imposes nothing on $\alpha$: the free slot absorbs the determinant condition. At the filled node $1$, instead, all $N$ slots lie on the chain, so $g_1=e^{i\alpha}\mathds{1}_N$ and there is no free phase left; the determinant condition becomes a genuine constraint on $\alpha$:
\begin{equation}\label{emaximalityandzn}
\det g_1=e^{iN\alpha}=1
\qquad\Longrightarrow\qquad
\alpha=\frac{2\pi k}{N}\,,\qquad k\in\mathbb{Z}\,.
\end{equation}

\medskip
\textit{Step 4: the stabilizer.} For $\alpha=2\pi k/N$, the fixed free phases satisfy $e^{i\beta_{1+j}}=e^{-i(N-1)\alpha}=e^{i\alpha}$, so that
\begin{equation}\label{estabresult_app}
g_a=e^{2\pi ik/N}\,\mathds{1}_N\quad \text{at every node}\,,
\end{equation}
i.e.\ the stabilizer of the boundary data is exactly the diagonal center
\begin{equation}
\mathbb{Z}_N\subset SU(N)^K\,.
\end{equation}
Notice that the maximality of the family length is crucial here: without it, no node would be completely filled and the constraint \eqref{emaximalityandzn} would be absent.
The maximal-winding solution attains this floor for every $(N,K)$.

The counting behind this result is worth isolating: the number of independent conditions on the chain phase $\alpha$ equals the number of nodes carrying no free slot. Here there is exactly one such node, giving the single condition $e^{iN\alpha}=1$. Had every node retained a free slot, every determinant condition would have been absorbed by a free phase, and a full $U(1)$ worth of $\alpha$ would have survived.

\subsubsection{An explicit example}\label{awindexample}
%==========================================================

To illustrate the structure concretely, take \(N=3\), \(K=2\) with couplings \(g_1^2=1\) and \(g_2^2=3\), so that
\begin{equation}
G=\frac43\,,\qquad H=\frac13\,,\qquad G_0=1\,,\qquad H_0=0\,,\qquad G_1=\frac43\,,\qquad H_1=\frac13\,.
\end{equation}
The spectrum is rigid: \eqref{ewindinghatphi_app} and \eqref{entheigenv} give $\hat\phi=2$ and $\phi_{2,3}=0$ independently of the couplings, so
\begin{equation}
\phi_1=\mathrm{diag}\big(2,\,0,\,-2\big)\,,
\qquad
\phi_2=\mathrm{diag}\big(1,\,-1,\,0\big)\,,
\end{equation}
where at node $1$ all three eigenvalues belong to the family ($w=0,1,2$), while at node $2$ the first two belong to the family ($w=0,1$) and the last is the free eigenvalue. The moduli \eqref{ewindingmoduli_app} are
\begin{equation}
x_0=\frac{4}{g_1^2}=4\,,\qquad
x_1=\frac{4}{g_1^2}+\frac{2}{g_2^2}=\frac{14}{3}\,,\qquad
x_2=\frac{4}{g_1^2}+\frac{2}{g_2^2}=\frac{14}{3}\,,\qquad
x_3=\frac{4}{g_1^2}=4\,,
\end{equation}
all strictly positive, as guaranteed for $K=2$ by \eqref{echamber_app}. It is immediate to verify the full matrix equation \eqref{efundeq1app}. At node $1$, whose slots sit at the positions $J=0,2,4$,
\begin{equation}
q_1\bar q_1-\bar q_2 q_2=\mathrm{diag}\big(x_0-x_{-1},\,x_2-x_1,\,x_4-x_3\big)=\mathrm{diag}\big(4,\,0,\,-4\big)\,,
\end{equation}
which is traceless, so the trace term in \eqref{efundeq1app} drops and $\tfrac{g_1^2}{2}\,\mathrm{diag}(4,0,-4)=\phi_1$. At node $2$, whose family slots sit at the positions $J=1,3$,
\begin{equation}
q_2\bar q_2-\bar q_1 q_1=\mathrm{diag}\big(x_1-x_0,\,x_3-x_2,\,0\big)=\mathrm{diag}\big(\tfrac23,\,-\tfrac23,\,0\big)\,,
\end{equation}
again traceless, and $\tfrac{g_2^2}{2}\,\mathrm{diag}(\tfrac23,-\tfrac23,0)=\phi_2$. As for the stabilizer: all four links are non-zero, so the five family slots (positions $J=0,\ldots,4$) carry a common phase $\alpha$; node $1$ is filled, so $g_1=e^{i\alpha}\mathds{1}_3$ and $\det g_1=1$ forces $\alpha\in\tfrac{2\pi}{3}\mathbb{Z}$, while at node $2$ the free phase is fixed to $\beta_2=-2\alpha=\alpha \bmod 2\pi$. The stabilizer is $g_1=g_2=e^{2\pi ik/3}\,\mathds{1}_3$, the diagonal center $\mathbb{Z}_3$, as in \eqref{estabresult_app}.

%----------------------------------------------------------
\subsection{The maximally symmetric case}\label{asymmetriccase}
%----------------------------------------------------------

The two solutions above are built from a single $\hat\phi$-family. We now consider the opposite extreme within the single-pole class: a configuration containing $K$ families, one starting at each node, so that every eigenvalue of every $\phi_a$ belongs to a family. Maximal symmetry is imposed as part of the ansatz: all the families are taken to start from the same value of $\hat\phi$, so that the discrete data are invariant under the cyclic relabeling of the nodes. This cyclic relabeling is a symmetry of the bulk theory on the equal-coupling
locus, and the ansatz is designed to respect it.

%==========================================================
\subsubsection{The moduli}\label{asymmmoduli}
%==========================================================

Just like in \eqref{ewindinghatphi_app}, the $\hat\phi$ values of all the $K$ families are fixed by tracelessness. For each $J=0,\ldots,N-1$, the node $a$ carries the slot at position $J$ of the
family starting at $a-J$ (node labels mod $K$, so that for $N>K$ a family may
visit the same node several times). Since all the families start from the same
$\hat\phi$, every node carries the same spectrum $\{\hat\phi-J\}_{J=0}^{N-1}$; imposing $\mathrm{Tr}\,\phi_a=0$ gives $N\hat\phi-N(N-1)/2=0$, that is
\begin{equation}\label{esymmhatphi_app}
\hat\phi=\frac{N-1}{2}\,,\qquad \phi_a=\mathrm{diag}\Big(\tfrac{N-1}{2},\ \tfrac{N-1}{2}-1,\ \ldots,\ -\tfrac{N-1}{2}\Big)\,,
\end{equation}
for every $N$ and $K$ and independently of the gauge couplings. The $N$ entries are distinct, and no node retains a free slot: the whole spectrum is fixed before any equation for the $q_a$ is imposed, and what remains to be solved are precisely those equations.

The remaining unknowns are the moduli of the blocks \eqref{eQpolar_app},
\begin{equation}
x^{(A)}_J\equiv\big|C^{(A)}_{A+J}\big|^2\,,\qquad 0\le J\le N-2\,,\qquad\text{with}\qquad x^{(A)}_{-1}=x^{(A)}_{N-1}=0\,.
\end{equation}
Because every node now carries $N$ family slots, the trace terms in \eqref{efundeq1app} are no longer proportional to a single modulus. At the node $a$,
\begin{equation}\label{esymmtraces_app}
\mathrm{Tr}\,(q_a\bar q_a)=\sum_{J=0}^{N-2}x^{(a-J)}_J\equiv T_a\,,\qquad \mathrm{Tr}\,(\bar q_{a-1}q_{a-1})=T_{a-1}\,,
\end{equation}
and it is convenient to introduce
\begin{equation}\label{esymmu_app}
u_a\equiv\frac{T_a-T_{a-1}}{N}\,,\qquad\text{which satisfies}\qquad \sum_{a=1}^{K}u_a=0\,,
\end{equation}
the last identity following because the sum telescopes around the quiver. Projecting \eqref{efundeq1app} onto the slot at position $J$ of the family $A$, that is at the node $a=A+J$, and using \eqref{esymmhatphi_app}, one finds
\begin{equation}\label{esymmrec_app}
x^{(A)}_J-x^{(A)}_{J-1}=\frac{N-1-2J}{g_{A+J}^2}+u_{A+J}\,.
\end{equation}
Summing over $J=0,\ldots,N-1$ and using the endpoint conditions $x^{(A)}_{-1}=x^{(A)}_{N-1}=0$ gives one linear relation per family,
\begin{equation}\label{esymmcirc_app}
\sum_{J=0}^{N-1}u_{A+J}=-\sum_{J=0}^{N-1}\frac{N-1-2J}{g_{A+J}^2}\,,\qquad A=1,\ldots,K\,.
\end{equation}

The relations \eqref{esymmcirc_app} are solved by differencing. Subtracting the relation at $A$ from the one at $A+1$, the two sums on the left overlap in $N-1$ terms and
\begin{equation}\label{esymmdiff_app}
u_{A+N}-u_A=R_{A+1}-R_A\,,\qquad\text{where}\qquad R_A\equiv-\sum_{J=0}^{N-1}\frac{N-1-2J}{g_{A+J}^2}\,.
\end{equation}
When $\gcd(N,K)=1$ the shift by $N$ generates the whole of $\mathbb{Z}_K$, so iterating \eqref{esymmdiff_app} determines every $u_a$ in terms of any one of them, and going once around the cycle is consistent because $\sum_A(R_{A+1}-R_A)=0$. The remaining additive constant is fixed by $\sum_a u_a=0$ in \eqref{esymmu_app}, which is itself what one gets by summing \eqref{esymmcirc_app} over $A$ and using $\sum_{J=0}^{N-1}(N-1-2J)=0$. The $u_a$ are therefore uniquely determined, and \eqref{esymmrec_app} integrates to
\begin{equation}\label{esymmmoduli_app}
x^{(A)}_J=\sum_{J'=0}^{J}\left[\frac{N-1-2J'}{g_{A+J'}^2}+u_{A+J'}\right]\,,\qquad 0\le J\le N-2\,.
\end{equation}
As in \eqref{ewindingmoduli_app}, the moduli are linear and homogeneous in the inverse couplings, and \eqref{esymmcirc_app} is precisely the statement that $x^{(A)}_{N-1}=0$, so that each chain closes. As always, the configuration defines a boundary datum only where all the $x^{(A)}_J$ are strictly positive.

When instead $\gcd(N,K)=d>1$ the shift by $N$ generates only a subgroup of index $d$, whose orbits are the $d$ residue classes modulo $d$, so \eqref{esymmdiff_app} determines the $u_a$ only within each class and up to $d$ constants. The system is then no longer solvable for arbitrary couplings, and the obstruction is visible directly in \eqref{esymmcirc_app}. Summing that relation over one residue class modulo $d$, the left-hand side meets each $u_a$ exactly $N/d$ times, because $d$ divides $N$, and therefore reduces to $(N/d)\sum_a u_a=0$ by \eqref{esymmu_app}. The right-hand side must then vanish as well, which gives $d$ conditions on the couplings, one of which is automatic since they sum to zero. Together they restrict the couplings to the subspace on which
\begin{equation}\label{esymmgcdlocus_app}
\sum_{a\,\equiv\,r\ (\mathrm{mod}\ d)}\frac{1}{g_a^2}\qquad\text{is independent of }r=0,\ldots,d-1\,,
\end{equation}
of codimension $d-1$. This subspace always contains the equal-coupling locus, where the bulk
theory has a $\mathbb{Z}_K$ symmetry. When $d=K$, that is when $K$ divides $N$,
it coincides with that locus; the common positive coupling remains arbitrary. On it the $u_a$ are not unique but come with $d-1$ flat directions, so that for $\gcd(N,K)>1$ the maximally symmetric configuration exists only there, and is not rigid: besides the $d-1$ flat directions of the $u_a$, $d-1$ gauge-invariant relative phases of the blocks survive (Appendix~\ref{astructureofsolution}).

%==========================================================
\subsubsection{The stabilizer}\label{asymmstabilizer}
%==========================================================

The spectrum \eqref{esymmhatphi_app} has $N$ distinct entries, so an element of $SU(N)^K$ preserving it must be diagonal at every node: before the hypermultiplets are taken into account the stabilizer is the maximal torus $U(1)^{N-1}$ at each node. The hypermultiplets then cut this down cyclically. Because every node carries the same spectrum, the unit-jump rule \eqref{ephinonzeroqmain} forces the non-zero blocks of $q_a$ to join the slot $i$ of the node $a$ to the slot $i+1$ of the node $a+1$, for $i=1,\ldots,N-1$; the last slot of each node carries no outgoing link, which is the endpoint condition $x^{(A)}_{N-1}=0$. Wherever the moduli are positive all these links are non-zero, and each of them forces the two phases it joins to agree, $\theta_{a,i}=\theta_{a+1,i+1}$. Iterating this identification chains together the $N$ slots of a given family, so that each family carries a single phase $\alpha_A$, and
\begin{equation}\label{esymmstabelt_app}
g_a=\mathrm{diag}\big(e^{i\alpha_a},\,e^{i\alpha_{a-1}},\ \ldots,\ e^{i\alpha_{a-N+1}}\big)\,.
\end{equation}
This leaves $K$ phases, one per family, on which the $K$ conditions $\det g_a=1$ read
\begin{equation}\label{esymmdet_app}
\sum_{J=0}^{N-1}\alpha_{a-J}\in 2\pi\mathbb{Z}\,,\qquad a=1,\ldots,K\,.
\end{equation}
These are solved by the same differencing used for the moduli: subtracting the condition at $a-1$ from the one at $a$ leaves $\alpha_a\equiv\alpha_{a-N}$ modulo $2\pi$. Given that $\alpha_a\sim\alpha_{a+K}$, if $\gcd(N,K)=1$ this set of equations links all the $\alpha_a$ together into a single $\alpha$, and $\det g_a=1$ then fixes the allowed transformations to be the diagonal centre
\begin{equation}\label{esymmstabres_app}
\mathbb{Z}_N\subset SU(N)^K\,.
\end{equation}
If instead $\gcd(N,K)=d>1$ the same equations split into $d$ independent systems, leaving $d$ independent $\alpha$-values, one per residue class modulo $d$. Since $d$ divides $N$, every window of $N$ consecutive slots meets each class exactly $N/d$ times, so the $K$ determinant conditions all collapse to the single relation $(N/d)\sum_{r=0}^{d-1}\alpha^{(r)}\in2\pi\mathbb{Z}$, and the stabilizer is
\begin{equation}\label{esymmstabgcd_app}
U(1)^{d-1}\times\mathbb{Z}_{N/d}\,,
\end{equation}
which reduces to \eqref{esymmstabres_app} at $d=1$. For $\gcd(N,K)=1$ this is the same minimal stabilizer found in Appendix~\ref{awindstabilizer} for the maximal-winding solution: minimality of the stabilizer does not distinguish between the two.

\subsubsection{Examples}\label{asymmexamples}
%==========================================================

For $N=2$ each family carries a single modulus, and $\gcd(2,K)=1$ requires $K$ odd. Solving \eqref{esymmcirc_app} gives
\begin{equation}\label{esymmN2_app}
x^{(A)}_0=\sum_{j=0}^{K-1}\frac{(-1)^j}{g_{A-j}^2}\,,
\end{equation}
which at $K=3$ reduces to $x^{(A)}_0=1/g_A^2+1/g_{A+1}^2-1/g_{A+2}^2$.

For $N=4$, $K=3$ one finds $u_A=1/g_{A-1}^2-1/g_{A+1}^2$ and
\begin{equation}\label{esymm43_app}
x^{(A)}_0=\frac{3}{g_A^2}+\frac{1}{g_{A-1}^2}-\frac{1}{g_{A+1}^2}\,,\qquad x^{(A)}_1=\frac{4}{g_A^2}\,,\qquad x^{(A)}_2=\frac{3}{g_A^2}+\frac{1}{g_{A+1}^2}-\frac{1}{g_{A-1}^2}\,,
\end{equation}
with $x^{(A)}_3=0$ closing each chain.

The two smallest cases with $\gcd(N,K)>1$ show the degeneration concretely.

For $N=4$, $K=2$ one has $d=K$, so each residue class is a single node and \eqref{esymmgcdlocus_app} reduces to $1/g_1^2=1/g_2^2$: the configuration exists only on the equal-coupling locus. Writing $t=1/g^2$ there, the flat direction is $u_1=-u_2\equiv c$ and
\begin{equation}\label{esymm42_app}
x^{(1)}=\big(3t+c,\ 4t,\ 3t+c\big)\,,\qquad x^{(2)}=\big(3t-c,\ 4t,\ 3t-c\big)\,,\qquad |c|<3t\,.
\end{equation}
For the stabilizer, $K=2$ makes $\alpha_{a-2}=\alpha_a$, so \eqref{esymmstabelt_app} becomes $g_a=\mathrm{diag}\big(e^{i\alpha_a},e^{i\alpha_{a-1}},e^{i\alpha_a},e^{i\alpha_{a-1}}\big)$ and both determinant conditions read $2(\alpha_1+\alpha_2)\in2\pi\mathbb{Z}$. The difference $\alpha_1-\alpha_2$ is left free, while $\alpha_1+\alpha_2$ is restricted to the two values $0$ and $\pi$, giving $U(1)\times\mathbb{Z}_2$ as in \eqref{esymmstabgcd_app}.

For $N=2$, $K=4$ the classes are $\{1,3\}$ and $\{2,4\}$, so \eqref{esymmgcdlocus_app} reads $1/g_1^2+1/g_3^2=1/g_2^2+1/g_4^2$, a hyperplane containing the equal-coupling locus; on that locus the flat direction gives
\begin{equation}\label{esymm24_app}
\big(x^{(1)}_0,x^{(2)}_0,x^{(3)}_0,x^{(4)}_0\big)=\big(t+c,\ t-c,\ t+c,\ t-c\big)\,,\qquad |c|<t\,.
\end{equation}
Here \eqref{esymmdet_app} reads $\alpha_a+\alpha_{a-1}\in2\pi\mathbb{Z}$; differencing gives $\alpha_3=\alpha_1$ and $\alpha_4=\alpha_2$, and the surviving condition $\alpha_1+\alpha_2\in2\pi\mathbb{Z}$ leaves the single free phase $\alpha_1$. The stabilizer is therefore $U(1)$, with no discrete factor because $N/d=1$.

In both cases the cyclic relabeling acts on the flat direction by $c\mapsto-c$, so only the subgroup generated by $a\mapsto a+d$ survives: the moduli are cyclically symmetric only at $c=0$, where $x^{(A)}_J=t\,(J+1)(N-1-J)$ for every $A$. Even there, the full datum is symmetric only for a compatible choice of the gauge-invariant relative phases of the blocks. For $N=2$, $K=4$ and $c=0$, writing the non-zero entry of $q_a$ as $\sqrt t\,e^{i\vartheta_a}$, the combination $e^{i(\vartheta_1-\vartheta_2+\vartheta_3-\vartheta_4)}$ is gauge invariant and is mapped to its inverse by the cyclic relabeling, so that only its values $\pm1$ give a symmetric datum.

%==========================================================
\subsection{The determinant of \texorpdfstring{$B$}{B} for saturating data}\label{adetB}
%==========================================================

We show that every saturating single-pole datum with all ranks one has
\(\det B\in\{0,N\}\). By \eqref{erankbound} and the saturation of
\eqref{ebudget}, every \(q_a\) has rank exactly \(N-1\). Exactly one slot of the
node \(a\) then has no outgoing block in \(q_a\), and exactly one slot of the node
\(a+1\) has no incoming one, so each node carries exactly one head and one tail
of a family, a free eigenvalue being both. We label each family by the node \(s\)
of its head, and denote by \(\sigma(s)\equiv s+L_s+1\) mod \(K\) the node that
follows its tail, so that \(\sigma\) is a permutation of the nodes. Since a
family occupies consecutive nodes,
\(B_{as}-B_{a-1,s}=\delta_{a,s}-\delta_{a,\sigma(s)}\), that is,
\begin{equation}\label{eDB_app}
D\,B=\mathds 1-P_\sigma\,,\qquad B\,\mathbf 1=N\,\mathbf 1\,,
\end{equation}
where \((Dv)_a\equiv v_a-v_{a-1}\), \(P_\sigma\) is the permutation matrix of
\(\sigma\), \(\mathbf 1=(1,\ldots,1)^T\), and the second relation is
\eqref{eBmatrix}. If \(\sigma\) has more than one cycle, the kernel of
\(\mathds 1-P_\sigma\) has dimension at least two, while that of \(D\) is spanned
by \(\mathbf 1\), so \(B\) is singular. If \(\sigma\) is a single cycle, all the
cofactors of \(\mathds 1-P_\sigma\) and of \(D\), which is of the same form, are
equal to one, so that
\(\operatorname{adj}(B)\,\mathbf 1\mathbf 1^T=\operatorname{adj}(DB)=\mathbf 1\mathbf 1^T\).
Hence \(\operatorname{adj}(B)\,\mathbf 1=\mathbf 1\), and
\(\det B\,\mathbf 1=B\operatorname{adj}(B)\,\mathbf 1=N\,\mathbf 1\), with the
columns of \(B\) ordered by the node of the head.

For the maximal-winding solution \(\sigma(s)=s+1\) is a single cycle, and
\(\det B=N\). For the maximally symmetric one \(\sigma(s)=s+N\) has
\(\gcd(N,K)\) cycles, so that \(\det B=0\) for \(\gcd(N,K)>1\), in agreement with
Appendices~\ref{awindstabilizer} and~\ref{asymmstabilizer}.

%==========================================================
\section{Existence chambers}\label{achambers}
%==========================================================

For the maximal-winding solution and for the maximally symmetric solution
with $\gcd(N,K)=1$, the moduli obtained in Appendix~\ref{ad6bc} are linear in
the inverse couplings, and a solution defines a boundary datum of the given
discrete type only where all of them are strictly positive. Here we determine
these regions. Throughout we write $t_a\equiv1/g_a^2>0$. Since the moduli do
not depend on the $\theta$-angles, the existence regions are determined by the
positive orthant $t\in\mathbb{R}^K_{>0}$, and statements about the conformal
manifold below refer to its image in these variables.

%==========================================================
\subsection{Maximal winding}\label{awindpositivity}
%==========================================================

The copy of the maximal-winding solution with filled node $A\in\{1,\ldots,K\}$
(node labels understood mod $K$) is obtained from \eqref{ewindingmoduli_app} by
the cyclic relabeling $g_{1+j}\to g_{A+j}$:
\begin{equation}\label{eframemoduli_app}
x^{(A)}_{w,j}\equiv|C_{A+j,w}|^2=(N-1-w)\Big[K(w-1)\,G+2H+2K\,G_j\Big]-2N\,H_j\,,
\end{equation}
with $G_j=\sum_{i=0}^{j}t_{A+i}$, $H_j=\sum_{i=0}^{j}i\,t_{A+i}$, $G=G_{K-1}$ and
$H=H_{K-1}$. Two combinations control everything: the partial sums
\begin{equation}\label{ePidef_app}
\Pi_j\equiv\sum_{i=0}^{j}(K-2i)\,t_{A+i}=K\,G_j-2H_j\,,\qquad \Pi_0=K\,t_A\,,
\end{equation}
and their full-cycle value
\begin{equation}\label{eDdef_app}
D_A\equiv\Pi_{K-1}=K\,G-2H=\sum_{i=0}^{K-1}(K-2i)\,t_{A+i}\,.
\end{equation}
In terms of the combination $S$ of \eqref{eSdef_app}, anchored at $A$, one has
$2S=D_A-K\,t_A$, and the two outermost links \eqref{eouterlinks_app} read
\begin{equation}\label{eouterD_app}
x^{(A)}_{0,0}=(N-1)\big(2K\,t_A-D_A\big)\,,\qquad x^{(A)}_{N-2,K-1}=(N-1)\,D_A\,.
\end{equation}

\paragraph{The chamber.}
The smallest modulus is always one of the two outermost links:
\begin{equation}\label{eminmoduli_app}
\min_{w,j}\,x^{(A)}_{w,j}=(N-1)\,\min\big(2K\,t_A-D_A,\;D_A\big)\,,
\end{equation}
for every $N\ge2$, $K\ge1$ and $t\in\mathbb{R}^K_{>0}$. Consequently the copy
with filled node $A$ exists, with all links non-zero, if and only if
\begin{equation}\label{echamber_app}
0<D_A<2K\,t_A\qquad\Longleftrightarrow\qquad |S|<\tfrac{K}{2}\,t_A\,,
\end{equation}
a convex cone cut out by two linear inequalities and independent of $N$.

To prove \eqref{eminmoduli_app}, substitute $2H_j=KG_j-\Pi_j$ and $2H=KG-D_A$ in
\eqref{eframemoduli_app}. Using
$w(N-1-w)+N-2-2w=(w+1)(N-2-w)$, this gives
\begin{equation}\label{erepresentation_app}
x^{(A)}_{w,j}=N\,\Pi_j-(N-1-w)\,D_A+K\Big[(w+1)(N-2-w)\,G_j+w(N-1-w)\big(G-G_j\big)\Big]\,.
\end{equation}
In the allowed range $0\le w\le N-2$ both terms in the square bracket are
non-negative, and since $G_j\ge t_A$ the bracket is at least
$(w+1)(N-2-w)\,t_A$. Moreover, the summands $(K-2i)\,t_{A+i}$ of $\Pi_j$ are
positive for $i<K/2$ and negative for $i>K/2$, so $\Pi_j$ first increases and
then decreases in $j$, and $\Pi_j\ge\min(\Pi_0,\Pi_{K-1})=\min(K\,t_A,D_A)$.
Together,
\begin{equation}\label{eboundmoduli_app}
x^{(A)}_{w,j}\ge N\min\big(K\,t_A,D_A\big)-(N-1-w)\,D_A+(w+1)(N-2-w)\,K\,t_A\,.
\end{equation}
If $D_A\le K\,t_A$, the right-hand side is
$(w+1)D_A+(w+1)(N-2-w)K\,t_A$. For $D_A\ge0$ this is at least
$(w+1)(N-1-w)D_A\ge(N-1)D_A$; for $D_A<0$, dropping the second term, it is at
least $(w+1)D_A\ge(N-1)D_A$. In both cases
$x^{(A)}_{w,j}\ge x^{(A)}_{N-2,K-1}$. If instead $D_A>K\,t_A$, the right-hand
side equals
$(N-1)\big(2K\,t_A-D_A\big)+w\big[(N-3-w)K\,t_A+D_A\big]$, and the square bracket
is at least $D_A-K\,t_A>0$, so $x^{(A)}_{w,j}\ge x^{(A)}_{0,0}$. This proves
\eqref{eminmoduli_app}.

For $K=1$ one has $D_A=t_A$, and the regular Nahm pole exists for all
couplings; for $K=2$ one has $D_A=2t_A$, and the solution exists for arbitrary
positive couplings. For $K\ge3$ the chamber of a given copy is a proper subset
of the conformal manifold: for $N=2$, $K=3$ and $A=1$ it is $|t_2-t_3|<3t_1$.
On a wall of the chamber an outermost link vanishes, and the wound chain
splits into two shorter families.

\paragraph{Covering of the conformal manifold.}
At every point $t\in\mathbb{R}^K_{>0}$ at least one of the $K$ copies exists with
all links non-zero. Passing from the copy $A$ to the copy $A+1$ in
\eqref{eDdef_app} gives
\begin{equation}\label{eincrement_app}
D_{A+1}-D_A=2G-2K\,t_A\,,
\qquad\text{hence}\qquad
D_A<2K\,t_A\ \Longleftrightarrow\ D_{A+1}<2G\,,
\end{equation}
and, since $\sum_{A=1}^{K}t_{A+i}=G$ for every $i$ and
$\sum_{i=0}^{K-1}(K-2i)=K$,
\begin{equation}\label{emean_app}
\sum_{A=1}^{K}D_A=K\,G\,,
\end{equation}
so the average of the $D_A$ is $G>0$. Let $A_0$ be a copy maximizing $D_A$, so
that $D_{A_0}\ge G$. Not all the $D_A$ can be $\ge2G$, so there is a first
index $A^\ast+1$ in the cyclic sequence $A_0+1,A_0+2,\ldots$ with
$D_{A^\ast+1}<2G$. All the indices between $A_0$ and $A^\ast+1$ have
$D_A\ge2G$, hence $D_{A^\ast}\ge G>0$, while $D_{A^\ast}<2K\,t_{A^\ast}$ by
\eqref{eincrement_app}. The copy $A^\ast$ therefore satisfies
\eqref{echamber_app}. All $K$ copies exist throughout the equal-coupling locus $t_a=t>0$,
where $D_A=K\,t$ for every $A$, while for $N=2$, $K=3$ and
$t=(1,\epsilon,\epsilon)$ with small $\epsilon$ only the copy $A=1$ survives.
Since the stabilizer computation of Appendix~\ref{awindstabilizer} only needs
all links to be non-zero, at every point of the conformal manifold some copy of
the maximal-winding solution exists with stabilizer exactly $\mathbb{Z}_N$.

%==========================================================
\subsection{Maximally symmetric case}\label{asympositivity}
%==========================================================

For \(\gcd(N,K)=1\), the moduli \eqref{esymmmoduli_app} are also linear in the
\(t_a\), and the configuration defines a boundary datum only where all the
\(K(N-1)\) moduli \(x^{(A)}_J\) are positive, a convex cone in the conformal
manifold. It contains the entire equal-coupling locus: for \(t_a=t>0\) one has
\(u_a=0\), and \eqref{esymmmoduli_app} gives
\begin{equation}\label{esymmsympoint_app}
x^{(A)}_J=t\,(J+1)(N-1-J)>0\,.
\end{equation}
Unlike the chambers of Appendix~\ref{awindpositivity}, however, the cone is not
enlarged by the cyclic relabeling, which maps the configuration to itself, so
there is no analogue of the covering argument. In all the cases with \(K\ge3\)
that we examined the cone is a proper subset of the conformal manifold. For
\(N=2\), \(K=3\), equation \eqref{esymmN2_app} gives
\(x^{(A)}_0=t_A+t_{A+1}-t_{A+2}\), so the cone is cut out by the triangle
inequalities on \((t_1,t_2,t_3)\); for \(N=4\), \(K=3\), the moduli
\(x^{(A)}_0\) of \eqref{esymm43_app} fail to be positive, for instance, at
\((t_1,t_2,t_3)=(1,\tfrac12,\tfrac1{10})\), where \(x^{(3)}_0=-\tfrac15\).

\addcontentsline{toc}{section}{Bibliography}
 
	\bibliography{references}
	\bibliographystyle{utphys}

\end{document}